\documentclass[manuscript]{aastex}
\usepackage{multirow}
\usepackage{rotating}

\def\lesssim{\mathrel{\hbox{\rlap{\hbox{\lower4pt\hbox{$\sim$}}}\hbox{$<$}}}}
\def\gtrsim{\mathrel{\hbox{\rlap{\hbox{\lower4pt\hbox{$\sim$}}}\hbox{$>$}}}}

\def\ltsima{$\;\buildrel < \over \sim \;$}
\def\simlt{\lower.5ex \hbox{\ltsima}}
\def\gtsima{$\;\buildrel > \over \sim \;$}
\def\simgt{\lower.5ex \hbox{\gtsima}}
\def\lesssim{\mathrel{\hbox{\rlap{\hbox{\lower4pt\hbox{$\sim$}}}\hbox{$<$}}}}
\def\gtrsim{\mathrel{\hbox{\rlap{\hbox{\lower4pt\hbox{$\sim$}}}\hbox{$>$}}}}
\def\gtrless{\mathrel{\hbox{\rlap{\hbox{\lower4pt\hbox{$<$}}}\hbox{$>$}}}}
\def\rightleftharpoons{\mathrel{\hbox{\rlap{\hbox{\raise2pt\hbox{$\rightharpoonup$}}}\hbox{$\leftharpoondown$}}}}
\def\notrightleftharpoons{\mathrel{\hbox{\rlap{\hbox{\raise1.5pt\hbox{$\;\mid$}}}\hbox{$\rightleftharpoons$}}}}
\def\dbar{\mathrel{\hbox{\rlap{\hbox{\raise3pt\hbox{$-$}}}\hbox{$d$}}}}
\def\hbar{\mathrel{\hbox{\rlap{\hbox{\raise3pt\hbox{$-$}}}\hbox{$h$}}}}
\def\nubar{\mathrel{\hbox{\rlap{\hbox{\raise2pt\hbox{$-$}}}\hbox{$\nu$}}}}
\def\lambdabar{\mathrel{\hbox{\rlap{\hbox{\raise2pt\hbox{$-$}}}\hbox{$\lambda$}}}}
\def\BbbV{\mathrel{\hbox{\rlap{\hbox{\raise2.5pt\hbox{${\rm v}$}}}\hbox{${\rm V}$}}}}
\def\BbbT{\mathrel{\hbox{\rlap{\hbox{\raise2pt\hbox{${\rm T}$}}}\hbox{${\rm T}$}}}}

\def\dddot{\hbox{\rlap{\hbox{\raise 8pt\hbox{${\bf ...}$}}}\hbox{$$}}}

\def\ltsima{$\;\buildrel < \over \sim \;$}
\def\simlt{\lower.5ex \hbox{\ltsima}}
\def\gtsima{$\;\buildrel > \over \sim \;$}
\def\simgt{\lower.5ex \hbox{\gtsima}}

\def\lsun{L$_\odot$}

\shorttitle{Strong C$^+$ emission in star forming galaxies at z\~1-2}
\shortauthors{Brisbin et al}

\begin{document}

\title{Strong C$^+$ emission in galaxies at z$\sim$1-2: \\
Evidence for cold flow accretion powered star formation in the early Universe}
\author{Drew Brisbin\altaffilmark{1,2}, Carl Ferkinhoff\altaffilmark{3,2}, Thomas Nikola\altaffilmark{2}, Stephen Parshley\altaffilmark{2}, Gordon J. Stacey\altaffilmark{4}, Henrik Spoon\altaffilmark{2}, Steven Hailey-Dunsheath\altaffilmark{5}, Aprajita Verma\altaffilmark{6}}

\email{dbrisbin@nrao.edu}

\altaffiltext{1}{National Radio Astronomy Observatory, Charlottesville, VA 22903, USA}
\altaffiltext{2}{Center for Radiophysics \& Space Research, Cornell University, Ithaca, NY 14853, USA}
\altaffiltext{3}{Max-Planck-Institut f\"{u}r Astronomie, Konigstuhl 17, D-69117 Heidelberg, Germany}
\altaffiltext{4}{Department of Astronomy, Cornell University, Ithaca, NY 14853, USA}
\altaffiltext{5}{California Institute of Technology, Mail Code 301-17, 1200 E. California Blvd., Pasadena, CA 91125, USA}
\altaffiltext{6}{University of Oxford, Oxford Astrophysics, Denys Wilkinson Building, Keble Road, Oxford, OX1 3RH, UK}

\begin{abstract}
We have recently detected the [CII] 157.7 $\mu$m line in eight star forming galaxies at redshifts 1 to 2 using the redshift(z) Early Universe Spectrometer (ZEUS). Our sample targets star formation dominant sources detected in PAH emission. This represents a significant addition to [CII] observations during the epoch of peak star formation. We have augmented this survey with observations of the [OI] 63 $\mu$m line and far infrared photometry from the PACS and SPIRE Herschel instruments as well as Spitzer IRS spectra from the literature showing PAH features. Our sources exhibit above average gas heating efficiency, many with both [OI]/FIR and [CII]/FIR $\sim$1\% or more. The relatively strong [CII] emission is consistent with our sources being dominated by star formation powered PDRs, extending to kpc scales. We suggest that the star formation mode in these systems follows a Schmidt-Kennicutt law similar to local systems, but at a much higher rate due to molecular gas surface densities 10 to 100 times that of local star forming systems.  The source of the high molecular gas surface densities may be the infall of neutral gas from the cosmic web. In addition to the high [CII]/FIR values, we also find high [CII]/PAH ratios and, in at least one source, a cool dust temperature. This source, SWIRE 4-5, bears a resemblance in these diagnostics to shocked regions of Stephan's Quintet, suggesting that another mode of [CII] excitation in addition to normal photoelectric heating may be contributing to the observed [CII] line.
\end{abstract}


\section{Introduction}
With recent developments in submillimeter spectroscopy including the conclusion of extensive observations with the Herschel Space Observatory\footnote{{\it Herschel} is an ESA space observatory with science instruments provided by European-led Principal Investigator consortia and with important participation from NASA.} \citep{pil2010}, ALMA beginning full science operations, and ongoing developments with other ground-based interferometers and large single disk telescopes, the study of the redshifted Universe in the far infrared (FIR) has come into its prime. 

Emission from ionized carbon is one important tool for FIR studies of early galaxies. Carbon is the fourth most abundant element in the Universe, and it takes 11.3 eV photons to form C+, so the low-lying (91 K above ground) 157.7 $\mu$m [CII] fine-structure line was long ago predicted to be the dominant coolant of the neutral ISM \citep{dal1972}.  The [CII] line is also usually optically thin and suffers very little extinction, so it is an excellent probe of the properties of the atomic gas heated by the far-UV (FUV) (6 - 13.6 eV) flux in galaxies.  Indeed, the first [CII] detections from local galaxies revealed that the [CII] line can be the brightest single emission line from star forming galaxies, amounting to between 0.1 and 1\% of the total FIR luminosity\footnote{In this paper FIR refers to 42.5-122.5 $\mu$m. This is one of the most common conventions in the literature. To compare to other samples and models we will also make use of several other conventional wavelength ranges including 30-1000 $\mu$m, 40-500 $\mu$m, and 8-1000 $\mu$m where explicitly noted.} \citep{cra1985,sta1991}.

The [CII] line luminosity is closely correlated with CO emission. Thus, while some [CII] does arise in ionized gas, a picture has emerged where most ($\sim$70\%) arises from the warm, dense, neutral gas of photo-dissociation regions (PDRs) \citep{sta1991,obe2006,vas2010}.  The PDR heating is dominated by FUV radiation from nearby early type stars. Nearly all of the UV intercepted by dust is absorbed and converted to long wavelength thermal emission, giving rise to the FIR continuum.  A small fraction ($\sim$1\%) of the UV photons eject hot electrons from dust and PAH grains heating the gas which, in turn, collisionally excites the [CII] line \citep{tie1985}.

For moderate gas densities (n$<$10$^4$ cm$^{-3}$) and FUV field strengths (G$_0<$10$^3$)\footnote{G$_0$, the Habing flux, parameterizes the far-UV flux in terms of a typical ISM FUV flux, G$_0$=F$_{FUV}$/1.6$\times$10$^{-6}$ W m$^{-2}$} that are typical in star forming galaxies averaged over large scales, the [CII] line is the primary coolant for gas in PDRs.
The ratio of power in this line to the FIR continuum represents the gas heating as a fraction of the total UV power. It is a first measure of the `gas heating efficiency'. 
A more complete measure includes the [OI] 63 $\mu$m line, which, due to its higher critical density for thermalization (n$_{crit}$$\sim$4.7$\times$10$^5$ cm$^{-3}$, vs. 2.8$\times$10$^3$ cm$^{-3}$ for [CII]) and greater energy above ground for the emitting level (228 K), becomes the dominant PDR coolant at higher densities and  FUV field intensities \citep[eg.][]{pou2008,kau2006}.  Together, the two fine structure lines constrain n and G$_0$ for PDRs. However, since the gas densities for most galaxies are between 10$^3$ and 10$^5$ cm$^{-3}$, the [CII]/FIR ratio is in and of itself a good tracer of the heating efficiency \citep{sta2010}. In PDR models with moderate densities, [CII]/FIR is inversely proportional to G$_0$. ISO-based studies showed that [CII]/FIR tends to be an order of magnitude smaller in local Ultra Luminous Infrared Galaxies (ULIRGs; L$_{FIR}$$>$10$^{12}$\lsun) relative to normal star forming galaxies - a characteristic sometimes referred to as the `[CII] deficit' \citep{luh2003}. 
This relationship shows that, in the local Universe, ULIRGs are not simply scaled up normal galaxies. Something is fundamentally different about their star formation to cause this deficient [CII] emission.

\citet{sta2010} demonstrated that the [CII] deficit does not hold throughout the Universe. The observed [CII] deficit in the local Universe is only indicative of the underlying star formation conditions in local ULIRGs. PDR models demonstrate that low [CII]/FIR on a galactic scale indicates very intense UV fields (G$_0\gtrsim$10$^4$) in star forming media. In local galaxies the star bursting episodes that give rise to such intense UV fields generally occur in ULIRGs. Recent major mergers leading to localized and very intense star formation are the source of extreme luminosity in these local systems. During the epoch of peak star formation, ULIRGs make up a larger fraction of the total star formation activity, and thus we might expect a continuation of the locally observed [CII] deficit. \citet{sta2010} showed, however, that at z=1-2 systems with extreme (ULIRG and HyLIRG) luminosities do not necessarily have suppressed [CII]/FIR ratios or extreme UV fields. \citet{sta2010} instead finds that the high luminosity star formation dominated systems in this epoch have very extended star formation regions with more moderate FUV fields. This is consistent with complementary lines of evidence supporting extended star formation \citep{pop2006,far2008,men2009} and has contributed to a paradigm shift in the accepted nature of star formation in the early Universe. Rather than being powered solely by major mergers, a significant population of ULIRGs in the early Universe are forming stars in a mode similar to normal local galaxies. The observational data could be explained by star formation proceeding through accretion of gas from the intergalactic medium which builds to surface densities $\sim$10-100 times that of the Milky Way. This is often seen in sub-galactic star formation regions nearby and is adequately described by the Schmidt-Kennicutt star formation law \citep{sch1959,ken1998}. We suggest that in these galaxies at high redshift, similar star formation is occurring on a near galaxy-wide scale leading to moderate UV field intensities (G$_0\sim$100-1000,) but very large luminosities due to their very large size. This is consistent with several recent findings by \citet{tac2010,dad2010,ivi2011,rie2011}, and \citet{hod2012}, which all show widespread CO emission $\sim$several kpc in extent and indicate high molecular gas fractions in normal star forming galaxies at a similar epoch.
Although they are certainly present in the early Universe \citep[eg.][]{fer2014}, major mergers are not required to explain the very large intrinsic luminosities (L$>$10$^{12}$L$_\odot$) in this epoch.

Motivated by the results of our previous work, we have undertaken an expanded [CII] survey of eight z=1-2 sources with the redshift (z) and Early Universe Spectrometer (ZEUS) on CSO; a survey which we have augmented with [OI] 63 $\mu$m observations from the Herschel PACS instrument \citep{pog2010}. This survey represents the continuation of the original work by \citet{sta2010}, which placed an equal emphasis on sources whose luminosities were AGN dominant, star formation dominant (SF-D), or of mixed nature. 
In this follow up survey we have focused on SF-D systems. We confirm that the [CII] deficit is not a ubiquitous trait in star formation powered ULIRGs in the redshift 1-2 epoch. We also find moderately intense (G$_0\sim$10$^{2-3}$) UV fields distributed over very large (several kpc) scales within galaxies.  A similar analysis of [OI] 63 $\mu$m in the sources from the original \citet{sta2010} sample will be found in \citet[][in prep.]{hai2014}.

The characterization of our sources as SF-D is based on evidence from the literature without consideration of our [CII] observations. It can generally be interpreted as a characterization of the dominant power source for the total infrared (TIR) (8-1000 $\mu$m) luminosity. AGN dominant sources have TIR power dominated by the mid-IR (MIR), while SF-D sources have TIR SEDs dominated by the FIR band. Due to the inhomogeneous nature of background data on our varied source set, however, the precise criteria for characterization is not uniform.   

In addition to [CII] and [OI], many of our sources have been observed photometrically with PACS or SPIRE \citep{gri2010}, either as part of our own OT2 program, the Herschel Multi-tiered Extragalactic Survey (HerMES) \citep{oli2012}, or {\it SEDs and energetics of lensed UV-bright high redshift galaxies} (PI Dieter Lutz, Obs ID 13422210503, 1342221289, 1342221290, 1342221291, 1342221292.) We have used these data along with photometry collected from the literature to provide uniformly processed SEDs. In choosing our survey sample we emphasized luminous sources with PAH emission detected in the MIR with the Spitzer Infrared Spectrograph (IRS)\footnote{The IRS was a collaborative venture between Cornell University and Ball Aerospace Corporation funded by NASA through the Jet Propulsion Laboratory and the Ames Research Center.} spectrometer \citep{hou2004}. This selection criteria strongly biases our sample towards SF-D systems. We have used these IRS spectra to analyze the presence and nature of PAHs.

The structure of the paper is as follows: in section \ref{sec:ciiinstrumentation} we present our observations, including the overall survey methodology; in section \ref{sec:ciinotes} we discuss notes on individual sources; in section \ref{sec:ciiresults} we present our data analysis and results for our global sample and individual sources; in section \ref{sec:distscale} we discuss the physical interpretation of our results and their implications for the z$>$1 Universe; and finally in section \ref{sec:ciiconclusions} we make concluding remarks.

\section{Instrumentation and Observations}\label{sec:ciiinstrumentation}
We conducted [CII] observations using ZEUS at the 10.4m Caltech Submillimeter Observatory (CSO) on Mauna Kea. ZEUS is well described in the literature \citep{sta2007,hai2009thesis} so we only briefly describe it here. ZEUS is an echelle grating spectrometer designed to operate over the 350 and 450 $\mu$m telluric transmission windows. It has a 1$\times$32 pixel thermistor sensed bolometer detector array that yields a 32 element spectrum split into the 350 and 450 $\mu$m bands for a single beam. The detector array and optics are designed to maximize sensitivity to broad lines of width $\sim$a few hundred km s$^{-1}$ - well matched for detecting emission lines from distant galaxies. The resolving power varies by wavelength but is $\sim$1000 (velocity resolution $\sim$300 km s$^{-1}$,)  with each pixel sampling one spectral resolution element.

Observations of eight sources took place over three observing runs in January 2010, January 2011, and March 2011. We used a standard chop-nod observing mode with a chopper frequency of 2 Hz, and a chopper throw of 30''. Calibration, including beam size, point-source coupling, and flux calibration, was determined by observations of Uranus, which is assumed to emit like a blackbody with temperatures 61 and 73 K within our 450 and 350 $\mu$m bands respectively \citep{hil1985}. We spectrally flat-fielded our observations based on observations of a cold chopped source. Source and system parameters relevant for each observation are given in Table \ref{tab:obslog}. The ZEUS/CSO beam size is 10.5'' at 350 $\mu$m and 11.5'' at 450 $\mu$m, and we estimate that typical pointing errors are less than 3'' and systematic calibration errors are less than 30\%. The wavelength calibration is good to about half a pixel, or 100-160 km s$^{-1}$.

We observed six sources in [OI] 63 $\mu$m  using the PACS spectrometer onboard the Herschel Space Observatory. We observed using line scans in chop-nod mode with a small chopper throw. Each [OI] observation had a full integration time (including on and off source integration but not including instrumental overheads) of just under an hour (57.3 minutes,) except for SWIRE 3-14 which is our faintest [CII] source, on which we integrated three times longer (172 minutes.) We reduced the data using the standard pipeline in the Herschel Interactive Processing Environment (HIPE v10.3.0) \citep{ott2010}, with minimal post-pipeline processing described in section \ref{sec:resultslinespectra}. Observations were carried out between May and October 2012 (ODs 1253, 1253, 1188, 1152, 1132, and 1118 covering Obs. IDs 1342253587, 1342253586, 1342249495, 1342247784, 1342247131, and 1342246639 respectively. The [OI] line was observed in a seventh source, SDSS J12, as part of the open time program {\it Herschel Extreme Lensing Line Observations (HELLO)} by PI Sangeeta Malhotra using the PACS chop-nod mapping mode (Obs ID 1342246395.) 

All sources except for SMM J03 lie in regions surveyed by either the HerMES or {\it SEDs and energetics of lensed UV-bright high redshift galaxies} projects from which we obtained photometric measurements. We conducted photometric observations of SMM J03 using the PACS blue and green cameras (which provide simultaneous red coverage resulting in observations at all three PACS wavebands - 70, 100, and 160 $\mu$m). We took two scan maps with each camera, one at an orientation of 110 and one at 70 degrees with respect to the camera array for optimal spatial coverage and minimal CCD latency effects after cross stitching. Scans were performed at medium speed, with 3 arcminute scan legs, a cross scan step of 4 arcseconds and a total of 10 scan legs. Total integration time including all four scan maps (without instrumental overhead,) was six minutes. The observations are noted in the Herschel archive as Obs IDs 1342249159, 1342249158, 1342249157, 1342249156 occurring on OD 1181.

\setlength\tabcolsep{0.1cm}
\begin{table}[htb]\small
\caption[ZEUS observing log]{ZEUS Observing log. Average line of sight transmission is indicated by t$_{los}$. In the text we refer to sources by their names in parentheses.}
\vspace{2.5mm}
\begin{center}
\begin{tabular}{ c c c c c c}
	\hline
		Source 				 &  	RA 			&	Dec.  		&	z$_{[CII]}$	&  Obs. Dates	& t$_{los}$     \\ \hline \hline
MIPS 22530					&	17h23m03.3s	&	59d16m00.2s 	&	1.9501	&	3/15/2011	&	15.4\%	\\ \hline
SWIRE3 J104343.93+571322.5	&	\multirow{2}{*}{10h43m43.9s}	&	\multirow{2}{*}{57d13m22.5s}	&	\multirow{2}{*}{1.7348}	&	\multirow{2}{*}{3/18/11}	&	\multirow{2}{*}{23.9\%}	\\
(SWIRE 3-9) \\ \hline
SWIRE3 J104514.38+575708.8	&	\multirow{2}{*}{10h45m14.4s}	&	\multirow{2}{*}{57d57m08.8s}	&	\multirow{2}{*}{1.7795}	&	1/07/10,					&	\multirow{2}{*}{31\%}	\\ 
(SWIRE 3-14) 				&								&								&							&	3/17/11					\\\hline
SWIRE3 J104632.93+563530.2	&	\multirow{2}{*}{10h46m32.9s}	&	\multirow{2}{*}{56d35m30s}	&	\multirow{2}{*}{1.771}	&	\multirow{2}{*}{12/31/10}&	\multirow{2}{*}{22.5\%}	 \\ 
(SWIRE 3-18) \\\hline
SMM J030227.73 +000653.5		&	\multirow{2}{*}{03h02m27.7s}	&	\multirow{2}{*}{00d06m52.0s}	&	\multirow{2}{*}{1.4076}	&	\multirow{2}{*}{1/4/11}	&	\multirow{2}{*}{16\%}	\\
(SMM J03)	\\ \hline
SWIRE4 J104427.52+584309.6	&	\multirow{2}{*}{10h44m27.52s}&	\multirow{2}{*}{58d43m09.6s}	&	\multirow{2}{*}{1.7560}	&	\multirow{2}{*}{1/24/11}	&	\multirow{2}{*}{36\%} 	\\ 
(SWIRE 4-5) \\\hline
SWIRE4 J104656.46+590235.5	&	\multirow{2}{*}{10h46m56.46s}&	\multirow{2}{*}{59d02m35.5s}	&	\multirow{2}{*}{1.8540}	&	\multirow{2}{*}{3/16/11}	&	\multirow{2}{*}{30\%}	\\ 
(SWIRE 4-15) \\\hline
SDSS J120602.09+514229.5		&	\multirow{2}{*}{12h06m01.71s}&	\multirow{2}{*}{51d42m27.6s} &	\multirow{2}{*}{1.9985}	&	\multirow{2}{*}{3/17/11}	&	\multirow{2}{*}{21.5\%}	\\
(SDSS J12)	\\ 
\end{tabular}
\end{center}
\label{tab:obslog}
\end{table}
\setlength\tabcolsep{2.12mm}

\section{Notes on observations of individual systems}\label{sec:ciinotes}

Several of our sources were initially discovered as part of the Spitzer Wide-area InfraRed Extragalactic (SWIRE) survey \citep{lon2003}, which undertook deep photometric observations of several different fields. High-z candidates showed bumps in the 4.5, or 5.8 $\mu$m IRAC bands or the MIPS 24 $\mu$m band, indicating a redshifted stellar photospheric 1.6 $\mu$m feature or PAH emission at z$\sim$2. Both types of features strongly suggest star formation. The 1.6 $\mu$m feature particularly selects for late type evolved stars, but it can be overwhelmed by a strong AGN contribution, so its presence in a FIR bright galaxy suggests star formation as the dominant power source \citep{far2006,far2008,lon2009,hua2009,des2009}.

All of our sources have MIR IRS spectra available which show strong PAH emission. The equivalent width (EW) of the 6.2 $\mu$m feature is often used as an AGN diagnostic. Sources with AGN dominating the MIR have a hot dust continuum which can overwhelm PAH emission and suppress the EW. Although our sources have PAH equivalent widths reported in various published works,  this measurement is highly sensitive to the details of the PAH spectrum fitting method used (especially so in spectra with low signal-to-noise and faint continuum as in our sources.) Therefore we cannot simply compare the published values of EW's in these sources, as they have been arrived at through diverse fitting methods.

Instead we adopt the method described by \citet{bra2006}, of integrating the PAH 6.2 $\mu$m feature above a spline-interpolated local continuum and using the spline continuum to determine the EW. This fit is performed in the rest frame of the source. The continuum spline is fit to anchor points located in regions of the spectra straddling, but relatively unaffected by, the PAH emission features. In general we used the same anchor points as those used by \citet{sti2013}, but we inspected each spectrum individually and adjusted the anchor points where necessary to ensure no PAH emission was included. Since the resulting EW is especially sensitive to the exact integration range for the feature as well as the spline fit, we fit each spectrum several times, allowing the anchor points and integration range (nominally 5.94 to 6.56 $\mu$m for the 6.2 $\mu$m feature) to vary slightly, and used the standard deviation of these fits to characterize the error.

This method has been used by several authors, including \citet{far2008} who included three of our sources among their sample (SWIRE 3-9, SWIRE 3-14, and SWIRE 3-18.) Below we note the published EW for those three sources and our own measured EW for the other five sources. Cutoffs of EW$_{6.2}\sim$0.3 - 0.5 $\mu$m have been used to identify star formation dominated systems as those with greater EW and AGN dominated systems as those with smaller EW \citep{sti2013,dia2013}. The generally large (though uncertain) EW's we observe in our sample are therefore consistent with, though not strong evidence for, star formation dominance over AGN. We caution that the EW's for many of our sources have uncertainties of order the EW itself. For two of the sources published in \citet{far2008}, the uncertainty actually exceeds the EW. In these cases the error is dominated by the uncertainty in identifying the faint continuum. Although we quote the published error, the true uncertainty is likely asymmetric skewing towards higher values since the continuum might easily be weaker than measured, buried underneath residual PAH emission, whereas a stronger continuum should be easier to detect and would lower the measured equivalent width.

\textit{SWIRE 4-15}
IRS observations in \citet{fio2010} (source L15) showed strong PAH features and indicated a (PAH determined) redshift, z=1.85 $\pm$ 0.01. We measure EW$_{6.2}$=0.97$\pm$0.60 $\mu$m. Photometry from Spitzer and SCUBA showing no significant presence of hot dust indicates a SF-D system with little to no contribution from AGN.

\textit{SWIRE 4-5}
IRS follow up of this source revealed strong PAH features and a redshift z=1.750$\pm$0.007 \citep[source L5,][]{fio2010}. We measure EW$_{6.2}$=0.51$\pm$0.30 $\mu$m.

\textit{SWIRE 3-14}
\citet{far2008} find a PAH determined redshift of z=1.78$\pm$0.02 and EW$_{6.2}$=1.26$\pm$1.52 $\mu$m.

\textit{SWIRE 3-9}
Another SWIRE source suspected of being at z$\sim$2, the MIR spectrum acquired by \citet{far2008} showed this to have strong PAH features with a gentle continuum slope characteristic of a SF-D system, with no contribution from an AGN. They find a PAH derived redshift of z=1.71$\pm$0.02 and EW$_{6.2}$=1.88$\pm$2.14 $\mu$m.

\textit{SWIRE 3-18} This source was selected for MIR spectral follow up by \citet{far2008}. It's MIR spectrum shows strong PAH features indicative of SF dominance. \citet{far2008} find EW$_{6.2}$=3.96$\pm$1.49 $\mu$m. Prior to our [CII] detection, only a PAH determined redshift of z=1.76$\pm$0.02 had been established. Our [CII] detection refines this to 1.771. Although it has been observed with IRS, little other data is known for this source. Our interest in it arose too late to include it in our [OI] Herschel survey.

\textit{SMM J03}
Based on optical and UV lines probing the ionized gas, SMM J03 is thought to have an AGN component \citep{swi2004,tak2006}. Looking at the UV spectral diagnostics along with the bolometric luminosity, however, \citet{cha2005} found that the AGN contributes insignificantly to the overall luminous energy. We therefore consider it tentatively to be SF-D, with a potential minor contribution from an AGN. This is a well studied submillimeter galaxy (SMG) with extensive photometry available in the literature in the optical, NIR, and radio regime. Rest frame optical spectroscopy reveals a redshift of z=1.4076$\pm$0.0002 \citep{swi2004}. We measure EW$_{6.2}$=0.41$\pm$0.28 $\mu$m.

\textit{SDSS J12} (`The Clone')
This is a UV bright lensed source with a magnification $\sim$ 27 \citep{lin2009}. It was discovered as the counterpart to the lensing galaxy at z=0.4 in the Sloan Digital Sky Survey (SDSS). An analysis of rest frame optical lines, [OIII], H$\beta$, [NII], and H$\alpha$ by \citet{hai2009} found SDSS J12 to share characteristics of local star forming systems, but with stronger ionized emission, indicating particularly vigorous star formation. MIR follow up by \citet{fad2010} showed strong PAH emission. SDSS J12 shows strong [S IV] in emission and has a steeply rising MIR continuum. Both of these features may be associated with AGN presence, however, based on the strong PAH features and lack of other highly ionized lines such as [Ne VI] in the MIR spectrum \citet{fad2010} conclude that AGN contribution is minimal. We measure EW$_{6.2}$=0.47$\pm$0.16 $\mu$m. We consider it tentatively to be SF-D with possible AGN contribution. Several spectral lines observed in emission from SDSS J12 have indicated redshifts in the range z=1.9967-2.0026 \citep{lin2009,hai2009}. 

\textit{MIPS 22530}
This source was selected as a potential z$\sim$2 galaxy from the Spitzer Extragalactic First Look Survey (XFLS) based on its 24 / 8 $\mu$m flux ratio, which indicated strong redshifted PAH emission \citep{yan2007}. Multiwavelength analysis by \citet{saj2008} finds a borderline indication of AGN powered radio emission, but shows a lack of AGN presence in optical lines. This, along with strong PAH emission, leads them to conclude it is a SF-D system. We measure EW$_{6.2}$=0.49$\pm$0.24 $\mu$m. We consider it to be SF-D with a possible modest AGN contribution. Keck spectroscopy by \citet{yan2007} reveal an optical redshift of z=1.9511.  

\section{Results and analysis}\label{sec:ciiresults}
Here we give an overview of the results that define several aspects of our sample as a whole, followed by analysis of the individual systems.

\subsection{Line spectra}\label{sec:resultslinespectra}
\begin{figure}[!htb]
	\centering
		\includegraphics[width=0.9\textwidth,trim=.2cm 0cm 0cm 0cm, clip=true]{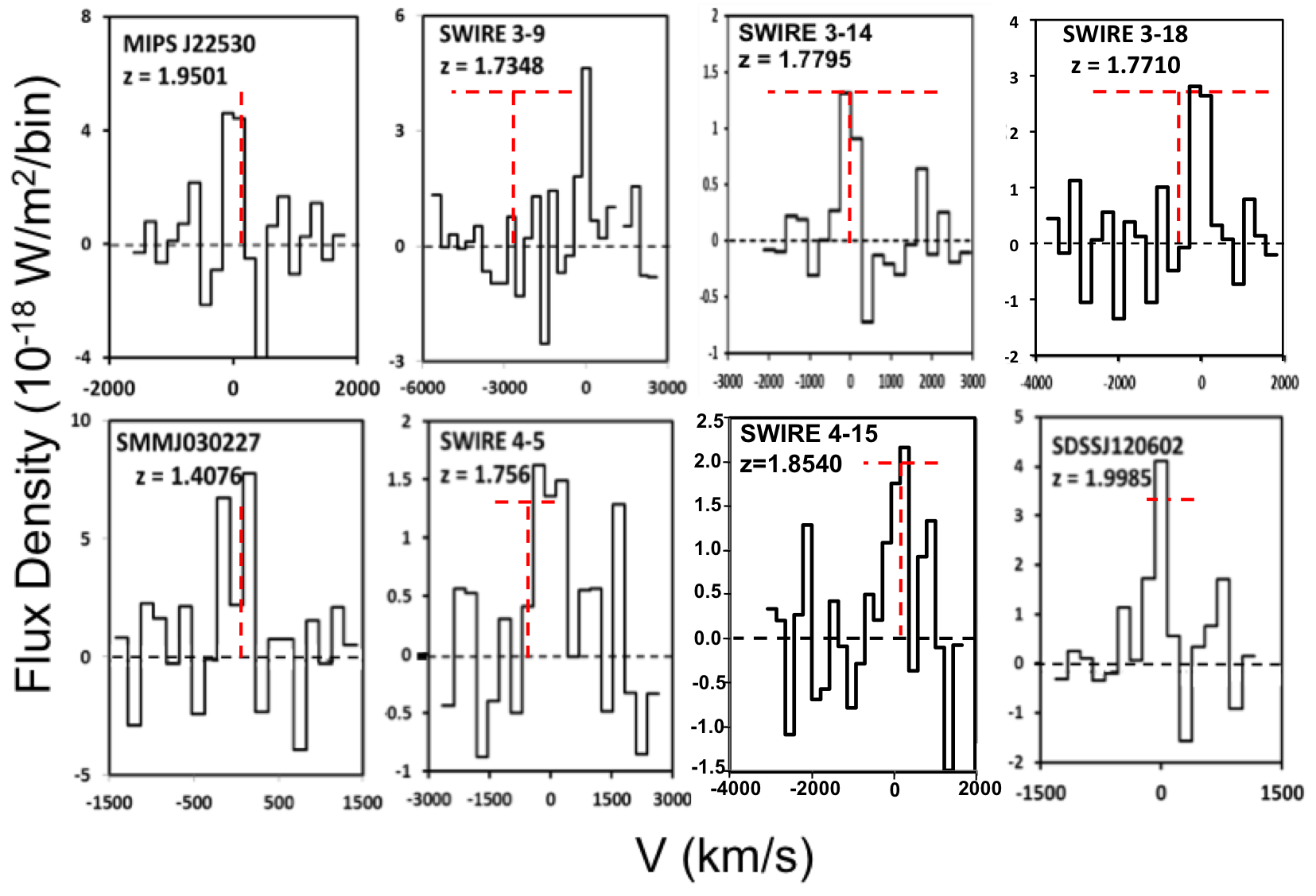}
	\caption[{ZEUS/CSO [CII] spectra}]{ZEUS/CSO [CII] spectra of the eight galaxies reported here.  Each spectral bin is one resolution element of the spectrometer and is statistically independent from its neighbors.  The velocity scale is centered on the [CII] line center. Red dashed lines indicate literature redshift values and uncertainty range where available. Note that the literature redshifts for MIPS J22530 and SMM J03 do not include uncertainty ranges, while for SDSS J12 a range of possible redshifts exist with no clearly preferred value.}
	\label{fig:ciispec}
\end{figure}

We have detected eight 1$<$z$<$2 sources in [CII] with ZEUS, representing a significant increase in the population of sources detected in this line during the epoch of peak star formation. In Figure \ref{fig:ciispec} we show the [CII] spectra from these eight sources.
Although some of our sources may be lensed, the apparent [CII] luminosities span a range of 0.73 - 5.44 $\times$10$^{10}$\lsun.

\begin{figure}[!htb]
	\centering
	\includegraphics[height=0.9\textwidth,angle=90,trim=.7cm 0cm 0cm 0cm, clip=true]{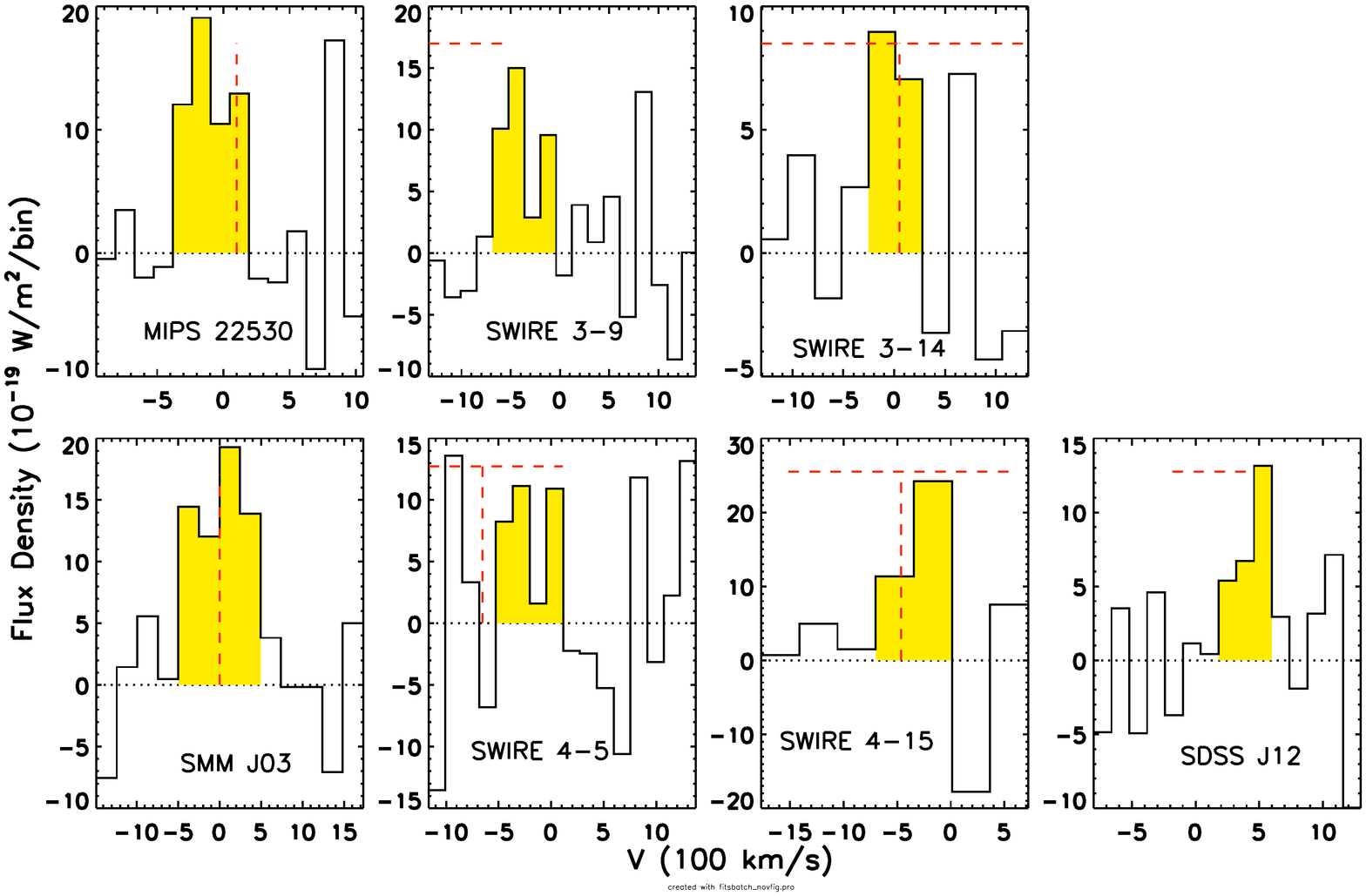}
	\caption[{Herschel PACS [OI] spectra}]{The PACS [OI] spectra from the six sources observed in our Herschel program and SDSS J12, observed by PI Sangeeta Malhotra. SWIRE 3-18 was not observed. Velocities are with respect to our [CII] line center. Region shaded in yellow indicates the area integrated to determine line flux. Red dashed lines indicate redshifts from the literature, as in Figure \ref{fig:ciispec}}
	\label{fig:oispec}
\end{figure}

The [OI] spectra for the six sources we observed in our Herschel PACS [OI] survey plus one observed by Sangeeta Malhotra (reduced here from archival data,) are plotted in Figure \ref{fig:oispec}. 
The [OI] spectra were reduced through standard methods using HIPE. In post-processing we rebinned them to resolutions $\sim$120 to 350 km s$^{-1}$ and fit a linear baseline to channels with no line emission. 
Line fluxes are tabulated in Table \ref{tab:sourceprops}.

The [OI] line is clearly detected in six of the seven sources, and marginally detected in SWIRE 4-5 (discussed in section \ref{sec:sw45}.) In a few sources we observe offsets from the [CII] line velocity by $\sim$250 km s$^{-1}$. Both the velocity resolution and signal to noise ratios in our [CII] spectra are modest, however, and a slight calibration error could introduce a velocity error of $\sim$200km s$^{-1}$, so the offset between [OI] and [CII] is acceptable and does not require a physical explanation.

\setlength\tabcolsep{2mm}
\begin{table}[htb]\scriptsize
\caption[Fine structure lines and SED properties.]{Fine structure lines and SED properties.}
\vspace{2.5mm}
\begin{center}
\begin{tabular}{ c c c c c c c c c c c}
\hline
\multirow{2}{*}{Source}	&	\multirow{2}{*}{log($\frac{L_{FIR}}{L_\odot}$)}	&	\multirow{2}{*}{$\frac{F_\nu(70)}{F_\nu(100)}$}	&	F$_{[CII]}$	&	F$_{[OI]}$	&	\multirow{2}{*}{$\frac{F_{[OI]}}{F_{[CII]}}$}	&	\multirow{2}{*}{$\frac{L_{[CII]}}{L_{FIR}}$} &	\multirow{2}{*}{$\frac{L_{[OI]}}{L_{FIR}}$}	\\
		&							&									&	\multicolumn{2}{c}{(10$^{-18}$ W m$^{-2}$)} &	\\
\hline

%  MIPS 22530 & 12.69$\pm$0.04 & 1.01$\pm^{0.10}_{0.05}$ &         9.3$\pm$ 2.1 &         5.4$\pm$ 1.1 &       0.59$\pm$ 0.18 &  (1.4$\pm$0.3)E -2 &    (8.0$\pm$1.6)E -3 \\ 
%  SWIRE 3-9 & 12.59$\pm$0.15 & 1.01$\pm^{0.26}_{0.29}$ &         6.4$\pm$ 1.4 &       3.75$\pm$ 0.78 &       0.59$\pm$ 0.18 &    (8.4$\pm$1.8)E -3 &    (4.9$\pm$1.0)E -3 \\ 
% SWIRE 3-14 & 12.21$\pm$0.13 & 0.61$\pm^{0.15}_{0.16}$ &         2.3$\pm$ 0.4 &       1.60$\pm$ 0.48 &       0.69$\pm$ 0.24 &    (8.0$\pm$1.4)E -3 &    (5.5$\pm$1.6)E -3 \\ 
% SWIRE 3-18 & 12.13$\pm$0.13 & 0.61$\pm^{0.15}_{0.17}$ &         5.5$\pm$ 1.0 &                  --- &                  --- &  (2.2$\pm$0.4)E -2 &                  --- \\ 
%    SMM J03 & 12.58$\pm$0.03 & 1.19$\pm^{0.08}_{0.09}$ &        16.9$\pm$ 3.5 &       5.97$\pm$ 0.86 &     0.353$\pm$ 0.089 &  (1.5$\pm$0.3)E -2 &  (5.11$\pm$0.74)E -3 \\ 
%  SWIRE 4-5 & 11.73$\pm$0.19 & 0.44$\pm^{0.27}_{0.03}$ &         4.5$\pm$ 1.1 &         3.2$\pm$ 1.0 &       0.72$\pm$ 0.28 &    (4.2$\pm$1.0)E -2 &  (3.03$\pm$0.96)E -2 \\ 
% SWIRE 4-15 & 12.36$\pm$0.15 & 0.85$\pm^{0.25}_{0.18}$ &         5.0$\pm$ 1.3 &       3.55$\pm$ 0.90 &       0.71$\pm$ 0.26 &  (1.3$\pm$0.3)E -2 &    (9.3$\pm$2.4)E -3 \\ 
%   SDSS J12 & 12.54$\pm$0.03 & 1.30$\pm^{0.06}_{0.15}$ &         5.6$\pm$ 1.4 &       2.52$\pm$ 0.47 &       0.45$\pm$ 0.14 &  (1.3$\pm$0.3)E -2 &    (5.6$\pm$1.0)E -3 \\ 

 MIPS 22530 & 12.69$\pm$0.04 & 1.01$\pm^{0.10}_{0.05}$ &         9.3$\pm$ 2.1 &         5.4$\pm$ 1.1 &       0.59$\pm$ 0.18 &  (1.4$\pm$0.3)E -2 &    (8.0$\pm$1.8)E -3 \\ 
  SWIRE 3-9 & 12.59$\pm$0.15 & 1.01$\pm^{0.26}_{0.29}$ &         6.4$\pm$ 1.4 &       3.75$\pm$ 0.78 &       0.59$\pm$ 0.18 &    (8.9$\pm$3.7)E -3 &    (5.2$\pm$2.1)E -3 \\ 
 SWIRE 3-14 & 12.21$\pm$0.13 & 0.61$\pm^{0.15}_{0.16}$ &         2.3$\pm$ 0.4 &       1.60$\pm$ 0.48 &       0.69$\pm$ 0.24 &    (8.3$\pm$2.9)E -3 &    (5.7$\pm$2.4)E -3 \\ 
 SWIRE 3-18 & 12.13$\pm$0.13 & 0.61$\pm^{0.15}_{0.17}$ &         5.5$\pm$ 1.0 &                  --- &                  --- &  (2.3$\pm$0.8)E -2 &                  --- \\ 
    SMM J03 & 12.58$\pm$0.03 & 1.19$\pm^{0.08}_{0.09}$ &        16.9$\pm$ 3.5 &       5.97$\pm$ 0.86 &     0.353$\pm$ 0.089 &  (1.5$\pm$0.3)E -2 &  (5.11$\pm$0.82)E -3 \\ 
  SWIRE 4-5 & 11.73$\pm$0.19 & 0.44$\pm^{0.27}_{0.03}$ &         4.5$\pm$ 1.1 &         3.2$\pm$ 1.0 &       0.72$\pm$ 0.28 &    (4.7$\pm$2.4)E -2 &    (3.3$\pm$1.8)E -2 \\ 
 SWIRE 4-15 & 12.36$\pm$0.15 & 0.85$\pm^{0.25}_{0.18}$ &         5.0$\pm$ 1.3 &       3.55$\pm$ 0.90 &       0.71$\pm$ 0.26 &  (1.4$\pm$0.6)E -2 &    (9.9$\pm$4.3)E -3 \\ 
   SDSS J12 & 12.54$\pm$0.03 & 1.30$\pm^{0.06}_{0.15}$ &         5.6$\pm$ 1.4 &       2.52$\pm$ 0.47 &       0.45$\pm$ 0.14 &  (1.3$\pm$0.3)E -2 &    (5.6$\pm$1.1)E -3 \\

\tablecomments{For the full set of SED properties that we use to compare our sample to other data sets and models see Table \ref{tab:appsourceprops} in the appendix. We summarize them here since many of the properties are highly correlated and encapsulate similar information (eg.  $\frac{F_\nu(70)}{F_\nu(100)}$ and $\frac{F_\nu(60)}{F_\nu(100)}$.)}

\end{tabular}
\end{center}
\label{tab:sourceprops}
\end{table}
\setlength\tabcolsep{2.12mm}

\clearpage

\subsection{Photometry and SEDs}\label{sec:sed_cii}
The photometry from our PACS program and the literature is compiled in Table \ref{tab:sourcephot} in the appendix. To estimate FIR properties such as L$_{FIR}$ and the 70 to 100 $\mu$m flux density ratio in a uniform manner (Table \ref{tab:sourceprops}) we fit star formation SED models from \citet{dal2002} to available photometry. 
The SED library consists of a set of 64 templates, each described by a different power law distribution of dust mass over heating intensity parameterized by $\alpha$:

\begin{equation}
dM_d(U)\propto U^{-\alpha}dU
\label{eq:sedalapha}
\end{equation}

where $M_d(U)$ is the dust mass heated by radiation field at intensity U. Higher (lower) values of alpha correspond to SEDs with dust distributions skewed toward cooler (warmer) temperatures averaged over the galaxy. The total luminosity of the system is a free parameter which can be scaled up or down.

\begin{figure}[htb]
	\centering
		\includegraphics[trim=7cm 0cm 0cm 0cm, clip=true, height=0.9\textwidth,angle=90]{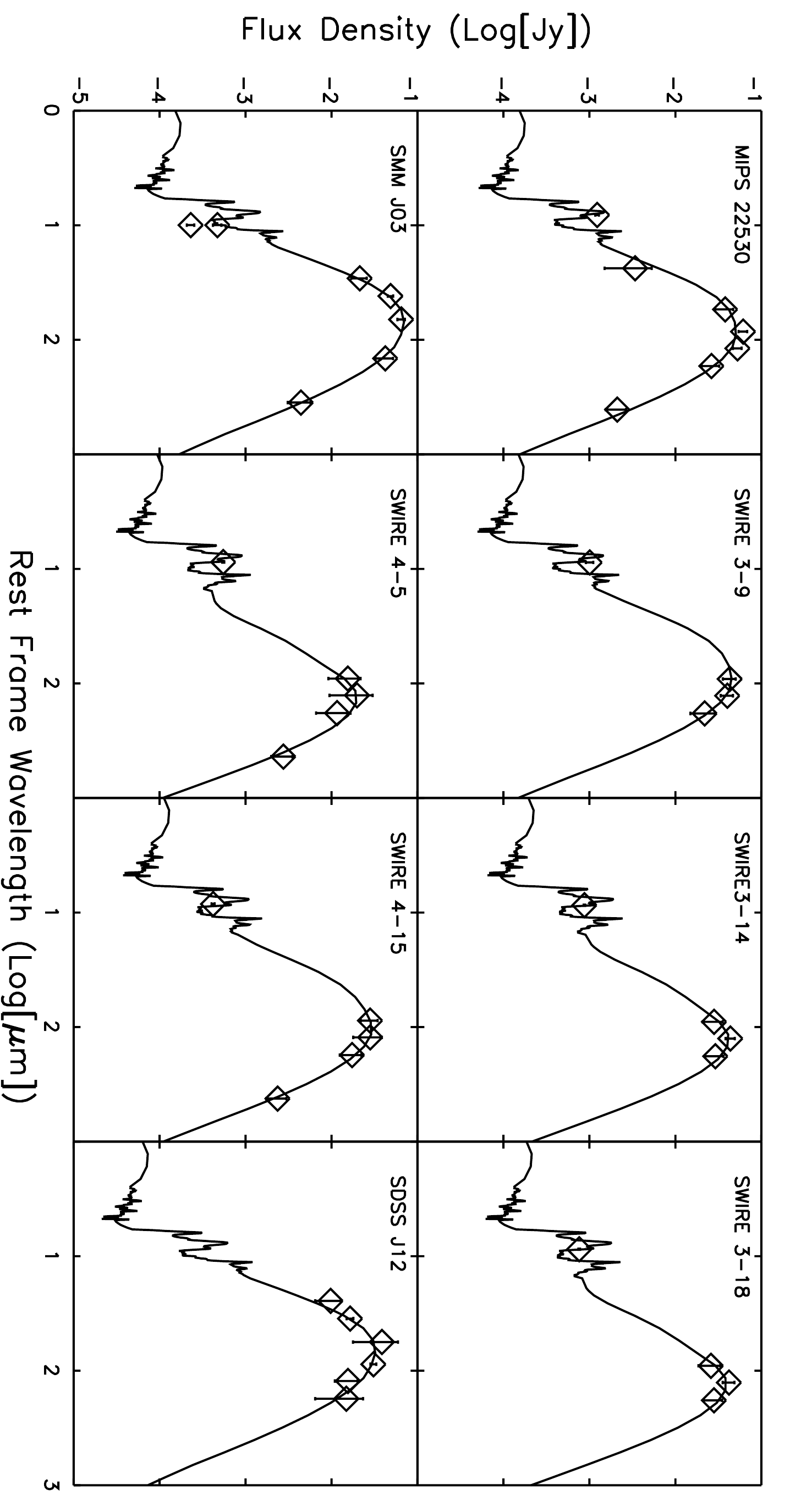}
	\caption[Best fit SEDs]{The best fit SED model from the library of \citet{dal2002} for each of our sources. Photometry (diamonds) is listed in Table \ref{tab:sourcephot}. Data from 24 $\mu$m observations is plotted but not used in SED fit.}
	\label{fig:allSED}
\end{figure}

We compared each template against available photometry in the 60-1200 $\mu$m range weighted by error and formed likelihood functions to determine several global properties including L$_{FIR}$, 70/100 and 60/100 $\mu$m flux density ratios along with their corresponding 68 percentile likelihood range.\footnote{In most sources 24 $\mu$m observations are also available. In the rest frames of our sources, 24 $\mu$m observations probe the PAH-dominated MIR and have little bearing on the FIR properties we hope to constrain with SED fits. Therefore, although we plot these data in the SEDs, we do not use them in performing the fit.} The best fit SED for each of our sources is shown in Figure \ref{fig:allSED}.

\begin{figure}[htb]
	\centering
		\includegraphics[trim=2cm 2cm 2cm 2.5cm, clip=true, angle=90,width=0.9\textwidth]{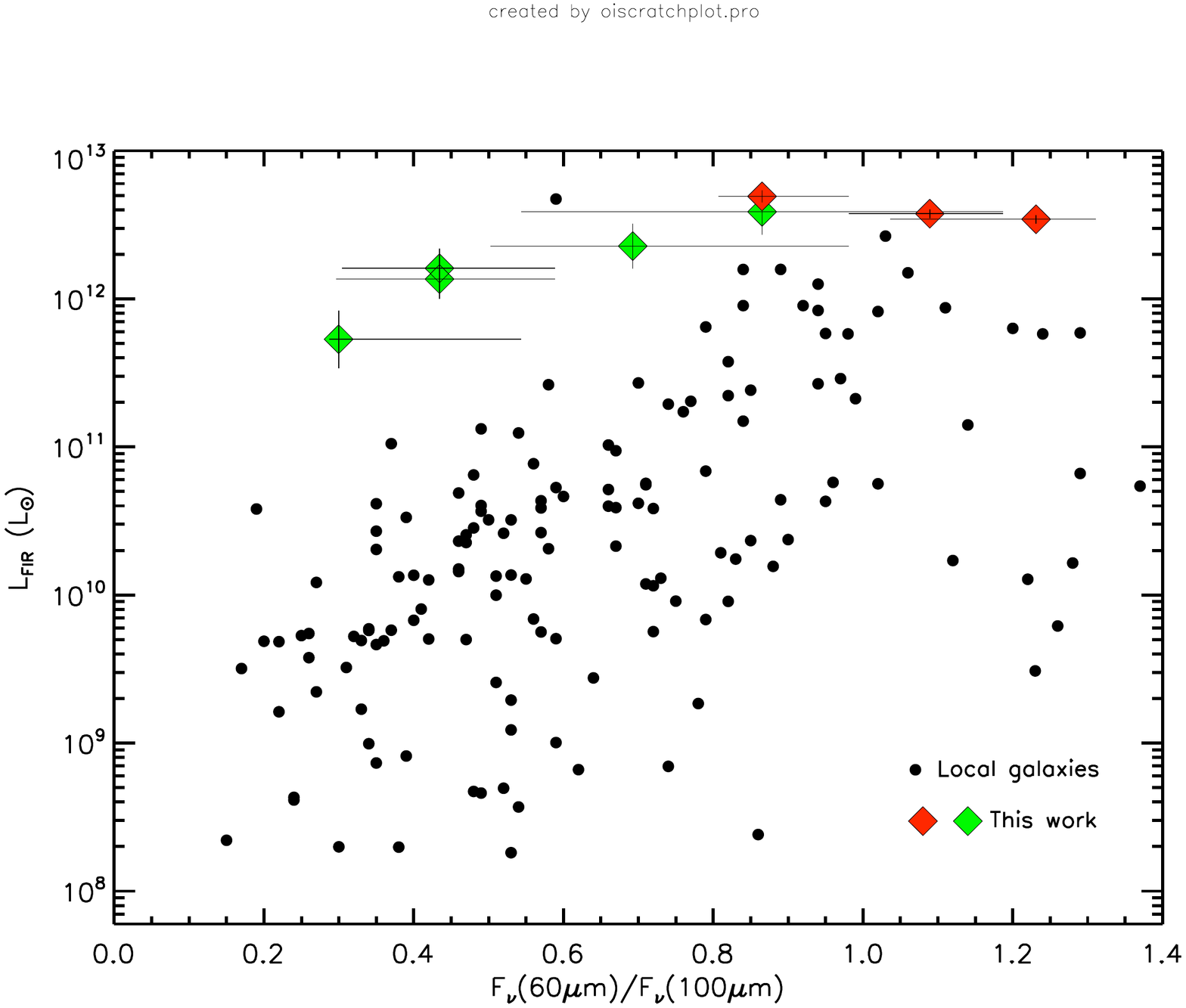}
	\caption{L$_{FIR}$ vs. F$_\nu$(60)/F$_\nu$(100). Local sources are from the sample of \citet{bra2008}. Our sources are indicated as diamonds. Although we believe the luminosity in all of our sample is dominated by star formation, we note that the three sources with marginal evidence for AGN in the literature (SMM J03, SDSS J12, and MIPS 22530 - noted by red diamonds,) tend to have hotter dust temperatures than those with no indication of an AGN present (green diamonds.) All of our sources have significantly larger luminosities than the average local galaxies of similar F$_{60}$/F$_{100}$ values. No corrections for gravitational lensing have been made. Magnification corrections (if applicable and known) affect luminosity only, and would bring our galaxy luminosities closer to those of the local galaxy group.}
	\label{fig:fircolor}
\end{figure}

The SED templates fit our observed photometry well. In Figure \ref{fig:fircolor} we compare the sources' luminosities and dust temperatures (indicated by the 60/100 $\mu$m flux ratio) with those of local sources \citep{bra2008}. In local sources there is a weak trend of increasing dust temperature with increasing luminosity. Our sources span the full range of dust temperatures seen in local sources. The three sources from our sample with marginal evidence for AGN contribution fall to the right, hot dust temperature side of the plot. One revealing aspect of this figure is that our sources with the lowest dust temperatures have luminosities many times greater than the equivalent dust temperature sources in the local Universe. No corrections have been made for gravitational lensing effects, though in our sample only one source, SDSS J12, is known to be gravitationally lensed. \citet{lin2009} find a magnification factor of 27.

The occurrence of highly luminous systems with cooler dust peaks at high-z is well established \citep{[eg],elb2011,sym2013}, and further confirms the general findings that SF-D systems at high-z can be represented as scaled up local star forming systems. The conditions of star formation in our z=1-2 sources give rise to cool dust peaks similar to those of local normal or LIRG class galaxies, but scaled up spatially to account for ULIRG (or LIRG in the case of SWIRE 4-5) luminosities.

To place these sources in context with the larger population of galaxies at these redshifts, we consider their bulk star formation rates (SFRs) and stellar masses. The TIR luminosity, largely emitted by dust heated by young stars, is a convenient star formation tracer with a long history of use. We use the TIR - SFR relationship established by \citet{ken1998} and adapted by \citet{nor2010} to adjust for a \citet{cha2003} IMF:

\begin{equation}
\frac{SFR}{M_\odot yr^{-1}}=\frac{L_{TIR}}{L_\odot} 1.015 \times 10^{-10}.
\label{eq:sfrtir}
\end{equation}

Several methods of estimating galaxy stellar mass based on various photometric recipes have been used in local galaxies. These methods are generally based on galaxy SED modeling and rely on multiple optical/NIR measurements to break degeneracies in star formation history. A crude but effective estimation can be arrived at based solely on the rest frame galaxy luminosity at $\sim$2 $\mu$m, a wavelength which yields nearly constant mass to luminosity ratios which are less dependent on star formation histories \citep{dej1996,bel2003}. The effectiveness of the 2 $\mu$m luminosity for the purpose of estimating stellar masses has already been demonstrated at z$\sim$3 using IRAC 8 $\mu$m measurements \citep{mag2010}, and here we extend the approach to z$\sim$1.8 where the appropriate rest wavelength coincides with the IRAC 5.8 $\mu$m band. We take this IRAC band to represent the rest frame 2 $\mu$m flux and directly use the relationship established by \citet{mag2010}:

\begin{equation}
log(M_*/M_\odot)=2.01(\pm0.65)-0.35(\pm0.03)\times M_{2\mu m},
\label{eq:stellarmass}
\end{equation}

where M$_{2\mu m}$ represents the rest frame 2 $\mu$m absolute magnitude. As is clear from the uncertainty in the coefficients to Equation \ref{eq:stellarmass}, the resulting stellar mass estimate should only be taken as an  order of magnitude approximation. Nonetheless, we are reassured that it is an appropriate order of magnitude estimate by the examples of SWIRE 4-5 and SWIRE 4-15. Both of these sources were examined in \citet{fio2009} where careful multi-band photometric stellar estimates yielded 1.36 and 3.21 $\times$10$^{11}$ M$_\odot$ respectively. Our simple stellar mass estimates of 1.35 and 2.05 $\times$10$^{11}$ M$_\odot$ are in satisfactory agreement.

\begin{table}[htb]\small
\caption[Stellar masses and star formation rates]{Stellar mass and SFR estimates for sources with IRAC measurements.}
\vspace{2.5mm}
\begin{center}
\begin{tabular}{ c c c c c }
	\hline

	Source 	&	log($\frac{SFR}{M_\odot yr^{-1}}$)						&	F$_{IRAC 5.8}$					&	M$_*$  \\ 
			&															&	mJy								&	10$^{11}$M$_\odot$	\\ \hline

MIPS 22530	&	2.96$\pm$0.04		&	(4.3$\pm$1.0)E -2	&	1.4	\\
SWIRE 3-9	&	2.83$\pm$0.16		&	(1.12$\pm$0.06)E -1	&	2.5	\\
SWIRE 3-14	&	2.55$\pm$0.08		&	(7.00$\pm$0.35)E -2	&	1.8	\\
SWIRE 3-18	&	2.49$\pm$0.09		&	(6.80$\pm$0.34)E -2	&	1.7	\\
SMM J03		&	2.85$\pm$0.03		&	(8.16$\pm$0.83)E -2	&	1.2	\\
SWIRE 4-5	&	2.11$\pm$0.15		&	(5.36$\pm$0.46)E -2	&	1.3	\\
SWIRE 4-15	&	2.62$\pm$0.15		&	(7.63$\pm$0.35)E -2	&	2.1	\\

\end{tabular}
\end{center}
\label{tab:sfrstellarmass}
\end{table}

\begin{figure}[!htb]
	\centering
		\includegraphics[width=0.7\textwidth,trim=0cm .3cm 0cm 0cm, clip=true]{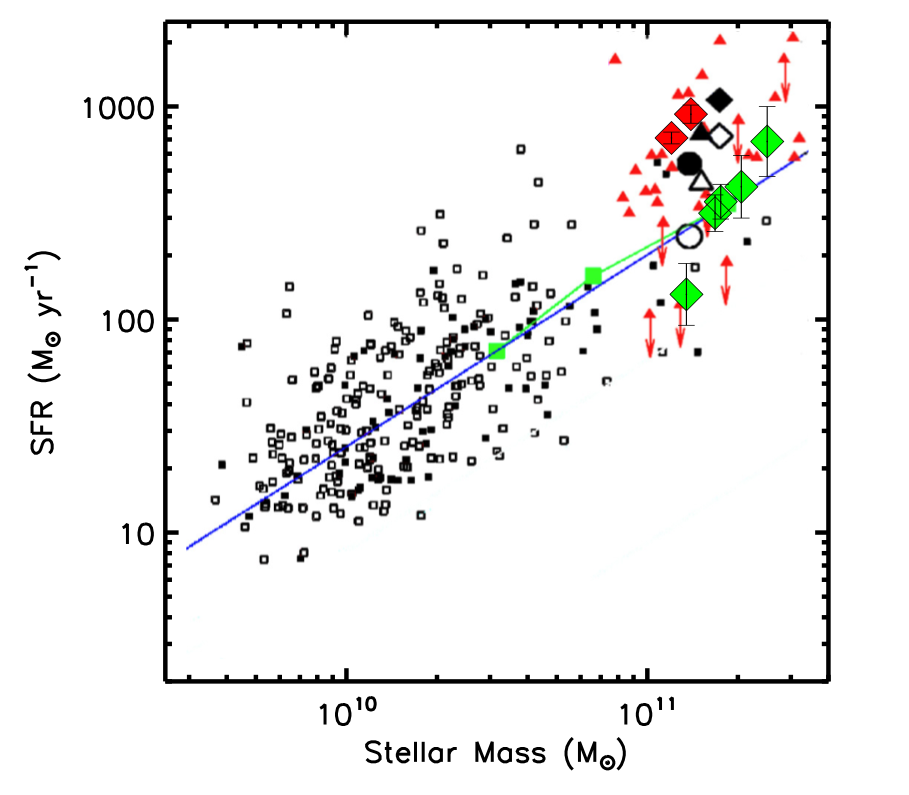}
	\caption[SFR vs. stellar mass]{SFR vs. stellar mass. Adapted from Figure 6 of \citet{fio2009}. The SFRs and stellar masses of our sample, represented by filled green and red diamonds, are determined by L$_{FIR}$ and the IRAC 5.8 $\mu$m flux density. One source, SDSS J12 is not included as no IRAC 5.8 $\mu$m data is available. The x error bars have been suppressed for clarity as we only have order of magnitude stellar mass estimates. Red triangles represent the z$\sim$2 sources presented by \citet{fio2009}, Large black symbols represent stacked subsets of the Fiolet sample, small black symbols represent z$\sim$2 sources from \citet{dad2007}. Green squares trace the average trend in GOODS-N. The `main sequence' of galaxies stands out as a strong correlation of SFR and stellar mass, clustering about the blue line.}
	\label{fig:mainsequence}
\end{figure}

In Table \ref{tab:sfrstellarmass} we note the derived SFRs and stellar masses of the seven sources for which there are IRAC 5.8 $\mu$m measurements and in Figure \ref{fig:mainsequence} we overplot these data on the sample presented by \citet{fio2009}. Although our sample consists of massive, highly star forming sources, their relative mass and SFRs (or alternatively their specific star formation rates) follow the trend exhibited by most galaxies, very close to the stacked results of the Fiolet sample. In other words, our sample falls on the galaxy `main sequence' \citep{noe2007}.

\subsection{[CII], [OI] and the FIR continuum}
\begin{figure}[!htb]
	\centering
		\includegraphics[trim=1.8cm 1.8cm 2.6cm 1.95cm, clip=true, height=.9\textwidth,angle=90]{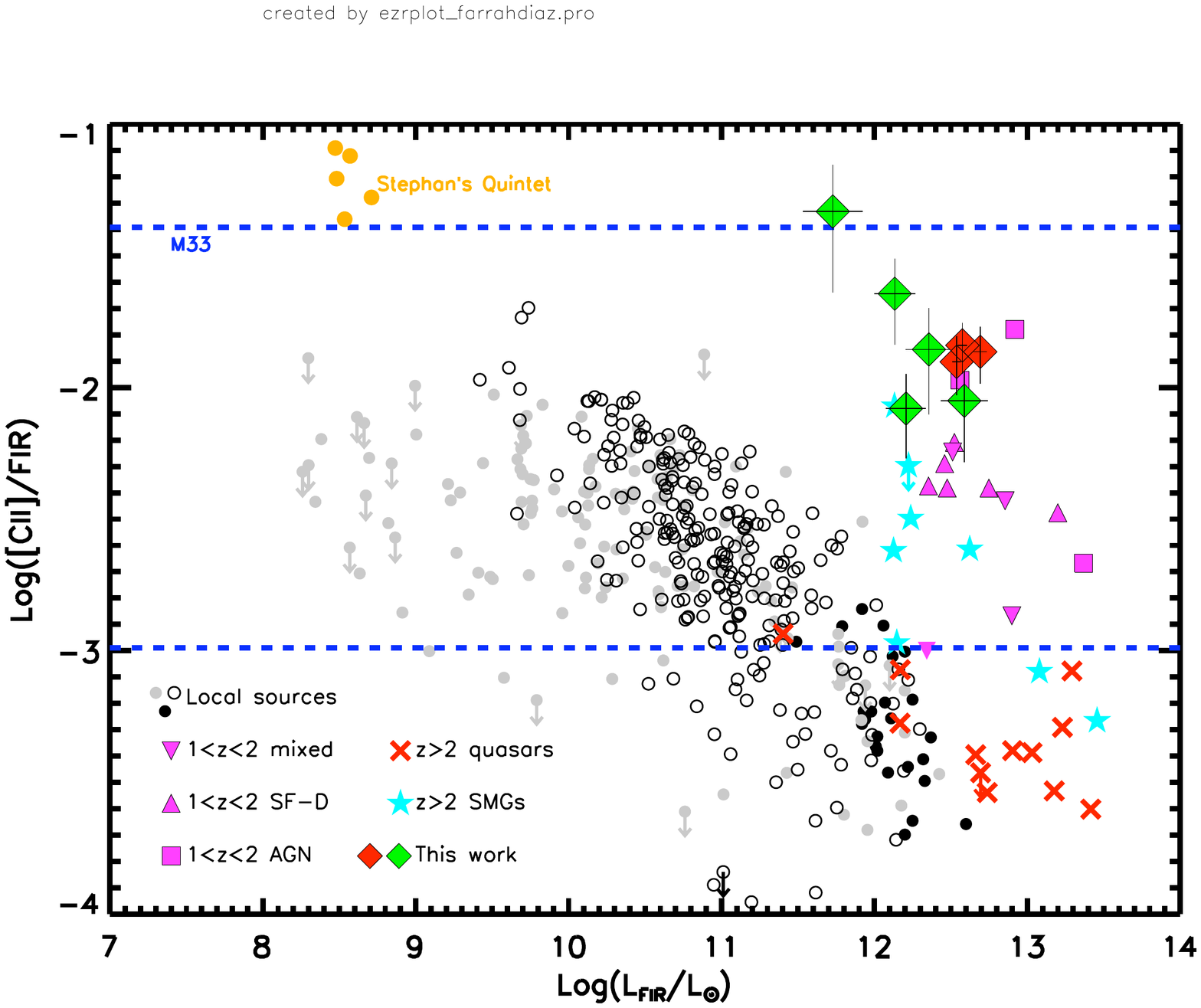}
	\caption[{[CII]/FIR vs. L$_{FIR}$}]{Observed [CII]/FIR vs. L$_{FIR}$. Our sample is shown as in previous Figures. We have also included local sources from \citet{bra2008} (gray circles,) the GOALS sample \citep{arm2009,dia2013} (open circles,) and \citet{far2013} (black filled circles.)  Magenta 1$<$z$<$2 sources are from \citet{sta2010} (updated by \citet{fer2014} and \citet{hai2014},) and several high redshift SMGs and QSOs are from \citep{pet2004,mar2005,mai2005,mai2009,ivi2010,val2011,deb2011,swi2012,wag2012,gal2012,ven2012,wil2013,wan2013b,rie2013}. We also note shock affected regions from Stephan's Quintet \citep{app2013} (orange circles,) and the ratios spanned by star forming regions within M33 \citep{moo2011}.}
	\label{fig:rplot}
\end{figure}

In Figure \ref{fig:rplot} we plot [CII]/FIR vs. L$_{FIR}$. While the absolute [CII] line luminosity and L$_{FIR}$ may be amplified by gravitational lensing, the FIR continuum and [CII] emitting regions will generally be cospatial in SF-D galaxies, so that the [CII]/FIR ratio is not very susceptible to differential magnification \citep{ser2012}. Local sources (black, white, and gray circles) clearly illustrate the historically reported [CII] deficit, showing [CII]/FIR decreasing with increasing L$_{FIR}$ \citep[largely established by the works][]{mal2001,luh2003}.

Many of our SF-D sources from this work show [CII] emission with [CII]/FIR ratios in excess of 1\%. Although such high [CII]/FIR is unusual, it is not unheard of. In section \ref{sec:individuals} we show that these observations push models for PDR origins of the [CII] emission to their limits. There are previous examples of sources with high [CII]/FIR ratios. For example, H-ATLAS SDP 81 \citep{val2011}, at z$\sim$2.3 and the nearby spiral galaxy M51 \citep{nik2001}, both show [CII]/FIR$\sim$0.01. Spatially resolved studies of star forming regions in M33 show [CII]/FIR luminosity ratios ranging from 0.001 up to 0.04 \citep{moo2011}. Assuming the [CII] and FIR arise in PDRs, [CII]/FIR$\sim$1\% indicates G$\sim$100 implying that the star formation regions in our sources must be very extended given their large FIR luminosities \citep{sta2010}. There are, however, hints that other mechanisms may sometimes contribute to [CII] emission. Several resolved regions of Stephan's Quintet, for instance, have extremely high [CII]/FIR ratios as a result of shocked PDRs \citep{app2013}.

\begin{figure}[!htb]
	\centering
		\includegraphics[height=0.9\textwidth,trim=2.2cm 1.5cm 1.75cm 1.75cm, clip=true,angle=90]{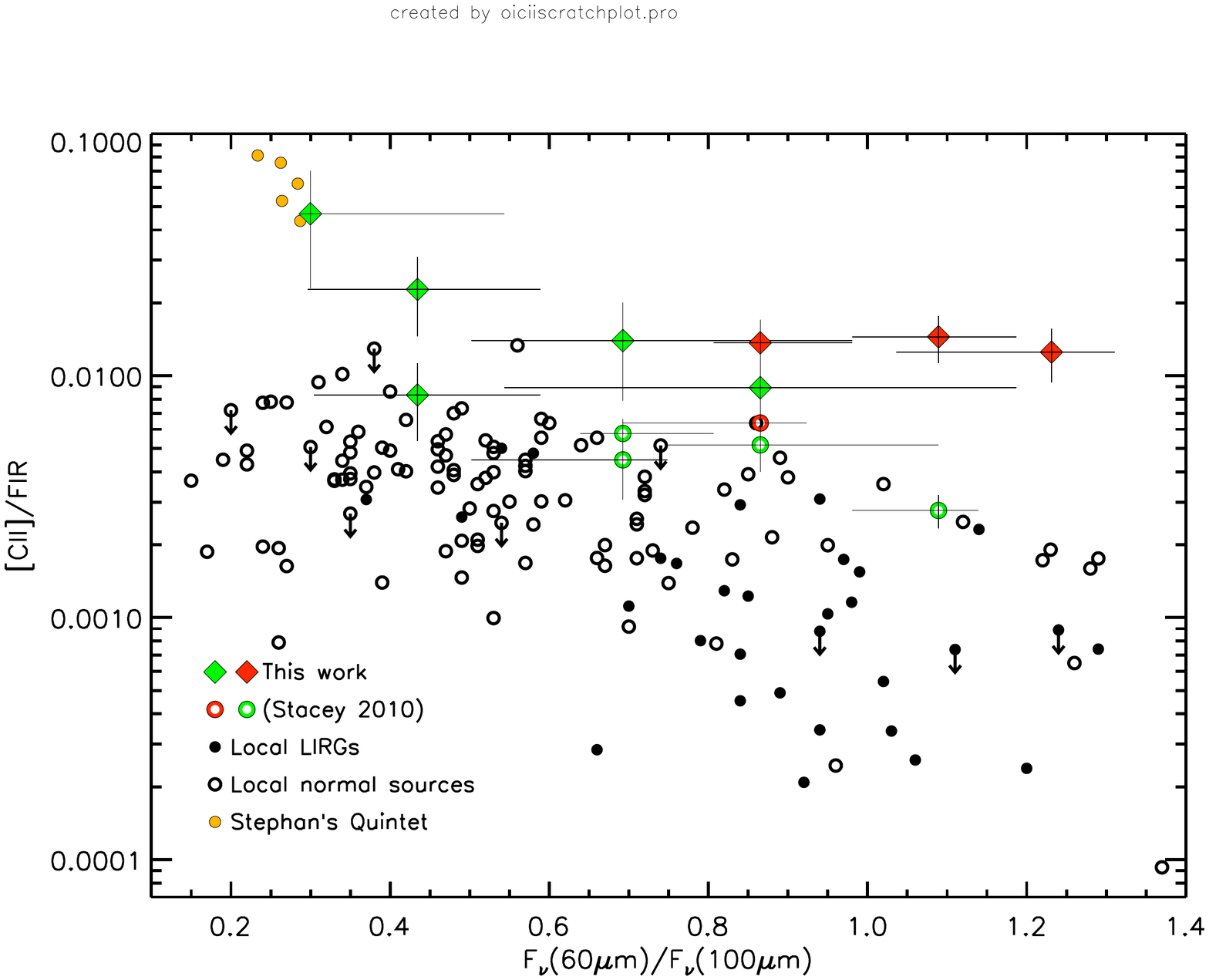}
	\caption[{[CII]/FIR vs. F$_\nu$(60)/F$_\nu$(60)}]{Observed [CII]/FIR ratio as a function of the 60 to 100 $\mu$m flux density ratio. Local sources are from the \citet{bra2008} sample and are plotted as small open circles (normal galaxies) and filled black circles (LIRGs). Our sources and Stephan's Quintet are plotted as in previous figures. We also plot sources revisited from \citet{sta2010} as large open circles. As with the sample from this work, the \citet{sta2010} sources are marked in green for star formation dominant, and red for a system with uncertain AGN contribution (SMM J22471.) Although the uncertainty in the dust temperature of our sources is large, our sample tends to decrease in [CII]/FIR as their dust temperature increases similar to the trend in local galaxies, though our sample lies above the local trend. Asymmetric error bars span the 68.27\% most likely range of 60/100 $\mu$m flux ratios in our SED fits.}
	\label{fig:rvscolor}
\end{figure}

Considering the [CII]/FIR ratio in context with the FIR dust temperature, in Figure \ref{fig:rvscolor}, we note that both local sources and our high redshift sample shows the [CII]/FIR ratio tending to decrease with increasing F$_\nu$(60)/F$_\nu$(100). Our sample, however, lies above the trend observed in local galaxies, and in the case of SWIRE 4-5 approaches the [CII]/FIR and dust temperature seen in regions of Stephan's Quintet. Although the uncertainties on the FIR colors in our SED fits are large, it is also worth noting that MIPS 22530, SDSS J12, and SMM J03, the three sources which have tentative evidence for AGN in the literature, have higher F$_\nu$(60)/F$_\nu$(100) ratios, consistent with the presence of an AGN contributing to a hot dust component.

\begin{figure}[!htb]
	\centering
		\includegraphics[width=0.7\textwidth,trim=4.575cm 13.5cm 6cm 3.45cm, clip=true]{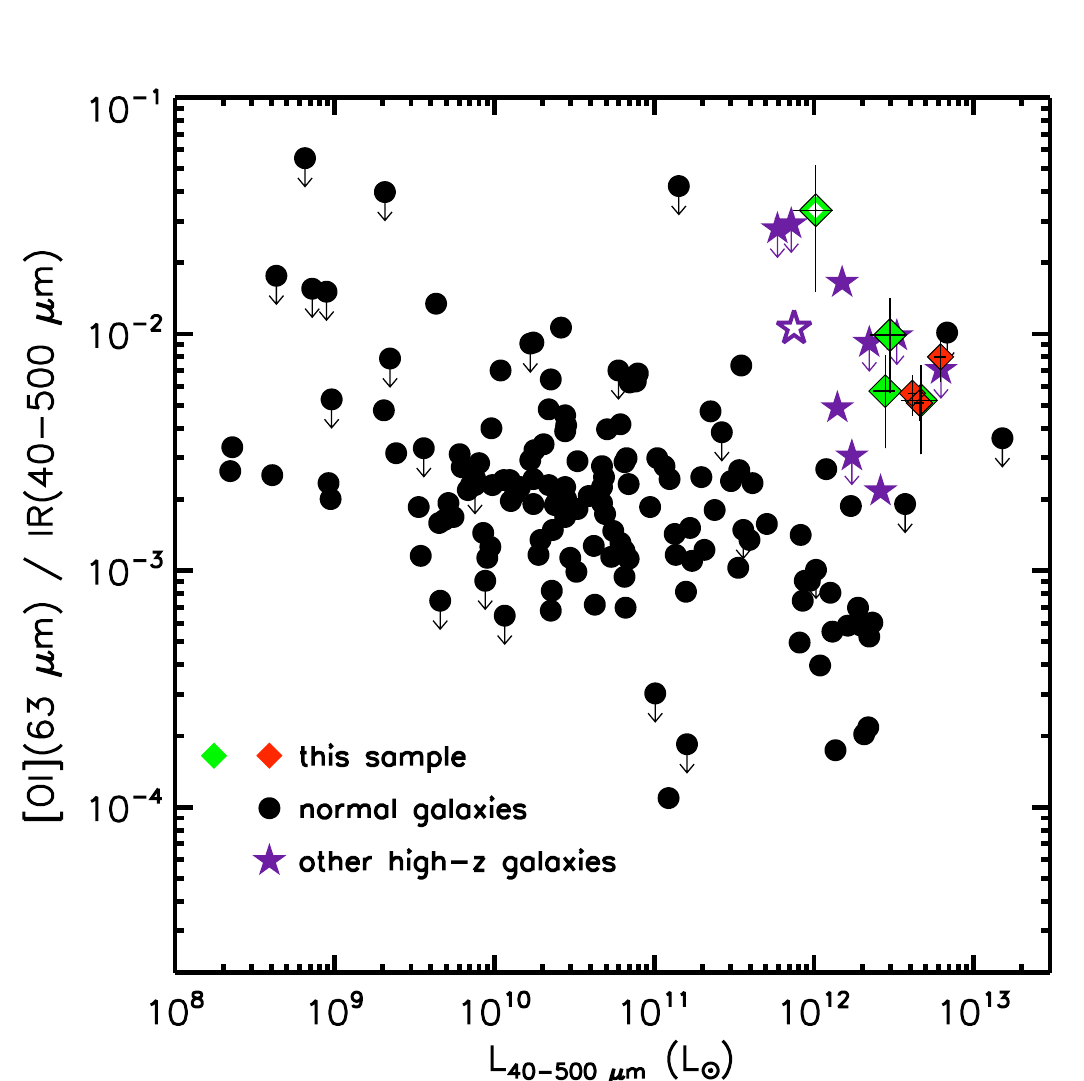}
	\caption[{[OI]/IR vs. L$_{IR}$}]{Observed [OI] 63 $\mu$m/IR ratio vs. L$_{IR}$. To compare to the sample of \citet{cop2012}, here we have used the infrared continuum range 40-500 $\mu$m. Local sources (black circles) are from the compilation by \citet{gra2011} which includes additional data from \citet{col1999,mal2001,neg2001,luh2003,dal2004,bra2008}. Literature high redshift sources are denoted with purple stars \citep{cop2012,bra2008,ivi2010,val2011,stu2010}. Our sample is marked with diamonds as before. In our sample and also the sample from \citet{cop2012}, there is one tentative [OI] detection which we mark as an open diamond and star respectively.}
	\label{fig:ofir}
\end{figure}

The [OI] 63 $\mu$m line is the other primary coolant in PDRs. As we show in Figure \ref{fig:ofir}, [OI] traces out a similar line to IR continuum deficit in local sources as the infrared luminosity increases. Although less pronounced than what we see in the [CII] line, our sources again show enhanced line emission compared to local sources of similar luminosity. Our sources have L$_{[OI]}$/L$_{IR}$ ratios similar to the galaxy sample of \citet{cop2012}. Their sample was a set of Large Apex Bolometer Camera (LABOCA) submillimeter selected galaxies with spectroscopic redshifts.  Based on their FIR photometry and [OI] line, \citet{cop2012} concluded their sample was most like scaled up `normal' star bursting systems with spatially extended star formation, rather than merging ULIRGs.

\begin{figure}[!htb]
	\centering
		\includegraphics[height=0.9\textwidth,trim=2.2cm 2.3cm 2.8cm 1.65cm, clip=true,angle=90]{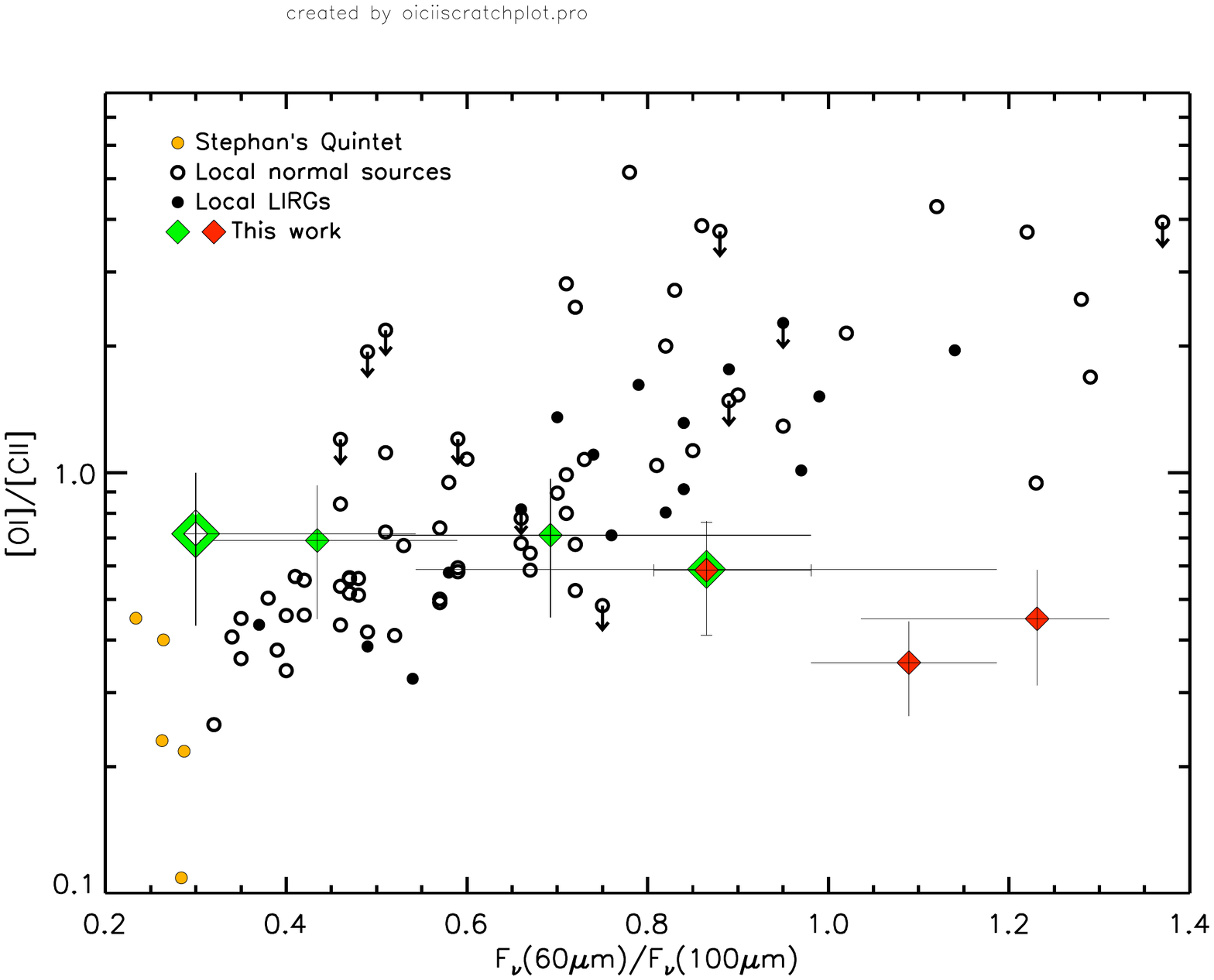}
	\caption[{[OI] 63 $\mu$m/[CII] vs. F$_\nu$(60)/F$_\nu$(60)}]{Observed [OI] 63 $\mu$m/[CII] line flux ratio as a function of the 60 to 100 $\mu$m color. Symbols are as in Figure \ref{fig:rvscolor} with the addition of an open diamond indicating our tentative [OI] detection of SWIRE 4-5. Note that SWIRE 3-9 and MIPS 22530 have the same line ratio and FIR color. We have enlarged the symbol for SWIRE 3-9 and added serifs to its error bars to differentiate the overlapping points, though they share the same line ratio error range. Local ULIRGs tend to have higher [OI]/[CII] ratios and warmer dust temperatures. The [OI]/[CII] ratios in our sample are similar to local normal sources. Even our sources with marginal AGN influence, which have warmer dust temperatures, have [OI]/[CII] ratios more similar to local normal galaxies rather than LIRGs.}
	\label{fig:ocratiocolor}
\end{figure}

Due to the higher critical density and excitation potential of [OI] 63 $\mu$m compared to [CII], higher [OI]/[CII] ratios correspond to more compact star forming regions and are correlated with warmer FIR colors \citep{mal2001,dia2013}. From Figure \ref{fig:ocratiocolor} we see that our high-z sample spans the full range of FIR color seen in local galaxies, but even in our sources with hot dust temperatures the [OI]/[CII] ratios are modest, similar to cool star forming galaxies in the local Universe. Possibly those sources in our sample with large F$_\nu$(60)/F$_\nu$(100) values have a hotter dust component related to a potential AGN contributing to the 60 $\mu$m continuum, while cool dusty PDRs contribute the bulk of the [OI] and [CII] emission.

\clearpage

\subsection{PAHs}\label{sec:pahs}
\begin{figure}[!htb]
	\centering
		\includegraphics[height=0.9\textwidth,angle=90,trim=0.1cm 0.4cm 1.2cm .2cm, clip=true]{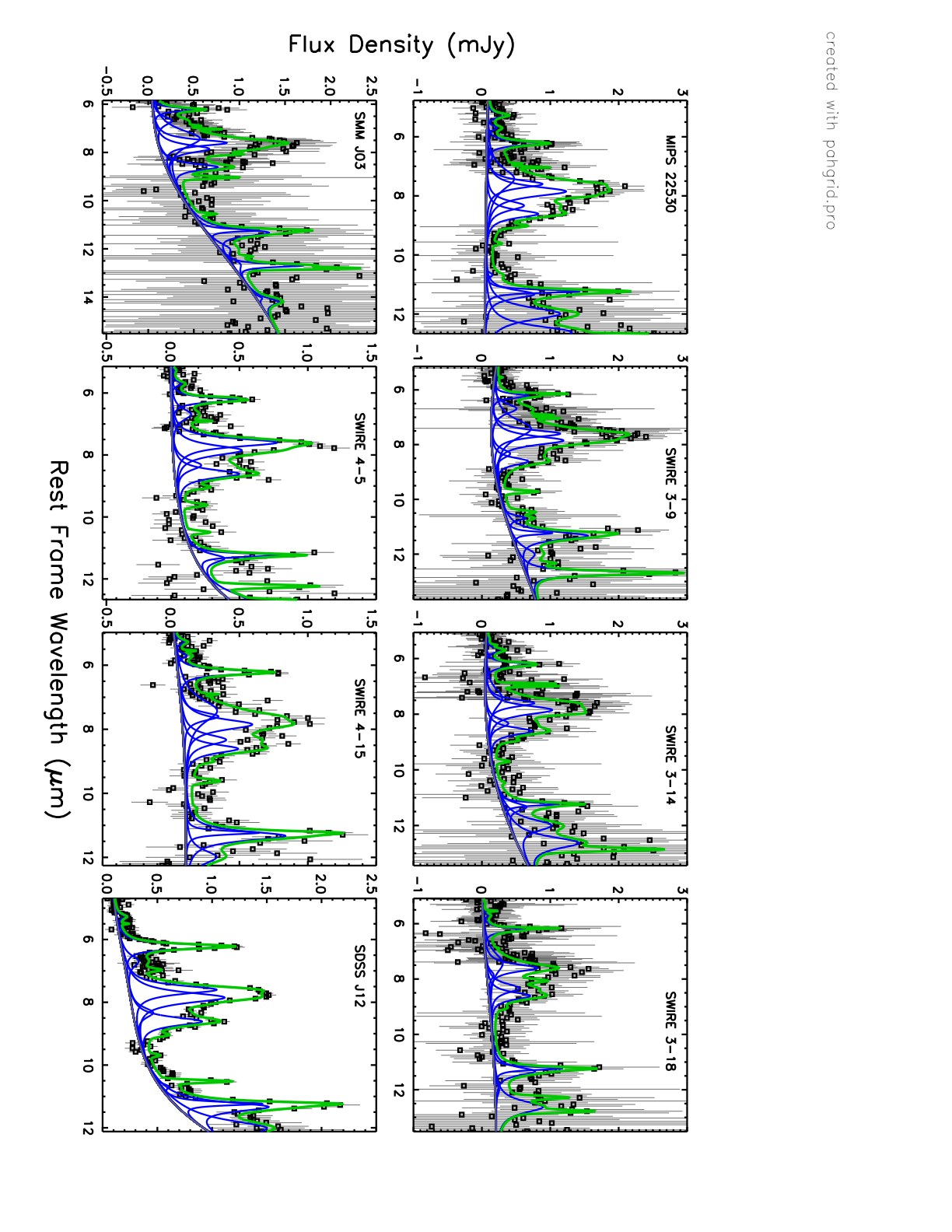}%trim LBRT
	\caption[Spitzer IRS mid infrared PAH spectra]{MIR IRS spectra fitted with PAHFIT \citep{smi2007}. The green curve is the overall fit, blue curves are PAH contributions, and the underlying gray curve is the continuum fit by stellar and dust contributions. Sources of PAH spectra are noted in Table \ref{tab:pahs}.}
	\label{fig:pahspectra}
\end{figure}

PAHs are the main source of photo-ejected electrons for neutral PDR gas heating \citep{wei2001}. As a simple diagnostic of heating (traced by PAHs) vs. cooling (traced by [CII]), the ratio of [CII] to PAH emission is of interest much in the same way as the [CII]/FIR ratio. 
The PAH features in several of our sources have been examined in other works, however, PAH fluxes are sensitive to the fitting method used.
Therefore, in order to obtain a uniform data set we fit the MIR spectra ourselves using PAHFIT \citep{smi2007}. The fitted spectra are shown in Figure \ref{fig:pahspectra} and the fluxes are listed in Table \ref{tab:pahs}. Most of the PAH spectra were acquired through the Cornell AtlaS of Spitzer/IRS Sources (CASSIS)\footnote{The Cornell Atlas of Spitzer/IRS Sources (CASSIS) is a product of the Infrared Science Center at Cornell University, supported by NASA and JPL.} version 6 using the optimal extraction for point sources. Since MIR spectra contain overlapping PAH and silicate features, it is difficult to independently estimate uncertainties. To get the best estimate we fit each spectrum repeatedly, adding a Gaussian distributed random error to each spectral data point in proportion to each point's error bar. We fit each spectrum a hundred times and report the mean PAH fluxes and their standard deviations. 

In all of our sources, only a portion of the full PAH spectrum is accessible (usually the range covered by the IRS long-low (LL) module.) At z$\sim$1.8 this covers wavelengths 5-14 $\mu$m. Although this range misses several PAH features, based on Figure 6d in \citet{cro2012}, we can estimate the total PAH emission as:

\begin{equation}
PAH_{tot}\approx\frac{8.47 PAH_{11.2}}{1.73-0.197 PAH_{7.7}/PAH_{11.2}}.
\label{eq:pahtotest}
\end{equation}

This PAH estimator uses both the 7.7 and 11.2 $\mu$m features, representative of the ionized and neutral PAH species respectively.

\begin{figure}[!htb]
	\centering
		\includegraphics[width=0.9\textwidth,trim=0.32cm 0.36cm 0.6cm 1.1cm, clip=true]{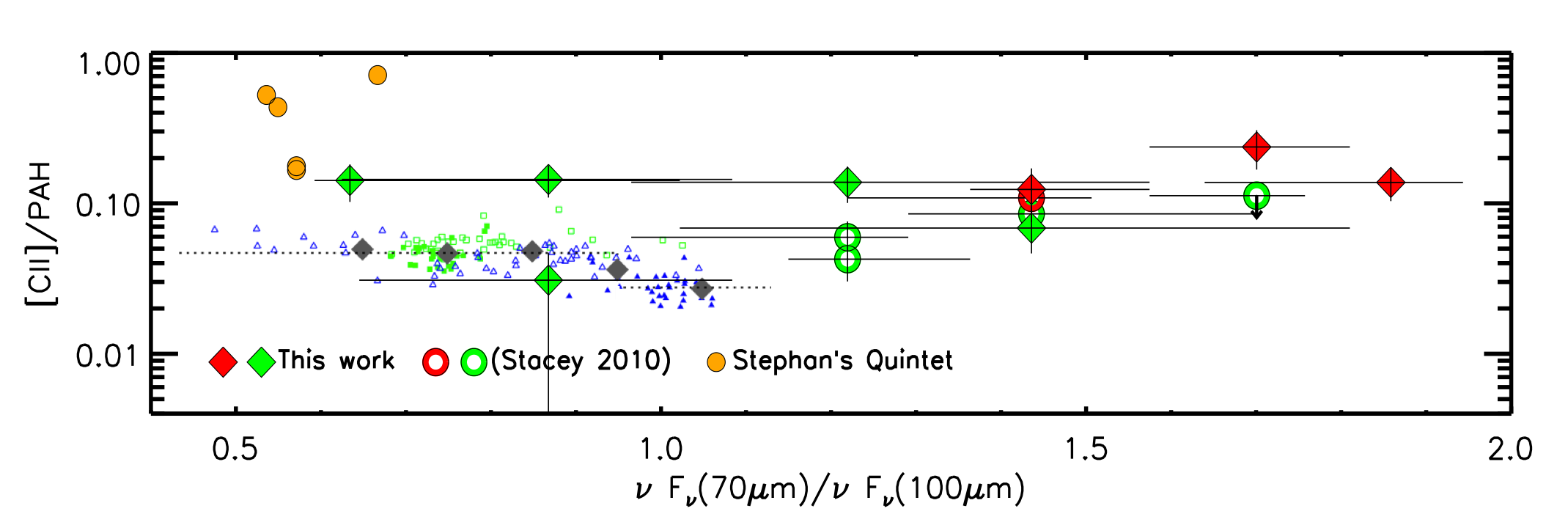}
	\caption[{[CII] to PAH ratio vs. $\nu$F$_\nu$(70)/$\nu$F$_\nu$(100)}]{Figure adapted from Figure 8 of \citet{cro2012}. [CII]/PAH vs. FIR color. [CII] is the observed line flux and PAH is the calculated total PAH flux based on Equation \ref{eq:pahtotest} with the qualifications noted in the text. Our sources (plotted as in previous figures) are plotted against sub-galactic star forming regions in NGC 1097 (light blue triangles) and NGC 4559 (light green squares) (see \citet{cro2012}.)
	We have also plotted regions from \citep{sta2010} and Stephan's Quintet as in previous Figures. Our sample tends to lie between the [CII]/PAH ratios in Stephan's Quintet and those in local star forming regions.}
	\label{fig:ciipahratio}
\end{figure}

We present PAH$_{tot}$ in Table \ref{tab:pahs}, but note that the PAH flux ratio in the denominator of Equation \ref{eq:pahtotest} can lead to very large errors, and in some cases the flux in the PAH features we observe is greater than the lower uncertainty bound on PAH$_{tot}$. Where this is the case we use the observed summed flux in PAH features between 6.2 and 11.2 $\mu$m to more tightly constrain the lower bound of the total PAH flux. Therefore, when we consider the [CII]/PAH ratio we often have asymmetric error bars. In Figure \ref{fig:ciipahratio} we plot the [CII]/PAH ratio with respect to the FIR color and compare to the sub-galactic sample of star forming regions from \cite{cro2012}. Our sources demonstrate a [CII] excess with respect to PAH emission compared to the sub-galactic local star forming regions. We have also revisited the star forming sources with available PAH spectra from \citet{sta2010} and included their [CII]/PAH$_{tot}$ values, which span a similar range. \citet{cro2012} found a negative correlation between [CII]/PAH and F$_\nu$(70)/F$_\nu$(100) (as well as the PAH$_{7.7}$/PAH$_{11.2}$ flux ratio.) This trend is not apparent in our sources, although the large error bars in F$_\nu$(70)/F$_\nu$(100) do not allow us to rule out a similar trend.

\begin{table}[htb]\tiny
\caption[PAH features]{PAH features.}

\vspace{2.5mm}
\begin{center}
\begin{tabular}{ c c c c c c c c c}
	\hline
			&	\multicolumn{4}{c}{PAH Flux (10$^{-18}$ W m$^{-2}$)}	& (Eq. \ref{eq:pahtotest})	&	&	&	 \\
Source		&	6.2 $\mu$m	&		7.7 $\mu$m	&		8.6 $\mu$m	&	11.2 $\mu$m	&	PAH$_{tot}$			&	[CII]/PAH$_{tot}$	& Ref. / AORkey	\\	\hline
  
   MIPS 22530 &  6.3$\pm$1.5 & 31.1$\pm$3.3 &  5.7$\pm$0.8 &  5.5$\pm$1.9 &  75$\pm$ 24 & 0.12$\pm^{0.07}_{0.07}$ &         AOR:11865856, 23632896 \\ 
  SWIRE 3-9 &  4.6$\pm$1.3 & 39.2$\pm$4.5 &  4.5$\pm$2.0 &  7.1$\pm$1.5 &  93$\pm$ 21 & 0.07$\pm^{0.02}_{0.02}$ &                   AOR:17414656 \\ 
 SWIRE 3-14 &  5.7$\pm$1.3 & 24.9$\pm$4.3 &  5.1$\pm$1.9 &  3.7$\pm$1.2 &  75$\pm$ 65 & 0.03$\pm^{0.02}_{0.03}$ &                   AOR:17415424 \\ 
 SWIRE 3-18 &  4.9$\pm$2.3 & 15.5$\pm$3.5 &  5.1$\pm$1.6 &  5.0$\pm$1.1 &  37$\pm$  6 & 0.19$\pm^{0.05}_{0.05}$ &                   AOR:17416960 \\ 
    SMM J03 &  4.3$\pm$1.3 & 31.3$\pm$3.7 &  5.0$\pm$1.9 &  6.2$\pm$2.4 &  71$\pm$ 14 & 0.24$\pm^{0.07}_{0.07}$ &         AOR:13983744, 14007040 \\ 
  SWIRE 4-5 &  4.1$\pm$0.3 & 13.2$\pm$0.6 &  3.7$\pm$0.4 &  2.4$\pm$0.4 &  31$\pm$  4 & 0.14$\pm^{0.04}_{0.04}$ &                \citep{fio2010} \\ 
 SWIRE 4-15 &  5.3$\pm$0.3 & 15.7$\pm$1.5 &  3.1$\pm$0.4 &  4.4$\pm$0.5 &  36$\pm$  2 & 0.17$\pm^{0.04}_{0.04}$ &                \citep{fio2010} \\ 
   SDSS J12 &  7.6$\pm$0.3 & 17.0$\pm$0.8 &  4.4$\pm$0.3 &  5.3$\pm$0.6 &  40$\pm$  2 & 0.14$\pm^{0.03}_{0.03}$ &                \citep{ruj2012} \\ 
\hline
\multicolumn{9}{l}{(Stacey et al. 2010) sources:}	 \\ \hline
 SMM J22471 & 10.3$\pm$3.1 & 38.2$\pm$5.0 & 15.9$\pm$5.0 &  9.1$\pm$2.8 &  85$\pm$ 10 & 0.11$\pm^{0.03}_{0.03}$ &                   AOR:16175616 \\ 
    SMM J12 & 10.7$\pm$0.2 & 46.1$\pm$2.0 &  6.1$\pm$0.6 &  6.7$\pm$0.5 & 152$\pm$ 38 & 0.04$\pm^{0.01}_{0.01}$ &                \citep{pop2008} \\ 
MIPS J14282 & 23.2$\pm$3.6 & 94.7$\pm$10.1 & 19.8$\pm$2.5 & 11.1$\pm$1.4 &  --- & $<$0.12$^1$  				&                AOR:12513536 \\ 
  SWIRE L25 &  4.4$\pm$0.3 & 12.6$\pm$0.9 &  2.8$\pm$0.6 &  3.2$\pm$0.9 &  28$\pm$  2 & 0.06$\pm^{0.02}_{0.02}$ &                \citep{fio2010} \\ 
  SWIRE L17 &  2.9$\pm$0.3 & 14.4$\pm$1.6 &  3.6$\pm$0.6 &  2.8$\pm$0.8 &  33$\pm$  6 & 0.09$\pm^{0.02}_{0.02}$ &                \citep{fio2010} \\

 \tablecomments{AOR refers to AORkey codes used to acquire spectra from CASSIS \citep{leb2011}. References refer to publications of PAH spectra. MIR spectra for the sources with references are published spectra for which electronic copies of the data were acquired through private communication with publication authors. \\
$^1$ 7.7/11.2 $\mu$m PAH feature flux ratio makes Equation \ref{eq:pahtotest} poorly defined for MIPS J14282. Upper limit set by total of 6.2, through 11.2 $\mu$m PAH features.}
\end{tabular}
\end{center}
\label{tab:pahs}
\end{table}

\subsection{PDRs}\label{sec:individuals}

Fine structure [CII] and [OI] line emission in normal star forming galaxies is generally explained in a PDR paradigm. To explore PDRs as a source of the observed line emission we use the PDR toolbox \citep{pou2008,kau2006} to analyze our source properties. The PDR models provide estimates of several PDR gas cooling lines over a large phase space of density, n (cm$^{-3}$), and FUV irradiation, G$_0$. For the majority of our sources the useful constraints on PDR characteristics are based on the [OI]/[CII] ratio, which rises with increasing density and G$_0$, and either [CII]/IR or ([CII]+[OI])/IR which characterizes the gas heating efficiency and decreases with G$_0$.\footnote{The infrared continuum used in the PDR toolbox corresponds to the wavelength range 30-1000 $\mu$m \citep{far2013}.} The output of the PDR models is the intrinsic line emission from the PDR. Before we can interpret the model outputs in terms of physical conditions we need to translate between the observed and intrinsic line emission. As a simple model, we assume a dual-slab molecular cloud geometry in which both faces of externally irradiated molecular clouds harbor PDRs. Cloud-to-cloud velocity variation generally allows us to observe emission from multiple clumps without optical depth effects. The [OI] line, however, is often optically thick on the surfaces of individual molecular clouds \citep{sta1983}, so we will only detect [OI] emission from the front surfaces of clouds.  To account for this, we multiply the observed [OI] flux by two to match the plane parallel models in the PDR toolbox. Geometry and velocity dispersion may vary, but results from this simple approximation are generally successful at characterizing observations on a galactic scale \citep[eg.][]{mal2001,vas2010}.

[CII], as previously mentioned, arises in both neutral and ionized gas. Without other observations characterizing the ionized medium we cannot be certain what fraction of the observed [CII] should be accounted for by the PDR models. Previous studies of many systems with [NII] and [CII] have shown the fraction of [CII] from PDRs generally ranges from $\sim$40-90\% \citep{mal2001,vas2010}. Although recent observations of extreme Lyman Alpha Emitters have revealed rare sources in which most of the [CII] arises in HII regions, \citet{dec2014} confirm that in SMGs [CII] is dominated by PDRs. In particular, in nearby starburst sources NGC 253 \citep{car1994} and M82 \citep{lor1996,col1999} combined HII and PDR modeling has shown $\sim$70\% of the [CII] emission comes from PDRs and we take these sources as representative analogs of our systems. This fraction is also very similar to the median contribution to [CII] from PDRs in the sample of \citet{vas2010}. As a representative model, we therefore assume PDR-derived [CII] emission is equal to 70\% of observed. Even if the actual amount of [CII] from PDRs in our systems is as small as 40\% (the minimum in the sample of \citet{vas2010},) this would only modestly effect our derived PDR conditions, raising the log(G$_0$) of our sample by an average of 22\%. Our conclusions based on the modest strength of the FUV fields are therefore relatively robust to the uncertainty in [CII] attributed to PDRs.

\begin{figure}[htb]
	\centering
		\includegraphics[width=0.85\textwidth,trim=.2cm 0cm 0.2cm 2.2cm, clip=true]{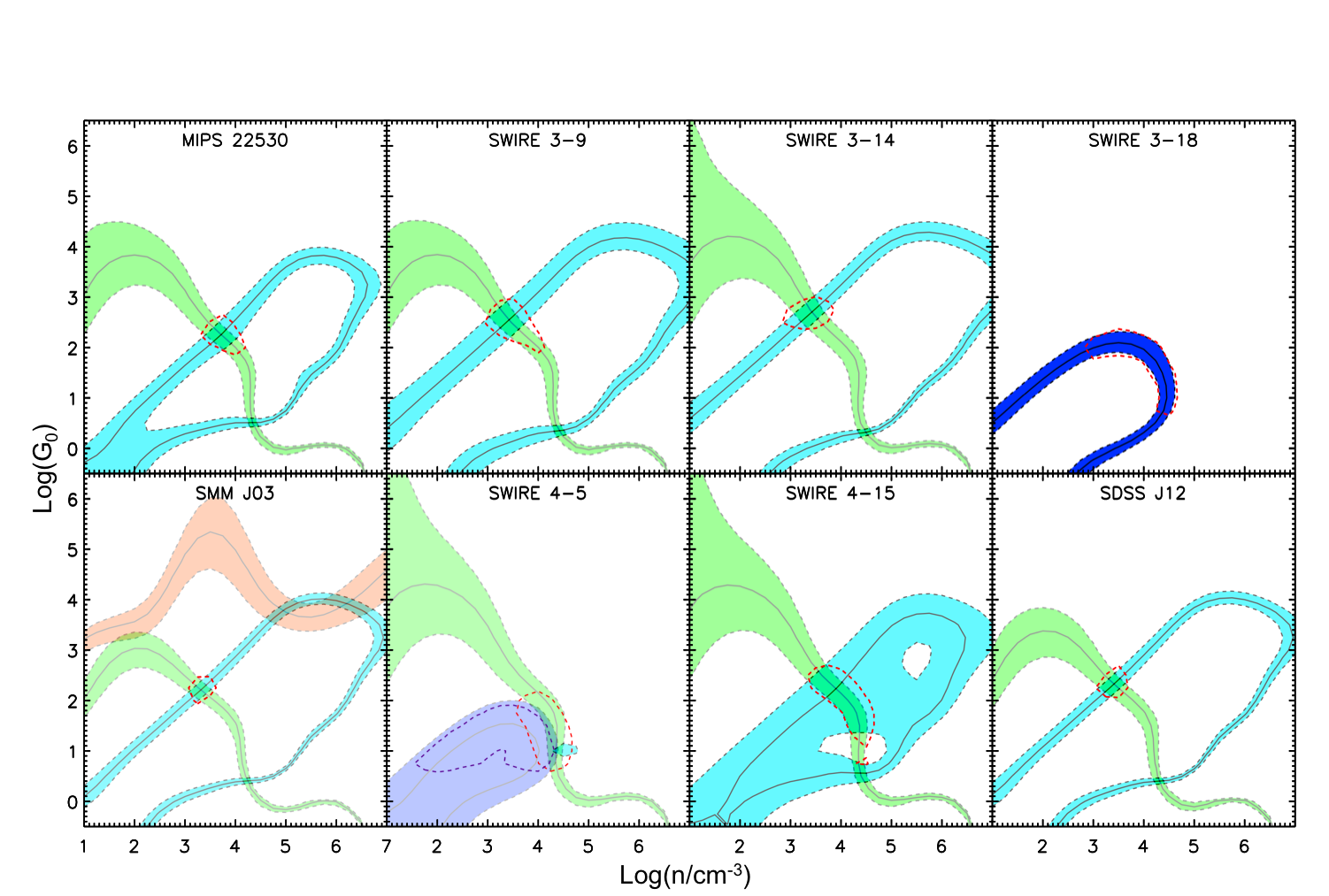}
	\caption[PDR diagnostic plots]{PDR diagnostic plots. Color shaded contours show constraints on n and G$_0$ based on the following ratios: green [OI]/[CII], cyan ([OI]+[CII])/IR, blue [CII]/IR, orange [OI]/CO(3$\rightarrow$2). Solid lines note ratio values and shaded areas represent $\pm$1$\sigma$ regions. We attribute 70\% of observed [CII] to the classic PDR model. Also overlaid as a dashed red line is the region containing 68.27\% of the volume in the two dimensional probability distribution of n and G$_0$. For source SWIRE 4-5, which has a tentative [OI] detection, we additionally show a purple dashed region indicating the 68.27\% region that would result if we take the [OI] detection as an upper limit instead. Note that in this source the UV field is constrained to low intensities largely due to the [CII]/IR ratio, and the [OI] measurement mostly effects the density determination. We have assumed G$_0>$10$^{0.5}$ as a prior to ignore the low G$_0$ high density solution which, as discussed in the text, is unrealistic for galaxy averaged properties. In SWIRE 3-18 we lack a useful [OI]/[CII] ratio to constrain density so we have additionally assumed n$\sim$10$^3$-10$^5$ cm$^{-3}$, consistent with other star forming sources at high redshift \citep{sta2010}.}
	\label{fig:pdrmegaplot}
\end{figure}

In the analysis that follows we compare PDR model grids of [CII], [OI], and IR continuum to our observations to create two-dimensional likelihood functions of the density and FUV irradiation strength that characterize each of our sources. We then integrate the likelihood function over one parameter at a time to create marginalized probability distributions for log(n) and log(G$_0$). We characterize these distributions using the most likely value with asymmetric error bars capturing 68.27\% of the total probability distribution. In many cases our observations support a double peaked two dimensional likelihood function with one maxima at modest n and G$_0$ and one at high n and very low G$_0$. We favor the modest n, G$_0$ solution as the density is more representative of a mixed phase galaxy average, and as we show in Section \ref{sec:distscale}, a very low G$_0$ value is difficult to reconcile with the large luminosities of our systems. We have therefore imposed a prior to rule out solutions with log(G$_0$)$\leq$0.5. Table \ref{tab:powersource} gives the PDR parameters required by the fitted models, and Figure \ref{fig:pdrmegaplot} shows the PDR parameter space for our systems with constraints imposed by our observed line ratios.

\begin{table}[htb]\small
\caption[PDR model parameter fits]{PDR model parameters.}
\vspace{2.5mm}
\begin{center}
\begin{tabular}{ c c c c c c c c}
	\hline
		
		Source 				&  		 			&	PDR parameters	&				 		\\ 
							&	log(n cm$^{3}$)	&	log(G$_0$)			&	size (kpc)		\\ \hline

MIPS 22530 & 3.75$\pm^{0.25}_{0.25}$ & 2.25$\pm^{0.25}_{0.25}$ &  2.9-9.1 \\
SWIRE 3-9 & 3.5$\pm^{0.5}_{0.5}$ & 2.5$\pm^{0.25}_{0.25}$ &  2.2-6.3 \\
SWIRE 3-14 & 3.5$\pm^{0.25}_{0.5}$ & 2.75$\pm^{0.25}_{0.25}$ &  1.4-3.0 \\
SWIRE 3-18 & 3-5 & 2.0$\pm^{0.25}_{0.5}$ &  2.3-6.5 \\
SMM J03 & 3.25$\pm^{0.25}_{0.25}$ & 2.25$\pm^{0.25}_{0.25}$ &  2.6-8.0 \\
SWIRE 4-5 & 4.25$\pm^{0.25}_{0.5}$ & 1.25$\pm^{0.5}_{0.5}$ &  3.1-10.0 \\
SWIRE 4-15 & 4.25$\pm^{0.25}_{0.5}$ & 2.25$\pm^{0.25}_{0.75}$ &  2.3-6.4 \\
SDSS J12 & 3.5$\pm^{0.25}_{0.25}$ & 2.25$\pm^{0.25}_{0.25}$ &  2.6-7.6 \\

		\tablecomments{The size scale represents the summed areal extent of star formation powered PDRs. It is a representative value only and its range is based on the most likely G$_0$ value under the separate assumptions of small and large mean free photon paths relative to cloud size as outlined in section \ref{sec:distscale} (and does not account for the uncertainty on G$_0$ or the uncertainty in the fraction of [CII] from PDRs). Representative n and G$_0$ values give the best PDR solution assuming 70\% of observed [CII] is due to PDRs. Error ranges are such that 68.27\% of the power in the marginalized probability distribution of each parameter is contained within.}

\end{tabular}
\end{center}
\label{tab:powersource}
\end{table}

\textit{MIPS 22530}
Our [CII] and [OI] detections of this source are blueshifted relative to the optical redshift by $\sim$100 km s$^{-1}$ and $\sim$200 km s$^{-1}$ respectively. MIPS 22530 is detected in all 3 MIPS bands and all three SPIRE bands, providing a well determined FIR SED. Its SED-derived luminosity is log(L$_{FIR}$/L$_\odot$)=12.69$\pm$0.04.
The ratio of the two fine structure lines is similar to normal local galaxies, while the 60/100 $\mu$m flux density ratio is slightly warmer (see Figure \ref{fig:ocratiocolor}.) This source may also have a marginal AGN contribution \citep{saj2008}. This could be giving rise to a minor hot dust region which contributes to the 60 $\mu$m emission without dominating the [OI] and [CII] lines.
The fine structure lines in this source can be well described with a PDR model. Figure \ref{fig:pdrmegaplot} shows the overlapping constraints placed on our PDR model from the [OI]/[CII] ratio and ([OI]+[CII])/IR. The multi-peaked probability distribution in our parameter space can be seen as the result of the two distinct intersections between the [OI]/[CII] contour and ([OI]+[CII])/IR. As discussed above we favor the solution with modest density and G$_0$ rather than high density and low G$_0$. The PDR model constrains density, n$\sim$10$^{3.75}$cm$^{-3}$ and UV flux G$_0$$\sim$10$^{2.25}$.

\textit{SWIRE 3-9}
Both [CII] and [OI] are strongly detected in this source, although there appears to be a slight offset between the line centers, likely due to calibration error in the [CII] observation. Given the greater velocity resolution in the PACS [OI] spectrum, we adopt its indicated redshift of z=1.732$\pm$0.003 with error bars encompassing our [CII] detection. The fine structure lines are significantly redward ($\sim$2500 km s$^{-1}$) of the PAH-derived redshift, but agree within 1$\sigma$ of the PAH redshift uncertainty. The PAH features are also well fit using our adopted redshift and the fitting quality is nearly equivalent to fits using the indicated PAH redshift. The fine structure lines and 60/100 $\mu$m flux density ratio are very similar to MIPS 22530. The fine structure lines are consistent with a star formation powered PDR source, and although the 60/100 $\mu$m flux density ratio is slightly warmer than our average source, the large uncertainty does not allow us to distinguish it from the normal local trend. Shown in Figure \ref{fig:pdrmegaplot}, [CII], [OI], and the FIR continuum constrain the PDR density, n$\sim$10$^{3.5}$ and UV flux G$_0$$\sim$10$^{2.5}$.

\textit{SWIRE 3-14}
Our [CII] line, detected at z=1.7795, is consistent with PAH observations. Our PACS observations of this source detect the [OI] line at the [CII] velocity at $\sim$3$\sigma$. The FIR SED of this source has a relatively low 60/100 $\mu$m flux density ratio. Although it has a high [CII]/FIR ratio, Figure \ref{fig:rvscolor} shows it is similar to local star forming systems with equivalent FIR colors, suggesting this system is forming stars like local normal galaxies scaled up in size (or lensed.) As shown in Figure \ref{fig:pdrmegaplot}, [CII], [OI], and the FIR continuum constrain the PDR density, n$\sim$10$^{3.5}$ and UV flux G$_0$$\sim$10$^{2.75}$.

\textit{SWIRE 3-18} is one of the strongest [CII] emitters with [CII]/FIR=0.022$\pm$0.004.  We have compiled SPIRE photometry from HerMES to determine it has a luminosity log(L$_{FIR}$/L$_\odot$)=12.13$\pm$0.13 and a cool dust temperature with F$_\nu$(60)/F$_\nu$(100)=0.43$\pm^{0.15}_{0.14}$, similar to SWIRE 4-5 and 4-15.  This is consistent with previous claims that it is a SF-D source. Without an [OI] observation we cannot fully disentangle density and G$_0$ degeneracies in the PDR model, but imposing an additional prior assumption that n$\sim$10$^{3-5}$ cm$^{-3}$ (a typical value for galaxies \citep{sta2010}) we find G$_0$$\sim$10$^{1.75}$. (Figure \ref{fig:pdrmegaplot}.)

\textit{SMM J03}
has a warm dust color, a high [CII]/FIR ratio, and an [OI]/[CII] ratio that falls below local sources with equivalent 60/100 $\mu$m flux density ratios. There are indications for marginal contribution from an AGN in this system \citep{swi2004,tak2006}. As we suggested for MIPS 22530, the [OI] and [CII] emission may be dominated by star formation spread throughout the disk, while an AGN fueled hot dust region may contribute to the 60 $\mu$m flux. In addition, CO 3$\rightarrow$2 has been detected in this source, F$_{CO(3\rightarrow2)}$=(6.4$\pm$1.9)$\times$10$^{-21}$ W m$^{-2}$ (Steve Hailey-Dunsheath, private communication.) Figure \ref{fig:pdrmegaplot} shows our PDR model for the fine structure lines ([CII], [OI]), FIR continuum, and CO emission. Our PACS photometry along with SCUBA data \citep{kov2006} provide a well constrained SED with log(L$_{FIR}$/L$_\odot$)=12.58$\pm$0.03.
Our [CII], [OI], and infrared continuum observations are consistent with a standard PDR model, the CO 3$\rightarrow$2 flux however, appears too low for a standard PDR given our observed fine structure lines. It would need to be a factor of ten higher for the [OI]/CO 3$\rightarrow$2 ratio (orange contour, Figure \ref{fig:pdrmegaplot}) to be in agreement with the other line ratios. This is probably evidence for low metallicity, as suppressed CO is a common feature in low metallicity systems \citep{sta1991,pog1995,smi1997}. Since we do not have other constraints on the metallicity of this system, we ignore the CO flux in fitting our PDR model. The [OI], [CII], and infrared continuum, are consistent with star formation powered PDRs with a modest UV intensity G$_0$$\sim$10$^{2.25}$, and density n$\sim$10$^{3.25}$cm$^{-3}$.

\textit{SWIRE 4-5}\label{sec:sw45}
The [CII] line is detected at z=1.756, consistent with the previous PAH determined redshift. The SED fit to FIR photometry reveals log(L$_{FIR}$/L$_\odot$)=11.73$\pm$0.19 and F$_\nu$(60)/F$_\nu$(100)=0.30$\pm^{0.24}_{0.01}$, making it the coolest dust temperature source in our sample. Furthermore, the resulting [CII]/FIR ratio, 0.042$\pm$0.01, makes this one of the most exceptional [CII] emitters known. 
The [OI] spectrum shows a marginal detection. The line is consistent with the [CII] velocity and a normal local [CII]/[OI] ratio, however the spectrum is taken from an off center PACS pixel corresponding to 9.3 arc seconds off the nominal position. We present it here as a tentative detection.

It is reasonable to conclude that the [CII] and [OI] are coming from the same region in this source, but their particularly high fluxes relative to the FIR make it difficult to rectify with a classic PDR model in which UV photons dominate the gas heating. In particular, the maximum ([CII]+[OI]) to infrared continuum ratio predicted by the PDR models is only just able to capture the lower 1$\sigma$ bound of our observations. In table \ref{tab:powersource} and Figure \ref{fig:pdrmegaplot} we show that this solution implies especially low FUV fields G$_0$$\sim$10$^{1.25}$. Since the [OI] detection is only tentative we also show the [CII] to infrared continuum ratio in Figure \ref{fig:pdrmegaplot} and note that if we treat the [OI] observation as an upper limit, PDR models still require an extremely low UV field (G$_0$$\sim$10$^{1.75}$.)

Although PDRs are capable of recreating the observed line emission, they require very low intensity UV fields to dominate over the galaxy, and in Figures \ref{fig:rplot} and \ref{fig:ocratiocolor}, SWIRE 4-5 is an outlier from trends seen in the local Universe regardless of its uncertain [OI] emission or assumptions over PDR models. We discuss the implications of these results below.

\textit{SWIRE 4-15}
Our [CII] detection at z=1.8540 is well within the PAH determined redshift range. FIR photometry shows an SED dominated by a modestly cool dust component (with large uncertainty on the 60/100 $\mu$m flux ratio) peaking at $\sim$100 $\mu$m, suggesting low FUV intensity extended star formation. We find log(L$_{FIR}$/L$_\odot$)=12.36$\pm$0.15 and F$_\nu$(60)/F$_\nu$(100)=0.69$\pm^{0.29}_{0.19}$. We have a broad line detection in [OI], (F$_{[OI]}$=4.4$\pm$1.1$\times$10$^{-18}$ W m$^{-2}$,) yielding a [CII]/[OI] ratio similar to normal galaxies. Our PDR model requires modest FUV fields densities to explain the emission (G$_0$$\sim$10$^{2.25}$, n$\sim$10$^{4.25}$.)

\textit{SDSS J12}
This source was observed and detected in [OI] by Sangeeta Malhotra and well constrained in the FIR continuum by PACS and SPIRE photometry measurements from the {\it SEDs and energetics of lensed UV-bright high redshift galaxies} project. We add to the discussion our [CII] observation, and the PAHfit MIR spectrum. The FIR photometry yields an apparent log(L$_{FIR}$/L$_\odot$)=12.54$\pm$0.03. Accounting for the system magnification ($\sim$27,) the intrinsic luminosity suggests that SDSS J12 is a relatively modest star forming source in terms of overall luminosity. SDSS J12 has the warmest FIR color in our sample, F$_\nu$(60)/F$_\nu$(100)=1.23$\pm^{0.08}_{0.20}$.  This may be due to exceptionally strong recent or ongoing star formation, which would cause increased abundance of hot dust along with ionized gas reserves \citep{hai2009}. Alternatively, this source may in fact have have an AGN that contributes significantly to the 60 $\mu$m emission. In Figure \ref{fig:pdrmegaplot} we show that a PDR model with log(n)$\sim$3.5 and log(G$_0$)$\sim$2.25 can explain the [CII] and [OI] emission along with the continuum.

\section{Discussion}\label{sec:distscale}
Our source sample is dominated by galaxies that lie along the galaxy main-sequence, powered by star formation. This is a reflection of our selection bias towards star forming sources due to our selection criteria requiring PAH emission. We have confirmed our previous result \citep{sta2010} that the [CII] deficit is not a ubiquitous phenomenon in star formation powered ULIRGs at redshifts 1-2.

If we interpret our observations within the PDR paradigm, as is the common practice for [CII] and [OI], then the relative strengths of [CII], [OI], and the infrared continuum characterize the density of the PDR media and the intensity of the local FUV field. In general, the densities that characterize our sources are of order or slightly greater than the [CII] critical density. This suggests that the [CII] line is cooling the PDR gas at nearly maximum efficiency, consistent with the high [CII]/FIR ratios we observe.

The FUV intensity we derive from the PDR models can be related to the integrated source luminosity. Essentially all of the FUV that impinges on neutral gas clouds is absorbed by dust and reradiated in the FIR continuum. Therefore, the observed infrared intensity measures the average FUV field intensity within our telescope beam. The PDR-derived FUV field intensity, (G$_0$,) and our beam-averaged field intensity are thus related by the source beam filling factor.

We can consider this beam filling factor in the form of a back of the envelope calculation. \citet{wol1990} found that a cloud of size D, and luminosity, L$_{IR}$, should have an average FUV field given by $G_0\propto \lambda L_{IR}/D^3$ where $\lambda$, the mean free path of a photon, is much smaller than the cloud extent, D; or $G_0\propto L_{IR}/D^2$ if $\lambda \gtrsim D$. Detailed models of the physical conditions of the molecular ISM in the star forming regions of the nearby starburst galaxy M82 are remarkably similar to the average values we find for our sources, namely, n$\sim$10$^4$ cm$^{-3}$ and G$\sim$10$^{3}$ \citep{lor1996}, so that we use it as a template by which we scale our redshift 1-2 sample. Note that strictly this comparison also requires that the relative distribution of star formation (extended vs nuclear) be similar between M82 and our sources. Without resolved observations of our sample it is unclear whether this is the case, so we proceed with the expectation that our findings will be indicative of the order of magnitude only.

For M82 if we take D$\sim$300 pc \citep{joy1987} and L$_{FIR}$$\sim$2.8$\times$10$^{10}$L$_\odot$ we find that (D/pc)$^3\sim$0.96 (L$_{FIR}$/L$_\odot$)/G$_0$ or (D/pc)$^2\sim$3.2e-3 (L$_{FIR}$/L$_\odot$)/G$_0$. As noted in Table \ref{tab:powersource}, we find the spatial extent of star formation in all of our sources is $\gtrsim$1 kpc, implying that star formation is occurring over a large fraction of the galactic disk. Note that a lower FUV field intensity, G$_0\sim$10$^{0.25}$, which often appears as a secondary solution in our PDR models comparing [CII] and [OI] (see Figure \ref{fig:pdrmegaplot}), would imply spatial scales an order of magnitude larger which is not reasonable, so we take the higher FUV field solution. Similar analysis by \citet{dia2014} of a sample of local LIRGs shows that there is indeed a strong correlation between the FUV intensity and emitting region size, but suggests that, on a galaxy-wide scale on average, $G_0\propto L_{IR}/D^\alpha$ where $\alpha < 2$. This implies that, if anything, our estimates of the emitting size regions is an underestimate.

Our finding of kpc scale star formation is consistent with the recent observations of spatially resolved CO in high redshift sources \citep{tac2010,dad2010,ivi2011,rie2011}. As \citet{tac2010} point out, the large spatial distribution of molecular gas does not necessarily imply a single coherent disk of star formation, but instead is very likely the observation of widely distributed clumps which are not individually resolved, but are all likely undergoing star formation obeying a Schmidt-Kennicutt law. 

\begin{figure}[htb]
	\centering
		\includegraphics[width=0.8\textwidth,trim=0.1cm 1.9cm 0.1cm 2.4cm, clip=true]{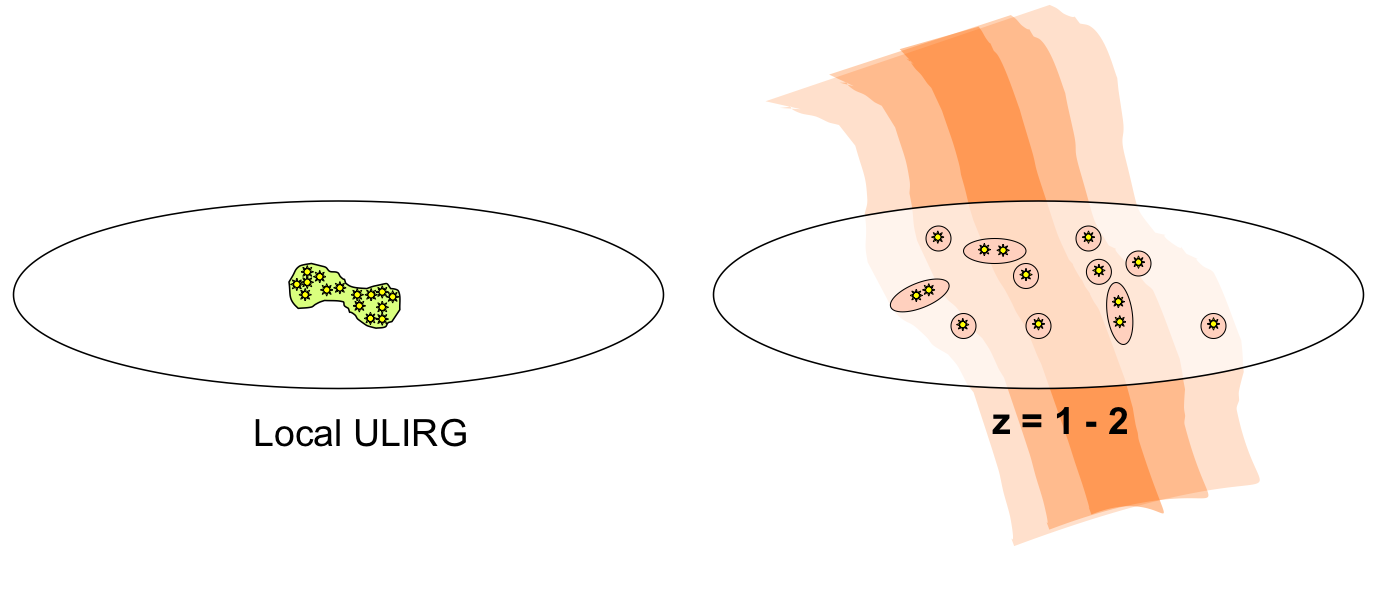}
	\caption[Star formation schematic representation]{A schematic representation of star formation in local ULIRGs (left) vs. star formation powered ULIRGs at z=1-2 (right.) Local ULIRGs may show one or more compact regions of intense star formation activity with strong UV fields and high dust temperatures, often powered by recent galaxy merging. At redshifts 1-2 the modest UV fields discussed here (as well as the spatially extended molecular gas observed by \citep{tac2010,dad2010,ivi2011,rie2011}) indicate star formation spread out over a significant fraction of the galaxy's disk. We suggest that intergalactic gas accretion from the cosmic web (indicated by shaded region) fuels such widespread star formation.}
	\label{fig:ulirgcartoon}
\end{figure}

As shown in Table \ref{tab:powersource}, the combined [CII], [OI], and and infrared continuum observations can only be fit within the PDR paradigm if the source size is, in all cases at least a kpc, and in many potentially much larger. This is in sharp contrast with local ULIRGs which have intense, and concentrated collision induced bursts of star formation occurring on sub-kpc scales. Figure \ref{fig:ulirgcartoon} presents a cartoon schematic of these different modes of star formation.

The large spatial scales of star formation are best understood as the expected star formation that results from large and abundant molecular reserves under a Schmidt-Kennicutt law. The large molecular gas aggregation likely results from gas accretion from the cosmic web, and not coalescence from major mergers, which models show produce very intense, but spatially concentrated star formation resulting in low [CII]/FIR ratios as seen in local ULIRGs. Under this interpretation we suggest that, at least in our class of SF-D galaxies detected in PAH emission, much of the star formation in the epoch of peak star formation resulted from gas accretion from the cosmic web, and not solely merger activity as was once assumed.

In addition to [CII] and [OI] 63 $\mu$m observations, we have also compiled IRS spectra which we have made use of here to consider PAH emission. In our sample the PAH features are strong but, as we show in Figure \ref{fig:ciipahratio}, the [CII] to PAH ratios in many of our sources exceed those of local star formation regions.

This accumulated wealth of data has allowed an unprecedented look at the nature of these galaxies, and in particular their dominant power sources for the far infrared cooling lines. The standard approach, using classic PDRs, can adequately produce the line and continuum emission we see. This does not fully explain why we would see such extensive low intensity UV PDRs in the case of SWIRE 4-5 (and SWIRE 3-18 to a lesser extent,) to produce a [CII]/FIR ratio that exceeds observations in the local Universe. Indeed, the fine structure line emission relative to the infrared continuum in SWIRE 4-5 is brushing up against the limits of what classic PDRs are capable of producing. It would require clumps of uniformly low FUV powered PDRs distributed throughout the galaxy to produce such a high [CII]/FIR ratio.

In addition to the difficulty in modeling the fine structure line emission in these systems with PDRs, many of our sources (including SWIRE 4-5) have high [CII]/PAH ratios compared to nearby star forming sub-galactic regions. This feature is particularly reminiscent of the work by \citet{gui2012} and \citet{app2013} on Stephan's Quintet, a nearby interacting compact group of galaxies. Their work shows enhanced CO and [CII] emission in filamentary structures between the main sources in the group. They have shown that this is a direct result of shock heated gas. In these shock powered regions \citet{app2013} finds [CII]/FIR and [CII]/PAH ratios a factor of a few higher than we see (as shown in Figures \ref{fig:rplot} and \ref{fig:ciipahratio},) as well as cool FIR dust temperatures, similar to SWIRE 4-5 (Figures \ref{fig:fircolor} and \ref{fig:rvscolor}.)

We have already seen cases where microturbulence contributes significantly to the overall heating budget of various lines on a galaxy-wide scale. In earlier work on NGC 253 and NGC 891 we found that microturbulence was needed in addition to classic PDRs to explain neutral and molecular gas emission \citep{hai2008,sta2010b,nik2011}. Implicating microturbulence to explain [CII] emission requires a source of radiation to ionize the carbon. Models by \citet{les2013} (and invoked in Stephan's Quintet by \citet{app2013},) showed that diffuse gas undergoing low velocity turbulent shocks that heat the gas can cool by emission in the [CII] line if C+ is formed within the cloud by modest FUV radiation fields. 

Although the resolved regions of Stephan's Quintet and our sources represent very different systems (shocked filamentary structures embedded in the IGM between interacting systems vs. unresolved (U)LIRGs,) given the displacement of SWIRE 4-5 from local galaxy trends, and the empirical similarity between it and shock powered regions of Stephan's Quintet, it may be that SWIRE 4-5 represents a hybrid systems involving significant contributions from both classic PDRs and shock powered regions. Shock heating may in fact be a common contributor to [CII] emission especially in the early Universe as IGM gas accretion becomes more ubiquitous.

This turbulent heating does not necessarily imply violent galaxy mergers, however. Indeed, low velocity shocks ($\sim$8 km s$^{-1}$) can efficiently heat gas and might contribute significantly to the observed [CII] line. Such low velocity shocks might be the result of galaxy interaction, as in the case of Stephan's Quintet, or might easily result from star formation feedback processes such as stellar outflow and supernovae, or impacting clumps and gas streams accreting from the intergalactic medium. With our spatially unresolved observations it is impossible to distinguish between these possibilities. To characterize the shock properties and the relative contributions from shocked diffuse gas and PDRs would either require spatially resolved observations or more diagnostic emission lines. MIR H$_2$ lines are particularly useful shocked gas diagnostics, unfortunately their typical luminosities are one or two orders of magnitude weaker than PAH features, putting them below the sensitivity of our MIR spectra (Figure \ref{fig:pahspectra}.) Upper limits on the H$_2$ S(3) 0-0 line are noted in the appendix Table \ref{tab:appsourceprops}  but are unfortunately too high to constrain shock or PDR models. Although in SWIRE 4-5 (and to a lesser extent SWIRE 3-18,) the very strong [CII] emission with respect to FIR and PAHs is a good indication that additional sources of heating are contributing to the [CII] emission, we cannot definitively say this is the case without further infrared and submillimeter observation. If similar sources are found to be common in the early Universe, characterizing their emission mechanisms will be paramount to understanding galaxy evolution.

\section{Conclusions}\label{sec:ciiconclusions}
We have detected eight new z=1-2 sources in the [CII] 157.7 $\mu$m line. Seven of them were also observed in the [OI] 63 $\mu$m line (six in our Herschel program and the seventh by Sangeeta Malhotra.) Far infrared photometry, made available through ours and other Herschel programs, has allowed us to establish SEDs providing reliable FIR luminosities and estimates of the FIR dust temperature.

Our sample selection required sources be detected in PAH emission, biasing the sample towards sources with FIR luminosities dominated by star formation rather than AGN. In general we have found that the relative [CII] and [OI] emission is comparable to local star forming systems and can largely be explained with classic PDR models. There is, however, very strong [CII] relative to both FIR continuum and PAH emission. Six of our eight sources exhibit both [CII]/FIR$>$10$^{-2}$ and [CII]/PAH$_{tot}>$0.1. In most of our sources their [CII]/FIR ratio is a factor of a few greater than found for local sources with comparable 60/100 $\mu$m flux density ratios.

We have used [CII]/FIR to determine G$_0$ based on a PDR paradigm, and with L$_{FIR}$ we determine the PDR filling factor in our beam. Based on this model, the extent of our star forming regions are very large, unlike local ULIRGs. This supports the scenario in which extended, moderate intensity, star formation pervades their disks, likely stimulated by large molecular gas reservoirs accreted from the cosmic web. Although drawing conclusions about the general z=1-2 galaxy population based on our sample is beyond the scope of this paper, the results from our sample are consistent with the idea that much of the star formation at z$\sim$1-2 is not driven by major mergers, but instead by cold flow accretion. Although there are several examples of major mergers in the early Universe, our sample suggests cold flow accretion fueled galaxy star formation abounds as well. Our [OI] and PAH detections further confirm this interpretation of [CII].

While the [CII] and [OI] likely arise from classic PDRs, we must qualify the standard PDR paradigm with the caveat that the exceptionally strong [CII] emission from SWIRE 4-5 may be powered by other heating mechanisms as well. The high [CII]/FIR ratio is significantly larger than what is seen locally and even approaches the limit of what PDRs can provide. It's cool FIR dust temperature suggests that any additional mechanisms that may be heating the gas must not be producing significant amounts of hot dust. This is consistent with directly heating the gas via low velocity shocks in a diffuse gas component as modeled by \citet{les2013} and invoked by \citet{app2013} to explain high [CII] ratios in regions of Stephan's Quintet.

The [CII] line remains a useful probe in the z$>$1 Universe. It was once assumed that large-scale mergers dominated the star formation in this epoch which would lead to compact and intense star formation regions with relatively weak [CII] emission like local ULIRG galaxies. The discovery of very extended [CII] emission is a key element of the new paradigm that in many systems, cold accretion from the cosmic web fuels very high gas surface densities leading to enhanced, wide-scale star formation following a Schmidt-Kennicutt law typical of normal galaxies.

With the observational avenues that have opened in the last few years, we are now able to acquire more diversity of data and work with better refined models than ever before. This presents us the welcome challenge of transitioning from studying broad brush characterization of star formation in general to examining individual sources. Going forward, the [CII] line, [OI], PAHs, and other multiwavelength data will allow us an exceptional look at the dynamics of individual galaxies, building on the picture presented here by further constraining the merger fraction of star forming galaxies throughout cosmological time.   

%I like tildes: {\~{n}} blah \~{n} blah \~n blah ~n blah `n blah 'n blah %$\tilde{n}$ blah $\~{n}$

\acknowledgements{We thank the anonymous referee for insightful comments on previous drafts of this paper.

ZEUS observations were supported by NSF grants AST-0705256, AST-0722220, and AST-1109476.  We would also like to thank the staff of the CSO for their support of ZEUS operations.

This research has made use of data from HerMES project (http://hermes.sussex.ac.uk/). HerMES is a Herschel Key Programme utilising Guaranteed Time from the SPIRE instrument team, ESAC scientists and a mission scientist. HerMES is described in \citep{oli2012}.

The HerMES data was accessed through the HeDaM database (http://hedam.oamp.fr) operated by CeSAM and hosted by the Laboratoire d'Astrophysique de Marseille.

The National Radio Astronomy Observatory is a facility of the National Science Foundation operated under cooperative agreement by Associated Universities, Inc.

}

\clearpage
\section{Appendix}

\begin{table}[htb]\tiny
\caption{Photometry used in SED fitting.}
\vspace{2.5mm}
\begin{center}
\begin{tabular}{ c c c c c c c c c c}
	\hline

		SWIRE 4-5		&	SWIRE 4-15			&	SWIRE 3-14			&	SWIRE 3-18			&	SWIRE 3-9		&	MIPS 22530		&	SDSS J12			&	SMM J03								\\ \hline
	MIPS 24				&	MIPS 24				&	MIPS 24				&	MIPS 24				&	MIPS 24			&	MIPS 24			&					&	MIPS 24		\\
	0.550$\pm$0.015 [1]	&	0.419$\pm$0.018 [1]	&	0.874$\pm$0.016 [2]	&	0.761$\pm$0.076 [2]	&	1.0$\pm$0.1 [2]	&	1.23$\pm$0.06 [3] &					&	0.23$\pm$0.02 [4] 					\\
						&						&						&						&					&					&					&	0.47$\pm$0.05 [5]					\\
						&						&						&						&					&					&					&													\\
						&						&						&						&					&	MIPS 70			&	PACS 70			&	PACS 70							\\
						&						&						&						&					&	3.4$\pm$1.9 [6]	&	9.7$\pm$3.3 [7]	&	21.3$\pm$4.4 [8]									\\
						&						&						&						&					&					&					&											\\
						&						&						&						&					&					&	PACS 100			&	PACS 100						\\
						&						&						&						&					&					&	16.5$\pm$1.5 [7]	&	48.6$\pm$3.5 [8]								\\
						&						&						&						&					&					&					&											\\
						&						&						&						&					&	MIPS 160			&	PACS 160			&	PACS 160							\\
						&						&						&						&					&	38.0$\pm$9.0 [9]	&	39$\pm$21 [7]	&	64.8$\pm$6.5 [8]								\\
						&						&						&						&					&					&					&											\\
	SPIRE 250			&	SPIRE 250			&	SPIRE 250			&	SPIRE 250			&	SPIRE 250		&	SPIRE 250		&	SPIRE 250		&										\\
	15.5$\pm$6.3 [10]	&	28.1$\pm$6.3 [10]	&	28.2$\pm$7.4 [10]	&	25.9$\pm$7.4 [10]	&	42.9$\pm$7.4 [10]&	61.4$\pm$7.1 [10]&	30.8$\pm$2.4 [7]	&										\\
						&						&						&						&					&					&					&										\\
	SPIRE 350			&	SPIRE 350			&	SPIRE 350			&	SPIRE 350			&	SPIRE 350		&	SPIRE 350		&	SPIRE 350		&	SHARC2 350							\\
	20$\pm$10. [10]	    &	28$\pm$10. [10]	    &	43.7$\pm$5.4 [10]	&	42.0$\pm$6.6 [10]	&	40.2$\pm$6.6 [10]&	52.8$\pm$6.1 [10]&	15.5$\pm$4.6 [7]	&	42.2$\pm$9.8 [11]							\\
						&						&						&						&					&					&					&											\\
	SPIRE 500			&	SPIRE 500			&	SPIRE 500			&	SPIRE 500			&	SPIRE 500		&	SPIRE 500		&	SPIRE 500		&	SCUBA 450									\\
	11.6	$\pm$5.0 [10]	&	17.4	$\pm$4.9 [10]	&	29.2$\pm$7.1 [10]	&	28.1$\pm$6.9 [10]	&	21.9$\pm$6.9 [10]& 26.3$\pm$6.1 [10]	&	14.9$\pm$8.4 [7]&	$<$63 [12]						\\
						&						&						&						&					&					&					&											\\
	MAMBO 1200			&	MAMBO 1200			&						&						&					&	MAMBO 1200		&					&	SCUBA 850							\\
	2.75$\pm$0.76 [1]	&	2.36$\pm$0.62 [1]	&						&						&					&	2.11$\pm$0.56 [9]&					&	4.4$\pm$1.3 [12]						\\

	\tablecomments{Each entry gives instrument and wave band on the first line, followed by measurement and reference code on the second line. SPIRE 250, 350, and 500 $\mu$m uncertainties are set to 30\% to account for potential confusion noise. In fitting the 24 $\mu$m photometry we used ten times the uncertainty quoted here to compensate for the inflexible treatment of PAH fluxes in the model grid. These photometry are not color corrected. In all cases, color correcting the PACS and SPIRE photometry resulted in changes less than 10\%.  References: [1]\citep{fio2009}, [2]\citep{far2008}, [3]\citep{fad2006}, [4]\citep{efs2009}, [5]\citep{hai2009b}, [6]\citep{saj2007}, [7](Data from Herschel program {\it SEDs and energetics of lensed UV-bright high redshift galaxies} - flux extracted using standard HIPE methods), [8](this work), [9]\citep{saj2008}, [10](HerMES source catalog), [11]\citep{kov2006}, [12]\citep{web2003}.}%\hspace{1.5\textwidth}
\end{tabular}
\end{center}
\label{tab:sourcephot}
\end{table}

\setlength\tabcolsep{1mm}%{1.5mm}
\begin{sidewaystable}[htb]\scriptsize
\caption[Fine structure lines and SED properties.]{Fine structure lines, SED properties, and H$_2$ S(3).}
\vspace{2.5mm}
\begin{center}
\begin{tabular}{ c c c c c c c c c c c c c}
\hline
	
\multirow{2}{*}{Source}	&	\multicolumn{4}{c}{log(L/L$_\odot$) ($\mu$m range)}	&	\multirow{2}{*}{$\frac{F_\nu(60)}{F_\nu(100)}$}	&	\multirow{2}{*}{$\frac{F_\nu(70)}{F_\nu(100)}$}	&		\multicolumn{3}{c}{Line flux (10$^{-18}$ W m$^{-2}$)}	&	\multirow{2}{*}{$\frac{F_{[OI]}}{F_{[CII]}}$}	&	\multirow{2}{*}{$\frac{L_{[CII]}}{L_{FIR}}$} &	\multirow{2}{*}{$\frac{L_{[OI]}}{L_{FIR}}$}	\\
	\cline{2-5} \cline{8-10}
	&	42.5-122.5	&	40-500	&	30-1000		&	8-1000	&		&		&	[CII]	&	[OI]	&	H$_2$S(3)\footnote{H2 S(3) measurements are 3$\sigma$ upper limits based on the noise in MIR IRS spectra at 9.615 $\mu$m.}	&		&		\\
\hline

 MIPS 22530 & 12.69$\pm$0.04 & 12.79$\pm$0.04 & 12.87$\pm$0.04 & 12.96$\pm$0.04 & 0.87$\pm^{0.12}_{0.06}$ & 1.01$\pm^{0.10}_{0.05}$ &         9.3$\pm$ 2.1 &         5.4$\pm$ 1.1 & $<$3.7 &       0.59$\pm$ 0.18 &    (1.4$\pm$0.3)E -2 &    (8.0$\pm$1.8)E -3 \\ 
  SWIRE 3-9 & 12.59$\pm$0.15 & 12.67$\pm$0.13 & 12.75$\pm$0.15 & 12.83$\pm$0.16 & 0.87$\pm^{0.32}_{0.32}$ & 1.01$\pm^{0.26}_{0.29}$ &         6.4$\pm$ 1.4 &         3.8$\pm$ 0.8 & $<$1.9 &       0.59$\pm$ 0.18 &    (8.9$\pm$3.7)E -3 &    (5.2$\pm$2.1)E -3 \\ 
 SWIRE 3-14 & 12.21$\pm$0.13 & 12.45$\pm$0.08 & 12.47$\pm$0.08 & 12.55$\pm$0.08 & 0.43$\pm^{0.15}_{0.13}$ & 0.61$\pm^{0.15}_{0.16}$ &         2.3$\pm$ 0.4 &         1.6$\pm$ 0.5 & $<$3.9 &       0.69$\pm$ 0.24 &    (8.3$\pm$2.9)E -3 &    (5.7$\pm$2.4)E -3 \\ 
 SWIRE 3-18 & 12.13$\pm$0.13 & 12.39$\pm$0.08 & 12.42$\pm$0.09 & 12.49$\pm$0.09 & 0.43$\pm^{0.15}_{0.14}$ & 0.61$\pm^{0.15}_{0.17}$ &         5.5$\pm$ 1.0 &                  --- & $<$2.4 &                  --- &    (2.3$\pm$0.8)E -2 &                  --- \\ 
    SMM J03 & 12.58$\pm$0.03 & 12.66$\pm$0.03 & 12.75$\pm$0.03 & 12.85$\pm$0.03 & 1.09$\pm^{0.10}_{0.11}$ & 1.19$\pm^{0.08}_{0.09}$ &        16.9$\pm$ 3.5 &         6.0$\pm$ 0.9 & $<$5.0 &     0.353$\pm$ 0.089 &    (1.5$\pm$0.3)E -2 &    (5.1$\pm$0.8)E -3 \\ 
  SWIRE 4-5 & 11.73$\pm$0.19 & 12.01$\pm$0.14 & 12.03$\pm$0.15 & 12.11$\pm$0.15 & 0.30$\pm^{0.24}_{0.01}$ & 0.44$\pm^{0.27}_{0.03}$ &         4.5$\pm$ 1.1 &         3.2$\pm$ 1.0 & $<$0.6 &       0.72$\pm$ 0.28 &    (4.7$\pm$2.4)E -2 &    (3.3$\pm$1.8)E -2 \\ 
 SWIRE 4-15 & 12.36$\pm$0.15 & 12.48$\pm$0.12 & 12.53$\pm$0.14 & 12.62$\pm$0.15 & 0.69$\pm^{0.29}_{0.19}$ & 0.85$\pm^{0.25}_{0.18}$ &         5.0$\pm$ 1.3 &         3.6$\pm$ 0.9 & $<$0.6 &       0.71$\pm$ 0.26 &    (1.4$\pm$0.6)E -2 &    (9.9$\pm$4.3)E -3 \\ 
   SDSS J12 & 12.54$\pm$0.03 & 12.62$\pm$0.03 & 12.72$\pm$0.03 & 12.82$\pm$0.03 & 1.23$\pm^{0.08}_{0.20}$ & 1.30$\pm^{0.06}_{0.15}$ &         5.6$\pm$ 1.4 &         2.5$\pm$ 0.5 & $<$0.3 &       0.45$\pm$ 0.14 &    (1.3$\pm$0.3)E -2 &    (5.6$\pm$1.1)E -3 \\

\tablecomments{H2 S(3) measurements are 3$\sigma$ upper limits based on the noise in MIR IRS spectra at 9.615 $\mu$m.}

\end{tabular}
%\tablecomments{H2 S(3) measurements are 3$\sigma$ upper limits based on the noise in MIR IRS spectra at 9.615 $\mu$m.}
\end{center}
%\tablecomments{H2 S(3) measurements are 3$\sigma$ upper limits based on the noise in MIR IRS spectra at 9.615 $\mu$m.}
\label{tab:appsourceprops}
%\tablecomments{H2 S(3) measurements are 3$\sigma$ upper limits based on the noise in MIR IRS spectra at 9.615 $\mu$m.}
\end{sidewaystable}
\setlength\tabcolsep{2.12mm}

\bibliography{mypsrvlb}

\begin{thebibliography}{113}
\expandafter\ifx\csname natexlab\endcsname\relax\def\natexlab#1{#1}\fi

\bibitem[{{Appleton} {et~al.}(2013){Appleton}, {Guillard}, {Boulanger},
  {Cluver}, {Ogle}, {Falgarone}, {Pineau des For{\^e}ts}, {O'Sullivan}, {Duc},
  {Gallagher}, {Gao}, {Jarrett}, {Konstantopoulos}, {Lisenfeld}, {Lord}, {Lu},
  {Peterson}, {Struck}, {Sturm}, {Tuffs}, {Valchanov}, {van der Werf}, \&
  {Xu}}]{app2013}
{Appleton}, P.~N., {Guillard}, P., {Boulanger}, F., {Cluver}, M.~E., {Ogle},
  P., {Falgarone}, E., {Pineau des For{\^e}ts}, G., {O'Sullivan}, E., {Duc},
  P.-A., {Gallagher}, S., {Gao}, Y., {Jarrett}, T., {Konstantopoulos}, I.,
  {Lisenfeld}, U., {Lord}, S., {Lu}, N., {Peterson}, B.~W., {Struck}, C.,
  {Sturm}, E., {Tuffs}, R., {Valchanov}, I., {van der Werf}, P., \& {Xu}, K.~C.
  2013, \apj, 777, 66

\bibitem[{{Armus} {et~al.}(2009){Armus}, {Mazzarella}, {Evans}, {Surace},
  {Sanders}, {Iwasawa}, {Frayer}, {Howell}, {Chan}, {Petric}, {Vavilkin},
  {Kim}, {Haan}, {Inami}, {Murphy}, {Appleton}, {Barnes}, {Bothun}, {Bridge},
  {Charmandaris}, {Jensen}, {Kewley}, {Lord}, {Madore}, {Marshall},
  {Melbourne}, {Rich}, {Satyapal}, {Schulz}, {Spoon}, {Sturm}, {U}, {Veilleux},
  \& {Xu}}]{arm2009}
{Armus}, L., {Mazzarella}, J.~M., {Evans}, A.~S., {Surace}, J.~A., {Sanders},
  D.~B., {Iwasawa}, K., {Frayer}, D.~T., {Howell}, J.~H., {Chan}, B., {Petric},
  A., {Vavilkin}, T., {Kim}, D.~C., {Haan}, S., {Inami}, H., {Murphy}, E.~J.,
  {Appleton}, P.~N., {Barnes}, J.~E., {Bothun}, G., {Bridge}, C.~R.,
  {Charmandaris}, V., {Jensen}, J.~B., {Kewley}, L.~J., {Lord}, S., {Madore},
  B.~F., {Marshall}, J.~A., {Melbourne}, J.~E., {Rich}, J., {Satyapal}, S.,
  {Schulz}, B., {Spoon}, H.~W.~W., {Sturm}, E., {U}, V., {Veilleux}, S., \&
  {Xu}, K. 2009, \pasp, 121, 559

\bibitem[{{Bell} {et~al.}(2003){Bell}, {McIntosh}, {Katz}, \&
  {Weinberg}}]{bel2003}
{Bell}, E.~F., {McIntosh}, D.~H., {Katz}, N., \& {Weinberg}, M.~D. 2003, \apjs,
  149, 289

\bibitem[{{Brandl} {et~al.}(2006){Brandl}, {Bernard-Salas}, {Spoon}, {Devost},
  {Sloan}, {Guilles}, {Wu}, {Houck}, {Weedman}, {Armus}, {Appleton}, {Soifer},
  {Charmandaris}, {Hao}, {Higdon}, {Marshall}, \& {Herter}}]{bra2006}
{Brandl}, B.~R., {Bernard-Salas}, J., {Spoon}, H.~W.~W., {Devost}, D., {Sloan},
  G.~C., {Guilles}, S., {Wu}, Y., {Houck}, J.~R., {Weedman}, D.~W., {Armus},
  L., {Appleton}, P.~N., {Soifer}, B.~T., {Charmandaris}, V., {Hao}, L.,
  {Higdon}, J.~A., {Marshall}, S.~J., \& {Herter}, T.~L. 2006, \apj, 653, 1129

\bibitem[{{Brauher} {et~al.}(2008){Brauher}, {Dale}, \& {Helou}}]{bra2008}
{Brauher}, J.~R., {Dale}, D.~A., \& {Helou}, G. 2008, \apjs, 178, 280

\bibitem[{{Carral} {et~al.}(1994){Carral}, {Hollenbach}, {Lord}, {Colgan},
  {Haas}, {Rubin}, \& {Erickson}}]{car1994}
{Carral}, P., {Hollenbach}, D.~J., {Lord}, S.~D., {Colgan}, S.~W.~J., {Haas},
  M.~R., {Rubin}, R.~H., \& {Erickson}, E.~F. 1994, \apj, 423, 223

\bibitem[{{Chabrier}(2003)}]{cha2003}
{Chabrier}, G. 2003, \apjl, 586, L133

\bibitem[{{Chapman} {et~al.}(2005){Chapman}, {Blain}, {Smail}, \&
  {Ivison}}]{cha2005}
{Chapman}, S.~C., {Blain}, A.~W., {Smail}, I., \& {Ivison}, R.~J. 2005, \apj,
  622, 772

\bibitem[{{Colbert} {et~al.}(1999){Colbert}, {Malkan}, {Clegg}, {Cox},
  {Fischer}, {Lord}, {Luhman}, {Satyapal}, {Smith}, {Spinoglio}, {Stacey}, \&
  {Unger}}]{col1999}
{Colbert}, J.~W., {Malkan}, M.~A., {Clegg}, P.~E., {Cox}, P., {Fischer}, J.,
  {Lord}, S.~D., {Luhman}, M., {Satyapal}, S., {Smith}, H.~A., {Spinoglio}, L.,
  {Stacey}, G., \& {Unger}, S.~J. 1999, \apj, 511, 721

\bibitem[{{Coppin} {et~al.}(2012){Coppin}, {Danielson}, {Geach}, {Hodge},
  {Swinbank}, {Wardlow}, {Bertoldi}, {Biggs}, {Brandt}, {Caselli}, {Chapman},
  {Dannerbauer}, {Dunlop}, {Greve}, {Hamann}, {Ivison}, {Karim}, {Knudsen},
  {Menten}, {Schinnerer}, {Smail}, {Spaans}, {Walter}, {Webb}, \& {van der
  Werf}}]{cop2012}
{Coppin}, K.~E.~K., {Danielson}, A.~L.~R., {Geach}, J.~E., {Hodge}, J.~A.,
  {Swinbank}, A.~M., {Wardlow}, J.~L., {Bertoldi}, F., {Biggs}, A., {Brandt},
  W.~N., {Caselli}, P., {Chapman}, S.~C., {Dannerbauer}, H., {Dunlop}, J.~S.,
  {Greve}, T.~R., {Hamann}, F., {Ivison}, R.~J., {Karim}, A., {Knudsen}, K.~K.,
  {Menten}, K.~M., {Schinnerer}, E., {Smail}, I., {Spaans}, M., {Walter}, F.,
  {Webb}, T.~M.~A., \& {van der Werf}, P.~P. 2012, \mnras, 427, 520

\bibitem[{{Crawford} {et~al.}(1985){Crawford}, {Genzel}, {Townes}, \&
  {Watson}}]{cra1985}
{Crawford}, M.~K., {Genzel}, R., {Townes}, C.~H., \& {Watson}, D.~M. 1985,
  \apj, 291, 755

\bibitem[{{Croxall} {et~al.}(2012){Croxall}, {Smith}, {Wolfire}, {Roussel},
  {Sandstrom}, {Draine}, {Aniano}, {Dale}, {Armus}, {Beir{\~a}o}, {Helou},
  {Bolatto}, {Appleton}, {Brandl}, {Calzetti}, {Crocker}, {Galametz}, {Groves},
  {Hao}, {Hunt}, {Johnson}, {Kennicutt}, {Koda}, {Krause}, {Li}, {Meidt},
  {Murphy}, {Rahman}, {Rix}, {Sauvage}, {Schinnerer}, {Walter}, \&
  {Wilson}}]{cro2012}
{Croxall}, K.~V., {Smith}, J.~D., {Wolfire}, M.~G., {Roussel}, H., {Sandstrom},
  K.~M., {Draine}, B.~T., {Aniano}, G., {Dale}, D.~A., {Armus}, L.,
  {Beir{\~a}o}, P., {Helou}, G., {Bolatto}, A.~D., {Appleton}, P.~N., {Brandl},
  B.~R., {Calzetti}, D., {Crocker}, A.~F., {Galametz}, M., {Groves}, B.~A.,
  {Hao}, C.-N., {Hunt}, L.~K., {Johnson}, B.~D., {Kennicutt}, R.~C., {Koda},
  J., {Krause}, O., {Li}, Y., {Meidt}, S.~E., {Murphy}, E.~J., {Rahman}, N.,
  {Rix}, H.-W., {Sauvage}, M., {Schinnerer}, E., {Walter}, F., \& {Wilson},
  C.~D. 2012, \apj, 747, 81

\bibitem[{{Daddi} {et~al.}(2007){Daddi}, {Alexander}, {Dickinson}, {Gilli},
  {Renzini}, {Elbaz}, {Cimatti}, {Chary}, {Frayer}, {Bauer}, {Brandt},
  {Giavalisco}, {Grogin}, {Huynh}, {Kurk}, {Mignoli}, {Morrison}, {Pope}, \&
  {Ravindranath}}]{dad2007}
{Daddi}, E., {Alexander}, D.~M., {Dickinson}, M., {Gilli}, R., {Renzini}, A.,
  {Elbaz}, D., {Cimatti}, A., {Chary}, R., {Frayer}, D., {Bauer}, F.~E.,
  {Brandt}, W.~N., {Giavalisco}, M., {Grogin}, N.~A., {Huynh}, M., {Kurk}, J.,
  {Mignoli}, M., {Morrison}, G., {Pope}, A., \& {Ravindranath}, S. 2007, \apj,
  670, 173

\bibitem[{{Daddi} {et~al.}(2010){Daddi}, {Bournaud}, {Walter}, {Dannerbauer},
  {Carilli}, {Dickinson}, {Elbaz}, {Morrison}, {Riechers}, {Onodera}, {Salmi},
  {Krips}, \& {Stern}}]{dad2010}
{Daddi}, E., {Bournaud}, F., {Walter}, F., {Dannerbauer}, H., {Carilli}, C.~L.,
  {Dickinson}, M., {Elbaz}, D., {Morrison}, G.~E., {Riechers}, D., {Onodera},
  M., {Salmi}, F., {Krips}, M., \& {Stern}, D. 2010, \apj, 713, 686

\bibitem[{{Dale} \& {Helou}(2002)}]{dal2002}
{Dale}, D.~A., \& {Helou}, G. 2002, \apj, 576, 159

\bibitem[{{Dale} {et~al.}(2004){Dale}, {Helou}, {Brauher}, {Cutri}, {Malhotra},
  \& {Beichman}}]{dal2004}
{Dale}, D.~A., {Helou}, G., {Brauher}, J.~R., {Cutri}, R.~M., {Malhotra}, S.,
  \& {Beichman}, C.~A. 2004, \apj, 604, 565

\bibitem[{{Dalgarno} \& {McCray}(1972)}]{dal1972}
{Dalgarno}, A., \& {McCray}, R.~A. 1972, \araa, 10, 375

\bibitem[{{De Breuck} {et~al.}(2011){De Breuck}, {Maiolino}, {Caselli},
  {Coppin}, {Hailey-Dunsheath}, \& {Nagao}}]{deb2011}
{De Breuck}, C., {Maiolino}, R., {Caselli}, P., {Coppin}, K.,
  {Hailey-Dunsheath}, S., \& {Nagao}, T. 2011, \aap, 530, L8

\bibitem[{{de Jong}(1996)}]{dej1996}
{de Jong}, R.~S. 1996, \aap, 313, 377

\bibitem[{{Decarli} {et~al.}(2014){Decarli}, {Walter}, {Carilli}, {Bertoldi},
  {Cox}, {Ferkinhoff}, {Groves}, {Maiolino}, {Neri}, {Riechers}, \&
  {Weiss}}]{dec2014}
{Decarli}, R., {Walter}, F., {Carilli}, C., {Bertoldi}, F., {Cox}, P.,
  {Ferkinhoff}, C., {Groves}, B., {Maiolino}, R., {Neri}, R., {Riechers}, D.,
  \& {Weiss}, A. 2014, \apjl, 782, L17

\bibitem[{{Desai} {et~al.}(2009){Desai}, {Soifer}, {Dey}, {Le Floc'h}, {Armus},
  {Brand}, {Brown}, {Brodwin}, {Jannuzi}, {Houck}, {Weedman}, {Ashby},
  {Gonzalez}, {Huang}, {Smith}, {Teplitz}, {Willner}, \& {Melbourne}}]{des2009}
{Desai}, V., {Soifer}, B.~T., {Dey}, A., {Le Floc'h}, E., {Armus}, L., {Brand},
  K., {Brown}, M.~J.~I., {Brodwin}, M., {Jannuzi}, B.~T., {Houck}, J.~R.,
  {Weedman}, D.~W., {Ashby}, M.~L.~N., {Gonzalez}, A., {Huang}, J., {Smith},
  H.~A., {Teplitz}, H., {Willner}, S.~P., \& {Melbourne}, J. 2009, \apj, 700,
  1190

\bibitem[{{D{\'{\i}}az-Santos} {et~al.}(2014){D{\'{\i}}az-Santos}, {Armus},
  {Charmandaris}, {Stacey}, {Murphy}, {Haan}, {Stierwalt}, {Malhotra},
  {Appleton}, {Inami}, {Magdis}, {Elbaz}, {Evans}, {Mazzarella}, {Surace}, {van
  der Werf}, {Xu}, {Lu}, {Meijerink}, {Howell}, {Petric}, {Veilleux}, \&
  {Sanders}}]{dia2014}
{D{\'{\i}}az-Santos}, T., {Armus}, L., {Charmandaris}, V., {Stacey}, G.,
  {Murphy}, E.~J., {Haan}, S., {Stierwalt}, S., {Malhotra}, S., {Appleton}, P.,
  {Inami}, H., {Magdis}, G.~E., {Elbaz}, D., {Evans}, A.~S., {Mazzarella},
  J.~M., {Surace}, J.~A., {van der Werf}, P.~P., {Xu}, C.~K., {Lu}, N.,
  {Meijerink}, R., {Howell}, J.~H., {Petric}, A.~O., {Veilleux}, S., \&
  {Sanders}, D.~B. 2014, \apjl, 788, L17

\bibitem[{{D{\'{\i}}az-Santos} {et~al.}(2013){D{\'{\i}}az-Santos}, {Armus},
  {Charmandaris}, {Stierwalt}, {Murphy}, {Haan}, {Inami}, {Malhotra},
  {Meijerink}, {Stacey}, {Petric}, {Evans}, {Veilleux}, {van der Werf}, {Lord},
  {Lu}, {Howell}, {Appleton}, {Mazzarella}, {Surace}, {Xu}, {Schulz},
  {Sanders}, {Bridge}, {Chan}, {Frayer}, {Iwasawa}, {Melbourne}, \&
  {Sturm}}]{dia2013}
{D{\'{\i}}az-Santos}, T., {Armus}, L., {Charmandaris}, V., {Stierwalt}, S.,
  {Murphy}, E.~J., {Haan}, S., {Inami}, H., {Malhotra}, S., {Meijerink}, R.,
  {Stacey}, G., {Petric}, A.~O., {Evans}, A.~S., {Veilleux}, S., {van der
  Werf}, P.~P., {Lord}, S., {Lu}, N., {Howell}, J.~H., {Appleton}, P.,
  {Mazzarella}, J.~M., {Surace}, J.~A., {Xu}, C.~K., {Schulz}, B., {Sanders},
  D.~B., {Bridge}, C., {Chan}, B.~H.~P., {Frayer}, D.~T., {Iwasawa}, K.,
  {Melbourne}, J., \& {Sturm}, E. 2013, \apj, 774, 68

\bibitem[{{Efstathiou} \& {Siebenmorgen}(2009)}]{efs2009}
{Efstathiou}, A., \& {Siebenmorgen}, R. 2009, \aap, 502, 541

\bibitem[{{Elbaz} {et~al.}(2011){Elbaz}, {Dickinson}, {Hwang},
  {D{\'{\i}}az-Santos}, {Magdis}, {Magnelli}, {Le Borgne}, {Galliano},
  {Pannella}, {Chanial}, {Armus}, {Charmandaris}, {Daddi}, {Aussel}, {Popesso},
  {Kartaltepe}, {Altieri}, {Valtchanov}, {Coia}, {Dannerbauer}, {Dasyra},
  {Leiton}, {Mazzarella}, {Alexander}, {Buat}, {Burgarella}, {Chary}, {Gilli},
  {Ivison}, {Juneau}, {Le Floc'h}, {Lutz}, {Morrison}, {Mullaney}, {Murphy},
  {Pope}, {Scott}, {Brodwin}, {Calzetti}, {Cesarsky}, {Charlot}, {Dole},
  {Eisenhardt}, {Ferguson}, {F{\"o}rster Schreiber}, {Frayer}, {Giavalisco},
  {Huynh}, {Koekemoer}, {Papovich}, {Reddy}, {Surace}, {Teplitz}, {Yun}, \&
  {Wilson}}]{elb2011}
{Elbaz}, D., {Dickinson}, M., {Hwang}, H.~S., {D{\'{\i}}az-Santos}, T.,
  {Magdis}, G., {Magnelli}, B., {Le Borgne}, D., {Galliano}, F., {Pannella},
  M., {Chanial}, P., {Armus}, L., {Charmandaris}, V., {Daddi}, E., {Aussel},
  H., {Popesso}, P., {Kartaltepe}, J., {Altieri}, B., {Valtchanov}, I., {Coia},
  D., {Dannerbauer}, H., {Dasyra}, K., {Leiton}, R., {Mazzarella}, J.,
  {Alexander}, D.~M., {Buat}, V., {Burgarella}, D., {Chary}, R.-R., {Gilli},
  R., {Ivison}, R.~J., {Juneau}, S., {Le Floc'h}, E., {Lutz}, D., {Morrison},
  G.~E., {Mullaney}, J.~R., {Murphy}, E., {Pope}, A., {Scott}, D., {Brodwin},
  M., {Calzetti}, D., {Cesarsky}, C., {Charlot}, S., {Dole}, H., {Eisenhardt},
  P., {Ferguson}, H.~C., {F{\"o}rster Schreiber}, N., {Frayer}, D.,
  {Giavalisco}, M., {Huynh}, M., {Koekemoer}, A.~M., {Papovich}, C., {Reddy},
  N., {Surace}, C., {Teplitz}, H., {Yun}, M.~S., \& {Wilson}, G. 2011, \aap,
  533, A119

\bibitem[{{Fadda} {et~al.}(2006){Fadda}, {Marleau}, {Storrie-Lombardi},
  {Makovoz}, {Frayer}, {Appleton}, {Armus}, {Chapman}, {Choi}, {Fang},
  {Heinrichsen}, {Helou}, {Im}, {Lacy}, {Shupe}, {Soifer}, {Squires}, {Surace},
  {Teplitz}, {Wilson}, \& {Yan}}]{fad2006}
{Fadda}, D., {Marleau}, F.~R., {Storrie-Lombardi}, L.~J., {Makovoz}, D.,
  {Frayer}, D.~T., {Appleton}, P.~N., {Armus}, L., {Chapman}, S.~C., {Choi},
  P.~I., {Fang}, F., {Heinrichsen}, I., {Helou}, G., {Im}, M., {Lacy}, M.,
  {Shupe}, D.~L., {Soifer}, B.~T., {Squires}, G.~K., {Surace}, J., {Teplitz},
  H.~I., {Wilson}, G., \& {Yan}, L. 2006, \aj, 131, 2859

\bibitem[{{Fadely} {et~al.}(2010){Fadely}, {Allam}, {Baker}, {Lin}, {Lutz},
  {Shapley}, {Shin}, {Allyn Smith}, {Strauss}, \& {Tucker}}]{fad2010}
{Fadely}, R., {Allam}, S.~S., {Baker}, A.~J., {Lin}, H., {Lutz}, D., {Shapley},
  A.~E., {Shin}, M.-S., {Allyn Smith}, J., {Strauss}, M.~A., \& {Tucker}, D.~L.
  2010, \apj, 723, 729

\bibitem[{{Farrah} {et~al.}(2013){Farrah}, {Lebouteiller}, {Spoon},
  {Bernard-Salas}, {Pearson}, {Rigopoulou}, {Smith}, {Gonz{\'a}lez-Alfonso},
  {Clements}, {Efstathiou}, {Cormier}, {Afonso}, {Petty}, {Harris}, {Hurley},
  {Borys}, {Verma}, {Cooray}, \& {Salvatelli}}]{far2013}
{Farrah}, D., {Lebouteiller}, V., {Spoon}, H.~W.~W., {Bernard-Salas}, J.,
  {Pearson}, C., {Rigopoulou}, D., {Smith}, H.~A., {Gonz{\'a}lez-Alfonso}, E.,
  {Clements}, D.~L., {Efstathiou}, A., {Cormier}, D., {Afonso}, J., {Petty},
  S.~M., {Harris}, K., {Hurley}, P., {Borys}, C., {Verma}, A., {Cooray}, A., \&
  {Salvatelli}, V. 2013, \apj, 776, 38

\bibitem[{{Farrah} {et~al.}(2006){Farrah}, {Lonsdale}, {Borys}, {Fang},
  {Waddington}, {Oliver}, {Rowan-Robinson}, {Babbedge}, {Shupe}, {Polletta},
  {Smith}, \& {Surace}}]{far2006}
{Farrah}, D., {Lonsdale}, C.~J., {Borys}, C., {Fang}, F., {Waddington}, I.,
  {Oliver}, S., {Rowan-Robinson}, M., {Babbedge}, T., {Shupe}, D., {Polletta},
  M., {Smith}, H.~E., \& {Surace}, J. 2006, \apjl, 641, L17

\bibitem[{{Farrah} {et~al.}(2008){Farrah}, {Lonsdale}, {Weedman}, {Spoon},
  {Rowan-Robinson}, {Polletta}, {Oliver}, {Houck}, \& {Smith}}]{far2008}
{Farrah}, D., {Lonsdale}, C.~J., {Weedman}, D.~W., {Spoon}, H.~W.~W.,
  {Rowan-Robinson}, M., {Polletta}, M., {Oliver}, S., {Houck}, J.~R., \&
  {Smith}, H.~E. 2008, \apj, 677, 957

\bibitem[{{Ferkinhoff} {et~al.}(2014){Ferkinhoff}, {Brisbin}, {Parshley},
  {Nikola}, {Stacey}, {Schoenwald}, {Higdon}, {Higdon}, {Verma}, {Riechers},
  {Hailey-Dunsheath}, {Menten}, {G{\"u}sten}, {Wei{\ss}}, {Irwin}, {Cho},
  {Niemack}, {Halpern}, {Amiri}, {Hasselfield}, {Wiebe}, {Ade}, \&
  {Tucker}}]{fer2014}
{Ferkinhoff}, C., {Brisbin}, D., {Parshley}, S., {Nikola}, T., {Stacey}, G.~J.,
  {Schoenwald}, J., {Higdon}, J.~L., {Higdon}, S.~J.~U., {Verma}, A.,
  {Riechers}, D., {Hailey-Dunsheath}, S., {Menten}, K.~M., {G{\"u}sten}, R.,
  {Wei{\ss}}, A., {Irwin}, K., {Cho}, H.~M., {Niemack}, M., {Halpern}, M.,
  {Amiri}, M., {Hasselfield}, M., {Wiebe}, D.~V., {Ade}, P.~A.~R., \& {Tucker},
  C.~E. 2014, \apj, 780, 142

\bibitem[{{Fiolet} {et~al.}(2010){Fiolet}, {Omont}, {Lagache}, {Bertincourt},
  {Fadda}, {Baker}, {Beelen}, {Berta}, {Boulanger}, {Farrah}, {Kov{\'a}cs},
  {Lonsdale}, {Owen}, {Polletta}, {Shupe}, \& {Yan}}]{fio2010}
{Fiolet}, N., {Omont}, A., {Lagache}, G., {Bertincourt}, B., {Fadda}, D.,
  {Baker}, A.~J., {Beelen}, A., {Berta}, S., {Boulanger}, F., {Farrah}, D.,
  {Kov{\'a}cs}, A., {Lonsdale}, C., {Owen}, F., {Polletta}, M., {Shupe}, D., \&
  {Yan}, L. 2010, \aap, 524, A33

\bibitem[{{Fiolet} {et~al.}(2009){Fiolet}, {Omont}, {Polletta}, {Owen},
  {Berta}, {Shupe}, {Siana}, {Lonsdale}, {Strazzullo}, {Pannella}, {Baker},
  {Beelen}, {Biggs}, {De Breuck}, {Farrah}, {Ivison}, {Lagache}, {Lutz},
  {Tacconi}, \& {Zylka}}]{fio2009}
{Fiolet}, N., {Omont}, A., {Polletta}, M., {Owen}, F., {Berta}, S., {Shupe},
  D., {Siana}, B., {Lonsdale}, C., {Strazzullo}, V., {Pannella}, M., {Baker},
  A.~J., {Beelen}, A., {Biggs}, A., {De Breuck}, C., {Farrah}, D., {Ivison},
  R., {Lagache}, G., {Lutz}, D., {Tacconi}, L.~J., \& {Zylka}, R. 2009, \aap,
  508, 117

\bibitem[{{Gallerani} {et~al.}(2012){Gallerani}, {Neri}, {Maiolino},
  {Mart{\'{\i}}n}, {De Breuck}, {Walter}, {Caselli}, {Krips}, {Meneghetti},
  {Nagao}, {Wagg}, \& {Walmsley}}]{gal2012}
{Gallerani}, S., {Neri}, R., {Maiolino}, R., {Mart{\'{\i}}n}, S., {De Breuck},
  C., {Walter}, F., {Caselli}, P., {Krips}, M., {Meneghetti}, M., {Nagao}, T.,
  {Wagg}, J., \& {Walmsley}, M. 2012, \aap, 543, A114

\bibitem[{{Graci{\'a}-Carpio} {et~al.}(2011){Graci{\'a}-Carpio}, {Sturm},
  {Hailey-Dunsheath}, {Fischer}, {Contursi}, {Poglitsch}, {Genzel},
  {Gonz{\'a}lez-Alfonso}, {Sternberg}, {Verma}, {Christopher}, {Davies},
  {Feuchtgruber}, {de Jong}, {Lutz}, \& {Tacconi}}]{gra2011}
{Graci{\'a}-Carpio}, J., {Sturm}, E., {Hailey-Dunsheath}, S., {Fischer}, J.,
  {Contursi}, A., {Poglitsch}, A., {Genzel}, R., {Gonz{\'a}lez-Alfonso}, E.,
  {Sternberg}, A., {Verma}, A., {Christopher}, N., {Davies}, R.,
  {Feuchtgruber}, H., {de Jong}, J.~A., {Lutz}, D., \& {Tacconi}, L.~J. 2011,
  \apjl, 728, L7

\bibitem[{{Griffin} {et~al.}(2010){Griffin}, {Abergel}, {Abreu}, {Ade},
  {Andr{\'e}}, {Augueres}, {Babbedge}, {Bae}, {Baillie}, {Baluteau}, {Barlow},
  {Bendo}, {Benielli}, {Bock}, {Bonhomme}, {Brisbin}, {Brockley-Blatt},
  {Caldwell}, {Cara}, {Castro-Rodriguez}, {Cerulli}, {Chanial}, {Chen},
  {Clark}, {Clements}, {Clerc}, {Coker}, {Communal}, {Conversi}, {Cox},
  {Crumb}, {Cunningham}, {Daly}, {Davis}, {de Antoni}, {Delderfield}, {Devin},
  {di Giorgio}, {Didschuns}, {Dohlen}, {Donati}, {Dowell}, {Dowell}, {Duband},
  {Dumaye}, {Emery}, {Ferlet}, {Ferrand}, {Fontignie}, {Fox}, {Franceschini},
  {Frerking}, {Fulton}, {Garcia}, {Gastaud}, {Gear}, {Glenn}, {Goizel},
  {Griffin}, {Grundy}, {Guest}, {Guillemet}, {Hargrave}, {Harwit}, {Hastings},
  {Hatziminaoglou}, {Herman}, {Hinde}, {Hristov}, {Huang}, {Imhof}, {Isaak},
  {Israelsson}, {Ivison}, {Jennings}, {Kiernan}, {King}, {Lange}, {Latter},
  {Laurent}, {Laurent}, {Leeks}, {Lellouch}, {Levenson}, {Li}, {Li},
  {Lilienthal}, {Lim}, {Liu}, {Lu}, {Madden}, {Mainetti}, {Marliani}, {McKay},
  {Mercier}, {Molinari}, {Morris}, {Moseley}, {Mulder}, {Mur}, {Naylor},
  {Nguyen}, {O'Halloran}, {Oliver}, {Olofsson}, {Olofsson}, {Orfei}, {Page},
  {Pain}, {Panuzzo}, {Papageorgiou}, {Parks}, {Parr-Burman}, {Pearce},
  {Pearson}, {P{\'e}rez-Fournon}, {Pinsard}, {Pisano}, {Podosek}, {Pohlen},
  {Polehampton}, {Pouliquen}, {Rigopoulou}, {Rizzo}, {Roseboom}, {Roussel},
  {Rowan-Robinson}, {Rownd}, {Saraceno}, {Sauvage}, {Savage}, {Savini},
  {Sawyer}, {Scharmberg}, {Schmitt}, {Schneider}, {Schulz}, {Schwartz},
  {Shafer}, {Shupe}, {Sibthorpe}, {Sidher}, {Smith}, {Smith}, {Smith},
  {Spencer}, {Stobie}, {Sudiwala}, {Sukhatme}, {Surace}, {Stevens}, {Swinyard},
  {Trichas}, {Tourette}, {Triou}, {Tseng}, {Tucker}, {Turner}, {Vaccari},
  {Valtchanov}, {Vigroux}, {Virique}, {Voellmer}, {Walker}, {Ward}, {Waskett},
  {Weilert}, {Wesson}, {White}, {Whitehouse}, {Wilson}, {Winter}, {Woodcraft},
  {Wright}, {Xu}, {Zavagno}, {Zemcov}, {Zhang}, \& {Zonca}}]{gri2010}
{Griffin}, M.~J., {Abergel}, A., {Abreu}, A., {Ade}, P.~A.~R., {Andr{\'e}}, P.,
  {Augueres}, J.-L., {Babbedge}, T., {Bae}, Y., {Baillie}, T., {Baluteau},
  J.-P., {Barlow}, M.~J., {Bendo}, G., {Benielli}, D., {Bock}, J.~J.,
  {Bonhomme}, P., {Brisbin}, D., {Brockley-Blatt}, C., {Caldwell}, M., {Cara},
  C., {Castro-Rodriguez}, N., {Cerulli}, R., {Chanial}, P., {Chen}, S.,
  {Clark}, E., {Clements}, D.~L., {Clerc}, L., {Coker}, J., {Communal}, D.,
  {Conversi}, L., {Cox}, P., {Crumb}, D., {Cunningham}, C., {Daly}, F.,
  {Davis}, G.~R., {de Antoni}, P., {Delderfield}, J., {Devin}, N., {di
  Giorgio}, A., {Didschuns}, I., {Dohlen}, K., {Donati}, M., {Dowell}, A.,
  {Dowell}, C.~D., {Duband}, L., {Dumaye}, L., {Emery}, R.~J., {Ferlet}, M.,
  {Ferrand}, D., {Fontignie}, J., {Fox}, M., {Franceschini}, A., {Frerking},
  M., {Fulton}, T., {Garcia}, J., {Gastaud}, R., {Gear}, W.~K., {Glenn}, J.,
  {Goizel}, A., {Griffin}, D.~K., {Grundy}, T., {Guest}, S., {Guillemet}, L.,
  {Hargrave}, P.~C., {Harwit}, M., {Hastings}, P., {Hatziminaoglou}, E.,
  {Herman}, M., {Hinde}, B., {Hristov}, V., {Huang}, M., {Imhof}, P., {Isaak},
  K.~J., {Israelsson}, U., {Ivison}, R.~J., {Jennings}, D., {Kiernan}, B.,
  {King}, K.~J., {Lange}, A.~E., {Latter}, W., {Laurent}, G., {Laurent}, P.,
  {Leeks}, S.~J., {Lellouch}, E., {Levenson}, L., {Li}, B., {Li}, J.,
  {Lilienthal}, J., {Lim}, T., {Liu}, S.~J., {Lu}, N., {Madden}, S.,
  {Mainetti}, G., {Marliani}, P., {McKay}, D., {Mercier}, K., {Molinari}, S.,
  {Morris}, H., {Moseley}, H., {Mulder}, J., {Mur}, M., {Naylor}, D.~A.,
  {Nguyen}, H., {O'Halloran}, B., {Oliver}, S., {Olofsson}, G., {Olofsson},
  H.-G., {Orfei}, R., {Page}, M.~J., {Pain}, I., {Panuzzo}, P., {Papageorgiou},
  A., {Parks}, G., {Parr-Burman}, P., {Pearce}, A., {Pearson}, C.,
  {P{\'e}rez-Fournon}, I., {Pinsard}, F., {Pisano}, G., {Podosek}, J.,
  {Pohlen}, M., {Polehampton}, E.~T., {Pouliquen}, D., {Rigopoulou}, D.,
  {Rizzo}, D., {Roseboom}, I.~G., {Roussel}, H., {Rowan-Robinson}, M., {Rownd},
  B., {Saraceno}, P., {Sauvage}, M., {Savage}, R., {Savini}, G., {Sawyer}, E.,
  {Scharmberg}, C., {Schmitt}, D., {Schneider}, N., {Schulz}, B., {Schwartz},
  A., {Shafer}, R., {Shupe}, D.~L., {Sibthorpe}, B., {Sidher}, S., {Smith}, A.,
  {Smith}, A.~J., {Smith}, D., {Spencer}, L., {Stobie}, B., {Sudiwala}, R.,
  {Sukhatme}, K., {Surace}, C., {Stevens}, J.~A., {Swinyard}, B.~M., {Trichas},
  M., {Tourette}, T., {Triou}, H., {Tseng}, S., {Tucker}, C., {Turner}, A.,
  {Vaccari}, M., {Valtchanov}, I., {Vigroux}, L., {Virique}, E., {Voellmer},
  G., {Walker}, H., {Ward}, R., {Waskett}, T., {Weilert}, M., {Wesson}, R.,
  {White}, G.~J., {Whitehouse}, N., {Wilson}, C.~D., {Winter}, B., {Woodcraft},
  A.~L., {Wright}, G.~S., {Xu}, C.~K., {Zavagno}, A., {Zemcov}, M., {Zhang},
  L., \& {Zonca}, E. 2010, \aap, 518, L3

\bibitem[{{Guillard} {et~al.}(2012){Guillard}, {Boulanger}, {Pineau des
  For{\^e}ts}, {Falgarone}, {Gusdorf}, {Cluver}, {Appleton}, {Lisenfeld},
  {Duc}, {Ogle}, \& {Xu}}]{gui2012}
{Guillard}, P., {Boulanger}, F., {Pineau des For{\^e}ts}, G., {Falgarone}, E.,
  {Gusdorf}, A., {Cluver}, M.~E., {Appleton}, P.~N., {Lisenfeld}, U., {Duc},
  P.-A., {Ogle}, P.~M., \& {Xu}, C.~K. 2012, \apj, 749, 158

\bibitem[{{Hailey-Dunsheath}(2014)}]{hai2014}
{Hailey-Dunsheath}, o. 2014, inprep, 622, 772

\bibitem[{{Hailey-Dunsheath}(2009)}]{hai2009thesis}
{Hailey-Dunsheath}, S. 2009, PhD thesis, Cornell University

\bibitem[{{Hailey-Dunsheath} {et~al.}(2008){Hailey-Dunsheath}, {Nikola},
  {Stacey}, {Oberst}, {Parshley}, {Bradford}, {Ade}, \& {Tucker}}]{hai2008}
{Hailey-Dunsheath}, S., {Nikola}, T., {Stacey}, G.~J., {Oberst}, T.~E.,
  {Parshley}, S.~C., {Bradford}, C.~M., {Ade}, P.~A.~R., \& {Tucker}, C.~E.
  2008, \apjl, 689, L109

\bibitem[{{Hainline} {et~al.}(2009{\natexlab{a}}){Hainline}, {Shapley},
  {Kornei}, {Pettini}, {Buckley-Geer}, {Allam}, \& {Tucker}}]{hai2009}
{Hainline}, K.~N., {Shapley}, A.~E., {Kornei}, K.~A., {Pettini}, M.,
  {Buckley-Geer}, E., {Allam}, S.~S., \& {Tucker}, D.~L. 2009{\natexlab{a}},
  \apj, 701, 52

\bibitem[{{Hainline} {et~al.}(2009{\natexlab{b}}){Hainline}, {Blain}, {Smail},
  {Frayer}, {Chapman}, {Ivison}, \& {Alexander}}]{hai2009b}
{Hainline}, L.~J., {Blain}, A.~W., {Smail}, I., {Frayer}, D.~T., {Chapman},
  S.~C., {Ivison}, R.~J., \& {Alexander}, D.~M. 2009{\natexlab{b}}, \apj, 699,
  1610

\bibitem[{{Hildebrand} {et~al.}(1985){Hildebrand}, {Loewenstein}, {Harper},
  {Orton}, {Keene}, \& {Whitcomb}}]{hil1985}
{Hildebrand}, R.~H., {Loewenstein}, R.~F., {Harper}, D.~A., {Orton}, G.~S.,
  {Keene}, J., \& {Whitcomb}, S.~E. 1985, \icarus, 64, 64

\bibitem[{{Hodge} {et~al.}(2012){Hodge}, {Carilli}, {Walter}, {de Blok},
  {Riechers}, {Daddi}, \& {Lentati}}]{hod2012}
{Hodge}, J.~A., {Carilli}, C.~L., {Walter}, F., {de Blok}, W.~J.~G.,
  {Riechers}, D., {Daddi}, E., \& {Lentati}, L. 2012, \apj, 760, 11

\bibitem[{{Houck} {et~al.}(2004){Houck}, {Roellig}, {Van Cleve}, {Forrest},
  {Herter}, {Lawrence}, {Matthews}, {Reitsema}, {Soifer}, {Watson}, {Weedman},
  {Huisjen}, {Troeltzsch}, {Barry}, {Bernard-Salas}, {Blacken}, {Brandl},
  {Charmandaris}, {Devost}, {Gull}, {Hall}, {Henderson}, {Higdon}, {Pirger},
  {Schoenwald}, {Sloan}, {Uchida}, {Appleton}, {Armus}, {Burgdorf},
  {Fajardo-Acosta}, {Grillmair}, {Ingalls}, {Morris}, \& {Teplitz}}]{hou2004}
{Houck}, J.~R., {Roellig}, T.~L., {Van Cleve}, J., {Forrest}, W.~J., {Herter},
  T.~L., {Lawrence}, C.~R., {Matthews}, K., {Reitsema}, H.~J., {Soifer}, B.~T.,
  {Watson}, D.~M., {Weedman}, D., {Huisjen}, M., {Troeltzsch}, J.~R., {Barry},
  D.~J., {Bernard-Salas}, J., {Blacken}, C., {Brandl}, B.~R., {Charmandaris},
  V., {Devost}, D., {Gull}, G.~E., {Hall}, P., {Henderson}, C.~P., {Higdon},
  S.~J.~U., {Pirger}, B.~E., {Schoenwald}, J., {Sloan}, G.~C., {Uchida}, K.~I.,
  {Appleton}, P.~N., {Armus}, L., {Burgdorf}, M.~J., {Fajardo-Acosta}, S.~B.,
  {Grillmair}, C.~J., {Ingalls}, J.~G., {Morris}, P.~W., \& {Teplitz}, H.~I.
  2004, in Society of Photo-Optical Instrumentation Engineers (SPIE) Conference
  Series, Vol. 5487, Optical, Infrared, and Millimeter Space Telescopes, ed.
  J.~C. {Mather}, 62--76

\bibitem[{{Huang} {et~al.}(2009){Huang}, {Faber}, {Daddi}, {Laird}, {Lai},
  {Omont}, {Wu}, {Younger}, {Bundy}, {Cattaneo}, {Chapman}, {Conselice},
  {Dickinson}, {Egami}, {Fazio}, {Im}, {Koo}, {Le Floc'h}, {Papovich},
  {Rigopoulou}, {Smail}, {Song}, {Van de Werf}, {Webb}, {Willmer}, {Willner},
  \& {Yan}}]{hua2009}
{Huang}, J.-S., {Faber}, S.~M., {Daddi}, E., {Laird}, E.~S., {Lai}, K.,
  {Omont}, A., {Wu}, Y., {Younger}, J.~D., {Bundy}, K., {Cattaneo}, A.,
  {Chapman}, S.~C., {Conselice}, C.~J., {Dickinson}, M., {Egami}, E., {Fazio},
  G.~G., {Im}, M., {Koo}, D., {Le Floc'h}, E., {Papovich}, C., {Rigopoulou},
  D., {Smail}, I., {Song}, M., {Van de Werf}, P.~P., {Webb}, T.~M.~A.,
  {Willmer}, C.~N.~A., {Willner}, S.~P., \& {Yan}, L. 2009, \apj, 700, 183

\bibitem[{{Ivison} {et~al.}(2011){Ivison}, {Papadopoulos}, {Smail}, {Greve},
  {Thomson}, {Xilouris}, \& {Chapman}}]{ivi2011}
{Ivison}, R.~J., {Papadopoulos}, P.~P., {Smail}, I., {Greve}, T.~R., {Thomson},
  A.~P., {Xilouris}, E.~M., \& {Chapman}, S.~C. 2011, \mnras, 412, 1913

\bibitem[{{Ivison} {et~al.}(2010){Ivison}, {Swinbank}, {Swinyard}, {Smail},
  {Pearson}, {Rigopoulou}, {Polehampton}, {Baluteau}, {Barlow}, {Blain},
  {Bock}, {Clements}, {Coppin}, {Cooray}, {Danielson}, {Dwek}, {Edge},
  {Franceschini}, {Fulton}, {Glenn}, {Griffin}, {Isaak}, {Leeks}, {Lim},
  {Naylor}, {Oliver}, {Page}, {P{\'e}rez Fournon}, {Rowan-Robinson}, {Savini},
  {Scott}, {Spencer}, {Valtchanov}, {Vigroux}, \& {Wright}}]{ivi2010}
{Ivison}, R.~J., {Swinbank}, A.~M., {Swinyard}, B., {Smail}, I., {Pearson},
  C.~P., {Rigopoulou}, D., {Polehampton}, E., {Baluteau}, J.-P., {Barlow},
  M.~J., {Blain}, A.~W., {Bock}, J., {Clements}, D.~L., {Coppin}, K., {Cooray},
  A., {Danielson}, A., {Dwek}, E., {Edge}, A.~C., {Franceschini}, A., {Fulton},
  T., {Glenn}, J., {Griffin}, M., {Isaak}, K., {Leeks}, S., {Lim}, T.,
  {Naylor}, D., {Oliver}, S.~J., {Page}, M.~J., {P{\'e}rez Fournon}, I.,
  {Rowan-Robinson}, M., {Savini}, G., {Scott}, D., {Spencer}, L., {Valtchanov},
  I., {Vigroux}, L., \& {Wright}, G.~S. 2010, \aap, 518, L35

\bibitem[{{Joy} {et~al.}(1987){Joy}, {Lester}, \& {Harvey}}]{joy1987}
{Joy}, M., {Lester}, D.~F., \& {Harvey}, P.~M. 1987, \apj, 319, 314

\bibitem[{{Kaufman} {et~al.}(2006){Kaufman}, {Wolfire}, \&
  {Hollenbach}}]{kau2006}
{Kaufman}, M.~J., {Wolfire}, M.~G., \& {Hollenbach}, D.~J. 2006, \apj, 644, 283

\bibitem[{{Kennicutt}(1998)}]{ken1998}
{Kennicutt}, Jr., R.~C. 1998, \apj, 498, 541

\bibitem[{{Kov{\'a}cs} {et~al.}(2006){Kov{\'a}cs}, {Chapman}, {Dowell},
  {Blain}, {Ivison}, {Smail}, \& {Phillips}}]{kov2006}
{Kov{\'a}cs}, A., {Chapman}, S.~C., {Dowell}, C.~D., {Blain}, A.~W., {Ivison},
  R.~J., {Smail}, I., \& {Phillips}, T.~G. 2006, \apj, 650, 592

\bibitem[{{Lebouteiller} {et~al.}(2011){Lebouteiller}, {Barry}, {Spoon},
  {Bernard-Salas}, {Sloan}, {Houck}, \& {Weedman}}]{leb2011}
{Lebouteiller}, V., {Barry}, D.~J., {Spoon}, H.~W.~W., {Bernard-Salas}, J.,
  {Sloan}, G.~C., {Houck}, J.~R., \& {Weedman}, D.~W. 2011, \apjs, 196, 8

\bibitem[{{Lesaffre} {et~al.}(2013){Lesaffre}, {Pineau des For{\^e}ts},
  {Godard}, {Guillard}, {Boulanger}, \& {Falgarone}}]{les2013}
{Lesaffre}, P., {Pineau des For{\^e}ts}, G., {Godard}, B., {Guillard}, P.,
  {Boulanger}, F., \& {Falgarone}, E. 2013, \aap, 550, A106

\bibitem[{{Lin} {et~al.}(2009){Lin}, {Buckley-Geer}, {Allam}, {Tucker},
  {Diehl}, {Kubik}, {Kubo}, {Annis}, {Frieman}, {Oguri}, \& {Inada}}]{lin2009}
{Lin}, H., {Buckley-Geer}, E., {Allam}, S.~S., {Tucker}, D.~L., {Diehl}, H.~T.,
  {Kubik}, D., {Kubo}, J.~M., {Annis}, J., {Frieman}, J.~A., {Oguri}, M., \&
  {Inada}, N. 2009, \apj, 699, 1242

\bibitem[{{Lonsdale} {et~al.}(2009){Lonsdale}, {Polletta}, {Omont}, {Shupe},
  {Berta}, {Zylka}, {Siana}, {Lutz}, {Farrah}, {Smith}, {Lagache}, {DeBreuck},
  {Owen}, {Beelen}, {Weedman}, {Franceschini}, {Clements}, {Tacconi},
  {Afonso-Luis}, {P{\'e}rez-Fournon}, {Cox}, \& {Bertoldi}}]{lon2009}
{Lonsdale}, C.~J., {Polletta}, M.~d.~C., {Omont}, A., {Shupe}, D., {Berta}, S.,
  {Zylka}, R., {Siana}, B., {Lutz}, D., {Farrah}, D., {Smith}, H.~E.,
  {Lagache}, G., {DeBreuck}, C., {Owen}, F., {Beelen}, A., {Weedman}, D.,
  {Franceschini}, A., {Clements}, D., {Tacconi}, L., {Afonso-Luis}, A.,
  {P{\'e}rez-Fournon}, I., {Cox}, P., \& {Bertoldi}, F. 2009, \apj, 692, 422

\bibitem[{{Lonsdale} {et~al.}(2003){Lonsdale}, {Smith}, {Rowan-Robinson},
  {Surace}, {Shupe}, {Xu}, {Oliver}, {Padgett}, {Fang}, {Conrow},
  {Franceschini}, {Gautier}, {Griffin}, {Hacking}, {Masci}, {Morrison},
  {O'Linger}, {Owen}, {P{\'e}rez-Fournon}, {Pierre}, {Puetter}, {Stacey},
  {Castro}, {Polletta}, {Farrah}, {Jarrett}, {Frayer}, {Siana}, {Babbedge},
  {Dye}, {Fox}, {Gonzalez-Solares}, {Salaman}, {Berta}, {Condon}, {Dole}, \&
  {Serjeant}}]{lon2003}
{Lonsdale}, C.~J., {Smith}, H.~E., {Rowan-Robinson}, M., {Surace}, J., {Shupe},
  D., {Xu}, C., {Oliver}, S., {Padgett}, D., {Fang}, F., {Conrow}, T.,
  {Franceschini}, A., {Gautier}, N., {Griffin}, M., {Hacking}, P., {Masci}, F.,
  {Morrison}, G., {O'Linger}, J., {Owen}, F., {P{\'e}rez-Fournon}, I.,
  {Pierre}, M., {Puetter}, R., {Stacey}, G., {Castro}, S., {Polletta},
  M.~d.~C., {Farrah}, D., {Jarrett}, T., {Frayer}, D., {Siana}, B., {Babbedge},
  T., {Dye}, S., {Fox}, M., {Gonzalez-Solares}, E., {Salaman}, M., {Berta}, S.,
  {Condon}, J.~J., {Dole}, H., \& {Serjeant}, S. 2003, \pasp, 115, 897

\bibitem[{{Lord} {et~al.}(1996){Lord}, {Hollenbach}, {Haas}, {Rubin}, {Colgan},
  \& {Erickson}}]{lor1996}
{Lord}, S.~D., {Hollenbach}, D.~J., {Haas}, M.~R., {Rubin}, R.~H., {Colgan},
  S.~W.~J., \& {Erickson}, E.~F. 1996, \apj, 465, 703

\bibitem[{{Luhman} {et~al.}(2003){Luhman}, {Satyapal}, {Fischer}, {Wolfire},
  {Sturm}, {Dudley}, {Lutz}, \& {Genzel}}]{luh2003}
{Luhman}, M.~L., {Satyapal}, S., {Fischer}, J., {Wolfire}, M.~G., {Sturm}, E.,
  {Dudley}, C.~C., {Lutz}, D., \& {Genzel}, R. 2003, \apj, 594, 758

\bibitem[{{Magdis} {et~al.}(2010){Magdis}, {Rigopoulou}, {Huang}, \&
  {Fazio}}]{mag2010}
{Magdis}, G.~E., {Rigopoulou}, D., {Huang}, J.-S., \& {Fazio}, G.~G. 2010,
  \mnras, 401, 1521

\bibitem[{{Maiolino} {et~al.}(2009){Maiolino}, {Caselli}, {Nagao}, {Walmsley},
  {De Breuck}, \& {Meneghetti}}]{mai2009}
{Maiolino}, R., {Caselli}, P., {Nagao}, T., {Walmsley}, M., {De Breuck}, C., \&
  {Meneghetti}, M. 2009, \aap, 500, L1

\bibitem[{{Maiolino} {et~al.}(2005){Maiolino}, {Cox}, {Caselli}, {Beelen},
  {Bertoldi}, {Carilli}, {Kaufman}, {Menten}, {Nagao}, {Omont}, {Wei{\ss}},
  {Walmsley}, \& {Walter}}]{mai2005}
{Maiolino}, R., {Cox}, P., {Caselli}, P., {Beelen}, A., {Bertoldi}, F.,
  {Carilli}, C.~L., {Kaufman}, M.~J., {Menten}, K.~M., {Nagao}, T., {Omont},
  A., {Wei{\ss}}, A., {Walmsley}, C.~M., \& {Walter}, F. 2005, \aap, 440, L51

\bibitem[{{Malhotra} {et~al.}(2001){Malhotra}, {Kaufman}, {Hollenbach},
  {Helou}, {Rubin}, {Brauher}, {Dale}, {Lu}, {Lord}, {Stacey}, {Contursi},
  {Hunter}, \& {Dinerstein}}]{mal2001}
{Malhotra}, S., {Kaufman}, M.~J., {Hollenbach}, D., {Helou}, G., {Rubin},
  R.~H., {Brauher}, J., {Dale}, D., {Lu}, N.~Y., {Lord}, S., {Stacey}, G.,
  {Contursi}, A., {Hunter}, D.~A., \& {Dinerstein}, H. 2001, \apj, 561, 766

\bibitem[{{Marsden} {et~al.}(2005){Marsden}, {Waite}, {Carter}, \&
  {Donati}}]{mar2005}
{Marsden}, S.~C., {Waite}, I.~A., {Carter}, B.~D., \& {Donati}, J.-F. 2005,
  \mnras, 359, 711

\bibitem[{{Men{\'e}ndez-Delmestre} {et~al.}(2009){Men{\'e}ndez-Delmestre},
  {Blain}, {Smail}, {Alexander}, {Chapman}, {Armus}, {Frayer}, {Ivison}, \&
  {Teplitz}}]{men2009}
{Men{\'e}ndez-Delmestre}, K., {Blain}, A.~W., {Smail}, I., {Alexander}, D.~M.,
  {Chapman}, S.~C., {Armus}, L., {Frayer}, D., {Ivison}, R.~J., \& {Teplitz},
  H. 2009, \apj, 699, 667

\bibitem[{{Mookerjea} {et~al.}(2011){Mookerjea}, {Kramer}, {Buchbender},
  {Boquien}, {Verley}, {Rela{\~{n}}o}, {Quintana-Lacaci}, {Aalto}, {Braine},
  {Calzetti}, {Combes}, {Garcia-Burillo}, {Gratier}, {Henkel}, {Israel},
  {Lord}, {Nikola}, {R{\"o}llig}, {Stacey}, {Tabatabaei}, {van der Tak}, \&
  {van der Werf}}]{moo2011}
{Mookerjea}, B., {Kramer}, C., {Buchbender}, C., {Boquien}, M., {Verley}, S.,
  {Rela{\~{n}}o}, M., {Quintana-Lacaci}, G., {Aalto}, S., {Braine}, J.,
  {Calzetti}, D., {Combes}, F., {Garcia-Burillo}, S., {Gratier}, P., {Henkel},
  C., {Israel}, F., {Lord}, S., {Nikola}, T., {R{\"o}llig}, M., {Stacey}, G.,
  {Tabatabaei}, F.~S., {van der Tak}, F., \& {van der Werf}, P. 2011, \aap,
  532, A152

\bibitem[{{Negishi} {et~al.}(2001){Negishi}, {Onaka}, {Chan}, \&
  {Roellig}}]{neg2001}
{Negishi}, T., {Onaka}, T., {Chan}, K.-W., \& {Roellig}, T.~L. 2001, \aap, 375,
  566

\bibitem[{{Nikola} {et~al.}(2001){Nikola}, {Geis}, {Herrmann}, {Madden},
  {Poglitsch}, {Stacey}, \& {Townes}}]{nik2001}
{Nikola}, T., {Geis}, N., {Herrmann}, F., {Madden}, S.~C., {Poglitsch}, A.,
  {Stacey}, G.~J., \& {Townes}, C.~H. 2001, \apj, 561, 203

\bibitem[{{Nikola} {et~al.}(2011){Nikola}, {Stacey}, {Brisbin}, {Ferkinhoff},
  {Hailey-Dunsheath}, {Parshley}, \& {Tucker}}]{nik2011}
{Nikola}, T., {Stacey}, G.~J., {Brisbin}, D., {Ferkinhoff}, C.,
  {Hailey-Dunsheath}, S., {Parshley}, S., \& {Tucker}, C. 2011, \apj, 742, 88

\bibitem[{{Noeske} {et~al.}(2007){Noeske}, {Faber}, {Weiner}, {Koo}, {Primack},
  {Dekel}, {Papovich}, {Conselice}, {Le Floc'h}, {Rieke}, {Coil}, {Lotz},
  {Somerville}, \& {Bundy}}]{noe2007}
{Noeske}, K.~G., {Faber}, S.~M., {Weiner}, B.~J., {Koo}, D.~C., {Primack},
  J.~R., {Dekel}, A., {Papovich}, C., {Conselice}, C.~J., {Le Floc'h}, E.,
  {Rieke}, G.~H., {Coil}, A.~L., {Lotz}, J.~M., {Somerville}, R.~S., \&
  {Bundy}, K. 2007, \apjl, 660, L47

\bibitem[{{Nordon} {et~al.}(2010){Nordon}, {Lutz}, {Shao}, {Magnelli}, {Berta},
  {Altieri}, {Andreani}, {Aussel}, {Bongiovanni}, {Cava}, {Cepa}, {Cimatti},
  {Daddi}, {Dominguez}, {Elbaz}, {F{\"o}rster Schreiber}, {Genzel}, {Grazian},
  {Magdis}, {Maiolino}, {P{\'e}rez Garc{\'{\i}}a}, {Poglitsch}, {Popesso},
  {Pozzi}, {Riguccini}, {Rodighiero}, {Saintonge}, {Sanchez-Portal}, {Santini},
  {Sturm}, {Tacconi}, {Valtchanov}, {Wetzstein}, \& {Wieprecht}}]{nor2010}
{Nordon}, R., {Lutz}, D., {Shao}, L., {Magnelli}, B., {Berta}, S., {Altieri},
  B., {Andreani}, P., {Aussel}, H., {Bongiovanni}, A., {Cava}, A., {Cepa}, J.,
  {Cimatti}, A., {Daddi}, E., {Dominguez}, H., {Elbaz}, D., {F{\"o}rster
  Schreiber}, N.~M., {Genzel}, R., {Grazian}, A., {Magdis}, G., {Maiolino}, R.,
  {P{\'e}rez Garc{\'{\i}}a}, A.~M., {Poglitsch}, A., {Popesso}, P., {Pozzi},
  F., {Riguccini}, L., {Rodighiero}, G., {Saintonge}, A., {Sanchez-Portal}, M.,
  {Santini}, P., {Sturm}, E., {Tacconi}, L., {Valtchanov}, I., {Wetzstein}, M.,
  \& {Wieprecht}, E. 2010, \aap, 518, L24

\bibitem[{{Oberst} {et~al.}(2006){Oberst}, {Parshley}, {Stacey}, {Nikola},
  {L{\"o}hr}, {Harnett}, {Tothill}, {Lane}, {Stark}, \& {Tucker}}]{obe2006}
{Oberst}, T.~E., {Parshley}, S.~C., {Stacey}, G.~J., {Nikola}, T., {L{\"o}hr},
  A., {Harnett}, J.~I., {Tothill}, N.~F.~H., {Lane}, A.~P., {Stark}, A.~A., \&
  {Tucker}, C.~E. 2006, \apjl, 652, L125

\bibitem[{{Oliver} {et~al.}(2012){Oliver}, {Bock}, {Altieri}, {Amblard},
  {Arumugam}, {Aussel}, {Babbedge}, {Beelen}, {B{\'e}thermin}, {Blain},
  {Boselli}, {Bridge}, {Brisbin}, {Buat}, {Burgarella},
  {Castro-Rodr{\'{\i}}guez}, {Cava}, {Chanial}, {Cirasuolo}, {Clements},
  {Conley}, {Conversi}, {Cooray}, {Dowell}, {Dubois}, {Dwek}, {Dye}, {Eales},
  {Elbaz}, {Farrah}, {Feltre}, {Ferrero}, {Fiolet}, {Fox}, {Franceschini},
  {Gear}, {Giovannoli}, {Glenn}, {Gong}, {Gonz{\'a}lez Solares}, {Griffin},
  {Halpern}, {Harwit}, {Hatziminaoglou}, {Heinis}, {Hurley}, {Hwang}, {Hyde},
  {Ibar}, {Ilbert}, {Isaak}, {Ivison}, {Lagache}, {Le Floc'h}, {Levenson},
  {Faro}, {Lu}, {Madden}, {Maffei}, {Magdis}, {Mainetti}, {Marchetti},
  {Marsden}, {Marshall}, {Mortier}, {Nguyen}, {O'Halloran}, {Omont}, {Page},
  {Panuzzo}, {Papageorgiou}, {Patel}, {Pearson}, {P{\'e}rez-Fournon}, {Pohlen},
  {Rawlings}, {Raymond}, {Rigopoulou}, {Riguccini}, {Rizzo}, {Rodighiero},
  {Roseboom}, {Rowan-Robinson}, {S{\'a}nchez Portal}, {Schulz}, {Scott},
  {Seymour}, {Shupe}, {Smith}, {Stevens}, {Symeonidis}, {Trichas}, {Tugwell},
  {Vaccari}, {Valtchanov}, {Vieira}, {Viero}, {Vigroux}, {Wang}, {Ward},
  {Wardlow}, {Wright}, {Xu}, \& {Zemcov}}]{oli2012}
{Oliver}, S.~J., {Bock}, J., {Altieri}, B., {Amblard}, A., {Arumugam}, V.,
  {Aussel}, H., {Babbedge}, T., {Beelen}, A., {B{\'e}thermin}, M., {Blain}, A.,
  {Boselli}, A., {Bridge}, C., {Brisbin}, D., {Buat}, V., {Burgarella}, D.,
  {Castro-Rodr{\'{\i}}guez}, N., {Cava}, A., {Chanial}, P., {Cirasuolo}, M.,
  {Clements}, D.~L., {Conley}, A., {Conversi}, L., {Cooray}, A., {Dowell},
  C.~D., {Dubois}, E.~N., {Dwek}, E., {Dye}, S., {Eales}, S., {Elbaz}, D.,
  {Farrah}, D., {Feltre}, A., {Ferrero}, P., {Fiolet}, N., {Fox}, M.,
  {Franceschini}, A., {Gear}, W., {Giovannoli}, E., {Glenn}, J., {Gong}, Y.,
  {Gonz{\'a}lez Solares}, E.~A., {Griffin}, M., {Halpern}, M., {Harwit}, M.,
  {Hatziminaoglou}, E., {Heinis}, S., {Hurley}, P., {Hwang}, H.~S., {Hyde}, A.,
  {Ibar}, E., {Ilbert}, O., {Isaak}, K., {Ivison}, R.~J., {Lagache}, G., {Le
  Floc'h}, E., {Levenson}, L., {Faro}, B.~L., {Lu}, N., {Madden}, S., {Maffei},
  B., {Magdis}, G., {Mainetti}, G., {Marchetti}, L., {Marsden}, G., {Marshall},
  J., {Mortier}, A.~M.~J., {Nguyen}, H.~T., {O'Halloran}, B., {Omont}, A.,
  {Page}, M.~J., {Panuzzo}, P., {Papageorgiou}, A., {Patel}, H., {Pearson},
  C.~P., {P{\'e}rez-Fournon}, I., {Pohlen}, M., {Rawlings}, J.~I., {Raymond},
  G., {Rigopoulou}, D., {Riguccini}, L., {Rizzo}, D., {Rodighiero}, G.,
  {Roseboom}, I.~G., {Rowan-Robinson}, M., {S{\'a}nchez Portal}, M., {Schulz},
  B., {Scott}, D., {Seymour}, N., {Shupe}, D.~L., {Smith}, A.~J., {Stevens},
  J.~A., {Symeonidis}, M., {Trichas}, M., {Tugwell}, K.~E., {Vaccari}, M.,
  {Valtchanov}, I., {Vieira}, J.~D., {Viero}, M., {Vigroux}, L., {Wang}, L.,
  {Ward}, R., {Wardlow}, J., {Wright}, G., {Xu}, C.~K., \& {Zemcov}, M. 2012,
  \mnras, 424, 1614

\bibitem[{{Ott}(2010)}]{ott2010}
{Ott}, S. 2010, in Astronomical Society of the Pacific Conference Series, Vol.
  434, Astronomical Data Analysis Software and Systems XIX, ed. Y.~{Mizumoto},
  K.-I. {Morita}, \& M.~{Ohishi}, 139

\bibitem[{{Pety} {et~al.}(2004){Pety}, {Beelen}, {Cox}, {Downes}, {Omont},
  {Bertoldi}, \& {Carilli}}]{pet2004}
{Pety}, J., {Beelen}, A., {Cox}, P., {Downes}, D., {Omont}, A., {Bertoldi}, F.,
  \& {Carilli}, C.~L. 2004, \aap, 428, L21

\bibitem[{{Pilbratt} {et~al.}(2010){Pilbratt}, {Riedinger}, {Passvogel},
  {Crone}, {Doyle}, {Gageur}, {Heras}, {Jewell}, {Metcalfe}, {Ott}, \&
  {Schmidt}}]{pil2010}
{Pilbratt}, G.~L., {Riedinger}, J.~R., {Passvogel}, T., {Crone}, G., {Doyle},
  D., {Gageur}, U., {Heras}, A.~M., {Jewell}, C., {Metcalfe}, L., {Ott}, S., \&
  {Schmidt}, M. 2010, \aap, 518, L1

\bibitem[{{Poglitsch} {et~al.}(1995){Poglitsch}, {Krabbe}, {Madden}, {Nikola},
  {Geis}, {Johansson}, {Stacey}, \& {Sternberg}}]{pog1995}
{Poglitsch}, A., {Krabbe}, A., {Madden}, S.~C., {Nikola}, T., {Geis}, N.,
  {Johansson}, L.~E.~B., {Stacey}, G.~J., \& {Sternberg}, A. 1995, \apj, 454,
  293

\bibitem[{{Poglitsch} {et~al.}(2010){Poglitsch}, {Waelkens}, {Geis},
  {Feuchtgruber}, {Vandenbussche}, {Rodriguez}, {Krause}, {Renotte}, {van
  Hoof}, {Saraceno}, {Cepa}, {Kerschbaum}, {Agn{\`e}se}, {Ali}, {Altieri},
  {Andreani}, {Augueres}, {Balog}, {Barl}, {Bauer}, {Belbachir}, {Benedettini},
  {Billot}, {Boulade}, {Bischof}, {Blommaert}, {Callut}, {Cara}, {Cerulli},
  {Cesarsky}, {Contursi}, {Creten}, {De Meester}, {Doublier}, {Doumayrou},
  {Duband}, {Exter}, {Genzel}, {Gillis}, {Gr{\"o}zinger}, {Henning},
  {Herreros}, {Huygen}, {Inguscio}, {Jakob}, {Jamar}, {Jean}, {de Jong},
  {Katterloher}, {Kiss}, {Klaas}, {Lemke}, {Lutz}, {Madden}, {Marquet},
  {Martignac}, {Mazy}, {Merken}, {Montfort}, {Morbidelli}, {M{\"u}ller},
  {Nielbock}, {Okumura}, {Orfei}, {Ottensamer}, {Pezzuto}, {Popesso},
  {Putzeys}, {Regibo}, {Reveret}, {Royer}, {Sauvage}, {Schreiber}, {Stegmaier},
  {Schmitt}, {Schubert}, {Sturm}, {Thiel}, {Tofani}, {Vavrek}, {Wetzstein},
  {Wieprecht}, \& {Wiezorrek}}]{pog2010}
{Poglitsch}, A., {Waelkens}, C., {Geis}, N., {Feuchtgruber}, H.,
  {Vandenbussche}, B., {Rodriguez}, L., {Krause}, O., {Renotte}, E., {van
  Hoof}, C., {Saraceno}, P., {Cepa}, J., {Kerschbaum}, F., {Agn{\`e}se}, P.,
  {Ali}, B., {Altieri}, B., {Andreani}, P., {Augueres}, J.-L., {Balog}, Z.,
  {Barl}, L., {Bauer}, O.~H., {Belbachir}, N., {Benedettini}, M., {Billot}, N.,
  {Boulade}, O., {Bischof}, H., {Blommaert}, J., {Callut}, E., {Cara}, C.,
  {Cerulli}, R., {Cesarsky}, D., {Contursi}, A., {Creten}, Y., {De Meester},
  W., {Doublier}, V., {Doumayrou}, E., {Duband}, L., {Exter}, K., {Genzel}, R.,
  {Gillis}, J.-M., {Gr{\"o}zinger}, U., {Henning}, T., {Herreros}, J.,
  {Huygen}, R., {Inguscio}, M., {Jakob}, G., {Jamar}, C., {Jean}, C., {de
  Jong}, J., {Katterloher}, R., {Kiss}, C., {Klaas}, U., {Lemke}, D., {Lutz},
  D., {Madden}, S., {Marquet}, B., {Martignac}, J., {Mazy}, A., {Merken}, P.,
  {Montfort}, F., {Morbidelli}, L., {M{\"u}ller}, T., {Nielbock}, M.,
  {Okumura}, K., {Orfei}, R., {Ottensamer}, R., {Pezzuto}, S., {Popesso}, P.,
  {Putzeys}, J., {Regibo}, S., {Reveret}, V., {Royer}, P., {Sauvage}, M.,
  {Schreiber}, J., {Stegmaier}, J., {Schmitt}, D., {Schubert}, J., {Sturm}, E.,
  {Thiel}, M., {Tofani}, G., {Vavrek}, R., {Wetzstein}, M., {Wieprecht}, E., \&
  {Wiezorrek}, E. 2010, \aap, 518, L2

\bibitem[{{Pope} {et~al.}(2008){Pope}, {Chary}, {Alexander}, {Armus},
  {Dickinson}, {Elbaz}, {Frayer}, {Scott}, \& {Teplitz}}]{pop2008}
{Pope}, A., {Chary}, R.-R., {Alexander}, D.~M., {Armus}, L., {Dickinson}, M.,
  {Elbaz}, D., {Frayer}, D., {Scott}, D., \& {Teplitz}, H. 2008, \apj, 675,
  1171

\bibitem[{{Pope} {et~al.}(2006){Pope}, {Scott}, {Dickinson}, {Chary},
  {Morrison}, {Borys}, {Sajina}, {Alexander}, {Daddi}, {Frayer}, {MacDonald},
  \& {Stern}}]{pop2006}
{Pope}, A., {Scott}, D., {Dickinson}, M., {Chary}, R.-R., {Morrison}, G.,
  {Borys}, C., {Sajina}, A., {Alexander}, D.~M., {Daddi}, E., {Frayer}, D.,
  {MacDonald}, E., \& {Stern}, D. 2006, \mnras, 370, 1185

\bibitem[{{Pound} \& {Wolfire}(2008)}]{pou2008}
{Pound}, M.~W., \& {Wolfire}, M.~G. 2008, in Astronomical Society of the
  Pacific Conference Series, Vol. 394, Astronomical Data Analysis Software and
  Systems XVII, ed. R.~W. {Argyle}, P.~S. {Bunclark}, \& J.~R. {Lewis}, 654

\bibitem[{{Riechers} {et~al.}(2013){Riechers}, {Bradford}, {Clements},
  {Dowell}, {P{\'e}rez-Fournon}, {Ivison}, {Bridge}, {Conley}, {Fu}, {Vieira},
  {Wardlow}, {Calanog}, {Cooray}, {Hurley}, {Neri}, {Kamenetzky}, {Aguirre},
  {Altieri}, {Arumugam}, {Benford}, {B{\'e}thermin}, {Bock}, {Burgarella},
  {Cabrera-Lavers}, {Chapman}, {Cox}, {Dunlop}, {Earle}, {Farrah}, {Ferrero},
  {Franceschini}, {Gavazzi}, {Glenn}, {Solares}, {Gurwell}, {Halpern},
  {Hatziminaoglou}, {Hyde}, {Ibar}, {Kov{\'a}cs}, {Krips}, {Lupu}, {Maloney},
  {Martinez-Navajas}, {Matsuhara}, {Murphy}, {Naylor}, {Nguyen}, {Oliver},
  {Omont}, {Page}, {Petitpas}, {Rangwala}, {Roseboom}, {Scott}, {Smith},
  {Staguhn}, {Streblyanska}, {Thomson}, {Valtchanov}, {Viero}, {Wang},
  {Zemcov}, \& {Zmuidzinas}}]{rie2013}
{Riechers}, D.~A., {Bradford}, C.~M., {Clements}, D.~L., {Dowell}, C.~D.,
  {P{\'e}rez-Fournon}, I., {Ivison}, R.~J., {Bridge}, C., {Conley}, A., {Fu},
  H., {Vieira}, J.~D., {Wardlow}, J., {Calanog}, J., {Cooray}, A., {Hurley},
  P., {Neri}, R., {Kamenetzky}, J., {Aguirre}, J.~E., {Altieri}, B.,
  {Arumugam}, V., {Benford}, D.~J., {B{\'e}thermin}, M., {Bock}, J.,
  {Burgarella}, D., {Cabrera-Lavers}, A., {Chapman}, S.~C., {Cox}, P.,
  {Dunlop}, J.~S., {Earle}, L., {Farrah}, D., {Ferrero}, P., {Franceschini},
  A., {Gavazzi}, R., {Glenn}, J., {Solares}, E.~A.~G., {Gurwell}, M.~A.,
  {Halpern}, M., {Hatziminaoglou}, E., {Hyde}, A., {Ibar}, E., {Kov{\'a}cs},
  A., {Krips}, M., {Lupu}, R.~E., {Maloney}, P.~R., {Martinez-Navajas}, P.,
  {Matsuhara}, H., {Murphy}, E.~J., {Naylor}, B.~J., {Nguyen}, H.~T., {Oliver},
  S.~J., {Omont}, A., {Page}, M.~J., {Petitpas}, G., {Rangwala}, N.,
  {Roseboom}, I.~G., {Scott}, D., {Smith}, A.~J., {Staguhn}, J.~G.,
  {Streblyanska}, A., {Thomson}, A.~P., {Valtchanov}, I., {Viero}, M., {Wang},
  L., {Zemcov}, M., \& {Zmuidzinas}, J. 2013, \nat, 496, 329

\bibitem[{{Riechers} {et~al.}(2011){Riechers}, {Hodge}, {Walter}, {Carilli}, \&
  {Bertoldi}}]{rie2011}
{Riechers}, D.~A., {Hodge}, J., {Walter}, F., {Carilli}, C.~L., \& {Bertoldi},
  F. 2011, \apjl, 739, L31

\bibitem[{{Rujopakarn} {et~al.}(2012){Rujopakarn}, {Rieke}, {Papovich},
  {Weiner}, {Rigby}, {Rex}, {Bian}, {Kuhn}, \& {Thompson}}]{ruj2012}
{Rujopakarn}, W., {Rieke}, G.~H., {Papovich}, C.~J., {Weiner}, B.~J., {Rigby},
  J.~R., {Rex}, M., {Bian}, F., {Kuhn}, O.~P., \& {Thompson}, D. 2012, \apj,
  755, 168

\bibitem[{{Sajina} {et~al.}(2007){Sajina}, {Yan}, {Armus}, {Choi}, {Fadda},
  {Helou}, \& {Spoon}}]{saj2007}
{Sajina}, A., {Yan}, L., {Armus}, L., {Choi}, P., {Fadda}, D., {Helou}, G., \&
  {Spoon}, H. 2007, \apj, 664, 713

\bibitem[{{Sajina} {et~al.}(2008){Sajina}, {Yan}, {Lutz}, {Steffen}, {Helou},
  {Huynh}, {Frayer}, {Choi}, {Tacconi}, \& {Dasyra}}]{saj2008}
{Sajina}, A., {Yan}, L., {Lutz}, D., {Steffen}, A., {Helou}, G., {Huynh}, M.,
  {Frayer}, D., {Choi}, P., {Tacconi}, L., \& {Dasyra}, K. 2008, \apj, 683, 659

\bibitem[{{Schmidt}(1959)}]{sch1959}
{Schmidt}, M. 1959, \apj, 129, 243

\bibitem[{{Serjeant}(2012)}]{ser2012}
{Serjeant}, S. 2012, \mnras, 424, 2429

\bibitem[{{Smith} \& {Madden}(1997)}]{smi1997}
{Smith}, B.~J., \& {Madden}, S.~C. 1997, \aj, 114, 138

\bibitem[{{Smith} {et~al.}(2007){Smith}, {Draine}, {Dale}, {Moustakas},
  {Kennicutt}, {Helou}, {Armus}, {Roussel}, {Sheth}, {Bendo}, {Buckalew},
  {Calzetti}, {Engelbracht}, {Gordon}, {Hollenbach}, {Li}, {Malhotra},
  {Murphy}, \& {Walter}}]{smi2007}
{Smith}, J.~D.~T., {Draine}, B.~T., {Dale}, D.~A., {Moustakas}, J.,
  {Kennicutt}, Jr., R.~C., {Helou}, G., {Armus}, L., {Roussel}, H., {Sheth},
  K., {Bendo}, G.~J., {Buckalew}, B.~A., {Calzetti}, D., {Engelbracht}, C.~W.,
  {Gordon}, K.~D., {Hollenbach}, D.~J., {Li}, A., {Malhotra}, S., {Murphy},
  E.~J., \& {Walter}, F. 2007, \apj, 656, 770

\bibitem[{{Stacey} {et~al.}(2010{\natexlab{a}}){Stacey}, {Charmandaris},
  {Boulanger}, {Wu}, {Combes}, {Higdon}, {Smith}, \& {Nikola}}]{sta2010b}
{Stacey}, G.~J., {Charmandaris}, V., {Boulanger}, F., {Wu}, Y., {Combes}, F.,
  {Higdon}, S.~J.~U., {Smith}, J.~D.~T., \& {Nikola}, T. 2010{\natexlab{a}},
  \apj, 721, 59

\bibitem[{{Stacey} {et~al.}(1991){Stacey}, {Geis}, {Genzel}, {Lugten},
  {Poglitsch}, {Sternberg}, \& {Townes}}]{sta1991}
{Stacey}, G.~J., {Geis}, N., {Genzel}, R., {Lugten}, J.~B., {Poglitsch}, A.,
  {Sternberg}, A., \& {Townes}, C.~H. 1991, \apj, 373, 423

\bibitem[{{Stacey} {et~al.}(2010{\natexlab{b}}){Stacey}, {Hailey-Dunsheath},
  {Ferkinhoff}, {Nikola}, {Parshley}, {Benford}, {Staguhn}, \&
  {Fiolet}}]{sta2010}
{Stacey}, G.~J., {Hailey-Dunsheath}, S., {Ferkinhoff}, C., {Nikola}, T.,
  {Parshley}, S.~C., {Benford}, D.~J., {Staguhn}, J.~G., \& {Fiolet}, N.
  2010{\natexlab{b}}, \apj, 724, 957

\bibitem[{{Stacey} {et~al.}(2007){Stacey}, {Hailey-Dunsheath}, {Nikola},
  {Oberst}, {Parshley}, {Benford}, {Staguhn}, {Moseley}, \& {Tucker}}]{sta2007}
{Stacey}, G.~J., {Hailey-Dunsheath}, S., {Nikola}, T., {Oberst}, T.~E.,
  {Parshley}, S.~C., {Benford}, D.~J., {Staguhn}, J.~G., {Moseley}, S.~H., \&
  {Tucker}, C. 2007, in Astronomical Society of the Pacific Conference Series,
  Vol. 375, From Z-Machines to ALMA: (Sub)Millimeter Spectroscopy of Galaxies,
  ed. A.~J. {Baker}, J.~{Glenn}, A.~I. {Harris}, J.~G. {Mangum}, \& M.~S.
  {Yun}, 52

\bibitem[{{Stacey} {et~al.}(1983){Stacey}, {Smyers}, {Kurtz}, \&
  {Harwit}}]{sta1983}
{Stacey}, G.~J., {Smyers}, S.~D., {Kurtz}, N.~T., \& {Harwit}, M. 1983, \apjl,
  268, L99

\bibitem[{{Stierwalt} {et~al.}(2013){Stierwalt}, {Armus}, {Surace}, {Inami},
  {Petric}, {Diaz-Santos}, {Haan}, {Charmandaris}, {Howell}, {Kim}, {Marshall},
  {Mazzarella}, {Spoon}, {Veilleux}, {Evans}, {Sanders}, {Appleton}, {Bothun},
  {Bridge}, {Chan}, {Frayer}, {Iwasawa}, {Kewley}, {Lord}, {Madore},
  {Melbourne}, {Murphy}, {Rich}, {Schulz}, {Sturm}, {Vavilkin}, \&
  {Xu}}]{sti2013}
{Stierwalt}, S., {Armus}, L., {Surace}, J.~A., {Inami}, H., {Petric}, A.~O.,
  {Diaz-Santos}, T., {Haan}, S., {Charmandaris}, V., {Howell}, J., {Kim},
  D.~C., {Marshall}, J., {Mazzarella}, J.~M., {Spoon}, H.~W.~W., {Veilleux},
  S., {Evans}, A., {Sanders}, D.~B., {Appleton}, P., {Bothun}, G., {Bridge},
  C.~R., {Chan}, B., {Frayer}, D., {Iwasawa}, K., {Kewley}, L.~J., {Lord}, S.,
  {Madore}, B.~F., {Melbourne}, J.~E., {Murphy}, E.~J., {Rich}, J.~A.,
  {Schulz}, B., {Sturm}, E., {Vavilkin}, T., \& {Xu}, K. 2013, \apjs, 206, 1

\bibitem[{{Sturm} {et~al.}(2010){Sturm}, {Verma}, {Graci{\'a}-Carpio},
  {Hailey-Dunsheath}, {Contursi}, {Fischer}, {Gonz{\'a}lez-Alfonso},
  {Poglitsch}, {Sternberg}, {Genzel}, {Lutz}, {Tacconi}, {Christopher}, \& {de
  Jong}}]{stu2010}
{Sturm}, E., {Verma}, A., {Graci{\'a}-Carpio}, J., {Hailey-Dunsheath}, S.,
  {Contursi}, A., {Fischer}, J., {Gonz{\'a}lez-Alfonso}, E., {Poglitsch}, A.,
  {Sternberg}, A., {Genzel}, R., {Lutz}, D., {Tacconi}, L., {Christopher}, N.,
  \& {de Jong}, J. 2010, \aap, 518, L36

\bibitem[{{Swinbank} {et~al.}(2012){Swinbank}, {Karim}, {Smail}, {Hodge},
  {Walter}, {Bertoldi}, {Biggs}, {de Breuck}, {Chapman}, {Coppin}, {Cox},
  {Danielson}, {Dannerbauer}, {Ivison}, {Greve}, {Knudsen}, {Menten},
  {Simpson}, {Schinnerer}, {Wardlow}, {Wei{\ss}}, \& {van der Werf}}]{swi2012}
{Swinbank}, A.~M., {Karim}, A., {Smail}, I., {Hodge}, J., {Walter}, F.,
  {Bertoldi}, F., {Biggs}, A.~D., {de Breuck}, C., {Chapman}, S.~C., {Coppin},
  K.~E.~K., {Cox}, P., {Danielson}, A.~L.~R., {Dannerbauer}, H., {Ivison},
  R.~J., {Greve}, T.~R., {Knudsen}, K.~K., {Menten}, K.~M., {Simpson}, J.~M.,
  {Schinnerer}, E., {Wardlow}, J.~L., {Wei{\ss}}, A., \& {van der Werf}, P.
  2012, \mnras, 427, 1066

\bibitem[{{Swinbank} {et~al.}(2004){Swinbank}, {Smail}, {Chapman}, {Blain},
  {Ivison}, \& {Keel}}]{swi2004}
{Swinbank}, A.~M., {Smail}, I., {Chapman}, S.~C., {Blain}, A.~W., {Ivison},
  R.~J., \& {Keel}, W.~C. 2004, \apj, 617, 64

\bibitem[{{Symeonidis} {et~al.}(2013){Symeonidis}, {Vaccari}, {Berta}, {Page},
  {Lutz}, {Arumugam}, {Aussel}, {Bock}, {Boselli}, {Buat}, {Capak}, {Clements},
  {Conley}, {Conversi}, {Cooray}, {Dowell}, {Farrah}, {Franceschini},
  {Giovannoli}, {Glenn}, {Griffin}, {Hatziminaoglou}, {Hwang}, {Ibar},
  {Ilbert}, {Ivison}, {Floc'h}, {Lilly}, {Kartaltepe}, {Magnelli}, {Magdis},
  {Marchetti}, {Nguyen}, {Nordon}, {O'Halloran}, {Oliver}, {Omont},
  {Papageorgiou}, {Patel}, {Pearson}, {P{\'e}rez-Fournon}, {Pohlen}, {Popesso},
  {Pozzi}, {Rigopoulou}, {Riguccini}, {Rosario}, {Roseboom}, {Rowan-Robinson},
  {Salvato}, {Schulz}, {Scott}, {Seymour}, {Shupe}, {Smith}, {Valtchanov},
  {Wang}, {Xu}, {Zemcov}, \& {Wuyts}}]{sym2013}
{Symeonidis}, M., {Vaccari}, M., {Berta}, S., {Page}, M.~J., {Lutz}, D.,
  {Arumugam}, V., {Aussel}, H., {Bock}, J., {Boselli}, A., {Buat}, V., {Capak},
  P.~L., {Clements}, D.~L., {Conley}, A., {Conversi}, L., {Cooray}, A.,
  {Dowell}, C.~D., {Farrah}, D., {Franceschini}, A., {Giovannoli}, E., {Glenn},
  J., {Griffin}, M., {Hatziminaoglou}, E., {Hwang}, H.-S., {Ibar}, E.,
  {Ilbert}, O., {Ivison}, R.~J., {Floc'h}, E.~L., {Lilly}, S., {Kartaltepe},
  J.~S., {Magnelli}, B., {Magdis}, G., {Marchetti}, L., {Nguyen}, H.~T.,
  {Nordon}, R., {O'Halloran}, B., {Oliver}, S.~J., {Omont}, A., {Papageorgiou},
  A., {Patel}, H., {Pearson}, C.~P., {P{\'e}rez-Fournon}, I., {Pohlen}, M.,
  {Popesso}, P., {Pozzi}, F., {Rigopoulou}, D., {Riguccini}, L., {Rosario}, D.,
  {Roseboom}, I.~G., {Rowan-Robinson}, M., {Salvato}, M., {Schulz}, B.,
  {Scott}, D., {Seymour}, N., {Shupe}, D.~L., {Smith}, A.~J., {Valtchanov}, I.,
  {Wang}, L., {Xu}, C.~K., {Zemcov}, M., \& {Wuyts}, S. 2013, \mnras, 431, 2317

\bibitem[{{Tacconi} {et~al.}(2010){Tacconi}, {Genzel}, {Neri}, {Cox}, {Cooper},
  {Shapiro}, {Bolatto}, {Bouch{\'e}}, {Bournaud}, {Burkert}, {Combes},
  {Comerford}, {Davis}, {Schreiber}, {Garcia-Burillo}, {Gracia-Carpio}, {Lutz},
  {Naab}, {Omont}, {Shapley}, {Sternberg}, \& {Weiner}}]{tac2010}
{Tacconi}, L.~J., {Genzel}, R., {Neri}, R., {Cox}, P., {Cooper}, M.~C.,
  {Shapiro}, K., {Bolatto}, A., {Bouch{\'e}}, N., {Bournaud}, F., {Burkert},
  A., {Combes}, F., {Comerford}, J., {Davis}, M., {Schreiber}, N.~M.~F.,
  {Garcia-Burillo}, S., {Gracia-Carpio}, J., {Lutz}, D., {Naab}, T., {Omont},
  A., {Shapley}, A., {Sternberg}, A., \& {Weiner}, B. 2010, \nat, 463, 781

\bibitem[{{Takata} {et~al.}(2006){Takata}, {Sekiguchi}, {Smail}, {Chapman},
  {Geach}, {Swinbank}, {Blain}, \& {Ivison}}]{tak2006}
{Takata}, T., {Sekiguchi}, K., {Smail}, I., {Chapman}, S.~C., {Geach}, J.~E.,
  {Swinbank}, A.~M., {Blain}, A., \& {Ivison}, R.~J. 2006, \apj, 651, 713

\bibitem[{{Tielens} \& {Hollenbach}(1985)}]{tie1985}
{Tielens}, A.~G.~G.~M., \& {Hollenbach}, D. 1985, \apj, 291, 722

\bibitem[{{Valtchanov} {et~al.}(2011){Valtchanov}, {Virdee}, {Ivison},
  {Swinyard}, {van der Werf}, {Rigopoulou}, {da Cunha}, {Lupu}, {Benford},
  {Riechers}, {Smail}, {Jarvis}, {Pearson}, {Gomez}, {Hopwood}, {Altieri},
  {Birkinshaw}, {Coia}, {Conversi}, {Cooray}, {de Zotti}, {Dunne}, {Frayer},
  {Leeuw}, {Marston}, {Negrello}, {Portal}, {Scott}, {Thompson}, {Vaccari},
  {Baes}, {Clements}, {Micha{\l}owski}, {Dannerbauer}, {Serjeant}, {Auld},
  {Buttiglione}, {Cava}, {Dariush}, {Dye}, {Eales}, {Fritz}, {Ibar}, {Maddox},
  {Pascale}, {Pohlen}, {Rigby}, {Rodighiero}, {Smith}, {Temi}, {Carpenter},
  {Bolatto}, {Gurwell}, \& {Vieira}}]{val2011}
{Valtchanov}, I., {Virdee}, J., {Ivison}, R.~J., {Swinyard}, B., {van der
  Werf}, P., {Rigopoulou}, D., {da Cunha}, E., {Lupu}, R., {Benford}, D.~J.,
  {Riechers}, D., {Smail}, I., {Jarvis}, M., {Pearson}, C., {Gomez}, H.,
  {Hopwood}, R., {Altieri}, B., {Birkinshaw}, M., {Coia}, D., {Conversi}, L.,
  {Cooray}, A., {de Zotti}, G., {Dunne}, L., {Frayer}, D., {Leeuw}, L.,
  {Marston}, A., {Negrello}, M., {Portal}, M.~S., {Scott}, D., {Thompson},
  M.~A., {Vaccari}, M., {Baes}, M., {Clements}, D., {Micha{\l}owski}, M.~J.,
  {Dannerbauer}, H., {Serjeant}, S., {Auld}, R., {Buttiglione}, S., {Cava}, A.,
  {Dariush}, A., {Dye}, S., {Eales}, S., {Fritz}, J., {Ibar}, E., {Maddox}, S.,
  {Pascale}, E., {Pohlen}, M., {Rigby}, E., {Rodighiero}, G., {Smith},
  D.~J.~B., {Temi}, P., {Carpenter}, J., {Bolatto}, A., {Gurwell}, M., \&
  {Vieira}, J.~D. 2011, \mnras, 415, 3473

\bibitem[{{Vasta} {et~al.}(2010){Vasta}, {Barlow}, {Viti}, {Yates}, \&
  {Bell}}]{vas2010}
{Vasta}, M., {Barlow}, M.~J., {Viti}, S., {Yates}, J.~A., \& {Bell}, T.~A.
  2010, \mnras, 404, 1910

\bibitem[{{Venemans} {et~al.}(2012){Venemans}, {McMahon}, {Walter}, {Decarli},
  {Cox}, {Neri}, {Hewett}, {Mortlock}, {Simpson}, \& {Warren}}]{ven2012}
{Venemans}, B.~P., {McMahon}, R.~G., {Walter}, F., {Decarli}, R., {Cox}, P.,
  {Neri}, R., {Hewett}, P., {Mortlock}, D.~J., {Simpson}, C., \& {Warren},
  S.~J. 2012, \apjl, 751, L25

\bibitem[{{Wagg} {et~al.}(2012){Wagg}, {Wiklind}, {Carilli}, {Espada}, {Peck},
  {Riechers}, {Walter}, {Wootten}, {Aravena}, {Barkats}, {Cortes}, {Hills},
  {Hodge}, {Impellizzeri}, {Iono}, {Leroy}, {Mart{\'{\i}}n}, {Rawlings},
  {Maiolino}, {McMahon}, {Scott}, {Villard}, \& {Vlahakis}}]{wag2012}
{Wagg}, J., {Wiklind}, T., {Carilli}, C.~L., {Espada}, D., {Peck}, A.,
  {Riechers}, D., {Walter}, F., {Wootten}, A., {Aravena}, M., {Barkats}, D.,
  {Cortes}, J.~R., {Hills}, R., {Hodge}, J., {Impellizzeri}, C.~M.~V., {Iono},
  D., {Leroy}, A., {Mart{\'{\i}}n}, S., {Rawlings}, M.~G., {Maiolino}, R.,
  {McMahon}, R.~G., {Scott}, K.~S., {Villard}, E., \& {Vlahakis}, C. 2012,
  \apjl, 752, L30

\bibitem[{{Wang} {et~al.}(2013){Wang}, {Wagg}, {Carilli}, {Walter}, {Lentati},
  {Fan}, {Riechers}, {Bertoldi}, {Narayanan}, {Strauss}, {Cox}, {Omont},
  {Menten}, {Knudsen}, {Neri}, \& {Jiang}}]{wan2013b}
{Wang}, R., {Wagg}, J., {Carilli}, C.~L., {Walter}, F., {Lentati}, L., {Fan},
  X., {Riechers}, D.~A., {Bertoldi}, F., {Narayanan}, D., {Strauss}, M.~A.,
  {Cox}, P., {Omont}, A., {Menten}, K.~M., {Knudsen}, K.~K., {Neri}, R., \&
  {Jiang}, L. 2013, \apj, 773, 44

\bibitem[{{Webb} {et~al.}(2003){Webb}, {Eales}, {Lilly}, {Clements}, {Dunne},
  {Gear}, {Ivison}, {Flores}, \& {Yun}}]{web2003}
{Webb}, T.~M., {Eales}, S.~A., {Lilly}, S.~J., {Clements}, D.~L., {Dunne}, L.,
  {Gear}, W.~K., {Ivison}, R.~J., {Flores}, H., \& {Yun}, M. 2003, \apj, 587,
  41

\bibitem[{{Weingartner} \& {Draine}(2001)}]{wei2001}
{Weingartner}, J.~C., \& {Draine}, B.~T. 2001, \apjs, 134, 263

\bibitem[{{Willott} {et~al.}(2013){Willott}, {Omont}, \& {Bergeron}}]{wil2013}
{Willott}, C.~J., {Omont}, A., \& {Bergeron}, J. 2013, \apj, 770, 13

\bibitem[{{Wolfire} {et~al.}(1990){Wolfire}, {Tielens}, \&
  {Hollenbach}}]{wol1990}
{Wolfire}, M.~G., {Tielens}, A.~G.~G.~M., \& {Hollenbach}, D. 1990, \apj, 358,
  116

\bibitem[{{Yan} {et~al.}(2007){Yan}, {Sajina}, {Fadda}, {Choi}, {Armus},
  {Helou}, {Teplitz}, {Frayer}, \& {Surace}}]{yan2007}
{Yan}, L., {Sajina}, A., {Fadda}, D., {Choi}, P., {Armus}, L., {Helou}, G.,
  {Teplitz}, H., {Frayer}, D., \& {Surace}, J. 2007, \apj, 658, 778

\end{thebibliography}

\end{document}